\documentclass[aps,twocolumn,nofootinbib,showpacs,superscriptaddress,hyperref]{revtex4-1}

\usepackage{amsfonts}
\usepackage{amsmath}
\usepackage{amssymb}
\usepackage{bm}
\usepackage{dcolumn}
\usepackage{graphicx}
\usepackage{graphics}
\usepackage[latin1]{inputenc}
\usepackage{latexsym}
\usepackage{rotating}
\usepackage{xspace} 
\usepackage[usenames]{color}
\usepackage{mathrsfs}

\usepackage{ulem}
\usepackage[colorlinks=true,citecolor={blue}]{hyperref}

\definecolor {darkgreen}{rgb}{0.2,0.7,0.2}

\newcommand\be{\begin{equation}}
\newcommand\ba{\begin{eqnarray}}
\newcommand\ee{\end{equation}}
\newcommand\ea{\end{eqnarray}}
\newcommand\bw{\begin{widetext}}
\newcommand\ew{\end{widetext}}

\newcommand{\nn}{\nonumber}

\newcommand{\MAT}{{\mbox{\tiny mat}}}

\newcommand{\GW}{{\mbox{\tiny GW}}}

\newcommand{\BH}{{\mbox{\tiny BH}}}

\newcommand{\NS}{*}
\newcommand{\ext}{\mathrm{ext}}
\newcommand{\inter}{\mathrm{int}}
\newcommand{\RNS}{\mathcal{R}_*}
\newcommand{\tid}{\mathrm{(tid)}}
\newcommand{\rot}{\mathrm{(rot)}}
\newcommand{\N}{{\mbox{\tiny N}}}
\newcommand{\mrm}{\mathrm}

\newcommand{\E}{{\mbox{\tiny E}}}

\newcommand{\lambdabartid}{\bar{\lambda}^\mathrm{(tid)}}
\newcommand{\Ibar}{\bar{I}}
\newcommand{\Qbar}{\bar{Q}}
\newcommand{\lambdabarrot}{\bar{\lambda}^\mathrm{(rot)}}

\begin{document}
\title{I-Love-Q Relations in Neutron Stars and their Applications to \\ Astrophysics, Gravitational Waves and Fundamental Physics}    

\author{Kent Yagi}
\affiliation{Department of Physics, Montana State University, Bozeman, MT 59717, USA.}

\author{Nicol\'as Yunes}
\affiliation{Department of Physics, Montana State University, Bozeman, MT 59717, USA.}


\begin{abstract} 

The exterior gravitational field of a slowly-rotating neutron star can be characterized by its multipole moments, the first few being the neutron star mass, moment of inertia, and quadrupole moment to quadratic order in spin.
In principle, all of these quantities depend on the neutron star's internal structure, and thus, on unknown nuclear physics at supra-nuclear energy densities, all of which is usually parameterized through an equation of state. 
We here find relations between the moment of inertia, the Love numbers and the quadrupole moment (I-Love-Q relations) that do not depend sensitively on the neutron star's internal structure.
Such universality may arise for two reasons: (i) these relations depend most sensitively on the internal structure far from the core, where all realistic equations of state mostly approach each other; (ii) as the NS compactness increases, the I-Love-Q trio approaches that of a BH, which does not have an internal-structure dependence. 
Three important consequences derive from these I-Love-Q relations.
On an observational astrophysics front, the measurement of a single member of the I-Love-Q trio would automatically provide information about the other two, even when the latter may not be observationally accessible. 
On a gravitational wave front, the I-Love-Q relations break the degeneracy between the quadrupole moment and the neutron-star spins in binary inspiral waveforms, allowing second-generation ground-based detectors to determine the (dimensionless) averaged spin to $\mathcal{O}(10)\%$, given a sufficiently large signal-to-noise ratio detection. 
On a fundamental physics front, the I-Love-Q relations allow for tests of General Relativity in the neutron-star strong-field that are both theory- and internal structure-independent. As an example, by combining gravitational-wave and electromagnetic observations, one may constrain dynamical Chern-Simons gravity in the future by more than 6 orders of magnitude more stringently than Solar System and table-top constraints.
\end{abstract}
\date{\today}
\maketitle
\tableofcontents

\section{Introduction}

Neutron-star (NS) astrophysics can provide crucial information about nuclear, gravitational-wave (GW) and fundamental physics that would be difficult to obtain by other means. On a nuclear physics front, NS observations allow us to probe the equation of state (EoS) of nuclear matter~\cite{lattimer-prakash-review} well beyond the densities available in Earth laboratories. For example, observations of the NS mass-radius relation and the mass-moment-of-inertia relation can be used to infer NS EoS within a certain observational uncertainty~\cite{steiner-lattimer-brown,ozel-baym-guver,ozel-review,guver,lattimer-schutz,kramer-wex}. 

On a GW physics front, the detection of GWs emitted during the late inspiral and merger of NS binaries could also be used to extract information about the EoS~\cite{hinderer-love,damour-nagar,binnington-poisson}. Binary NSs are, in fact, one of the most promising GW sources~\cite{thorne-source,schutz-source,abadie} for second-generation, ground-based detectors, such as Adv.~LIGO~\cite{ligo}, Adv.~VIRGO~\cite{virgo} and KAGRA~\cite{kagra}. Since NSs are tidally deformed in the late inspiral and merger, violating the test-particle approximation, NS binary waveforms must include corrections induced by the NS internal structure, for example in terms of NS tidal Love number~\cite{hinderer-love,damour-nagar,binnington-poisson}. In turn, this implies that a sufficiently high signal-to-noise ratio (SNR) detection of such a GW could be used to extract information about the NS EoS~\cite{flanagan-hinderer-love,read-love,hinderer-lackey-lang-read,kiuchi,kyutoku,hotokezaka,vines1,vines2,lackey,damour-nagar-villain,baiotti,pannarale,bernuzzi,hotokezaka2}.  

On a fundamental physics front, NSs are ideal to test General Relativity (GR) since they produce strong gravitational fields~\cite{stairs,psaltis-review,dedeo-psaltis}. Currently, GR has passed all Solar System tests with flying colors, but these only sample the weak field regime~\cite{TEGP,will-living}, where gravitational fields are stationary and weak, and all characteristic velocities are much smaller than the speed of light. Electromagnetic binary pulsar observations can test GR in a regime where the gravitational field is much stronger than in the Solar System, but still sufficiently non-dynamical that one can expand in the ratio of the orbital velocity to the speed of light to leading-order~\cite{stairs}. 

The exterior gravitational field of NSs, however, is not just determined by their mass and radius, but also by higher multipole moments, like the moment of inertia and the quadrupole moment, and ignorance of the NS EoS can hinder the extraction of the latter from observations. On the GW physics front, degeneracies between the NS spin and the quadrupole moment prevent future detections from separately measuring these quantities. On the fundamental physics front, degeneracies between the effect of the NS EoS and modified gravity corrections on observables prevent robust tests of GR that are internal-structure independent.  

\begin{figure*}[htb]
\begin{center}
\includegraphics[width=8.5cm,clip=true]{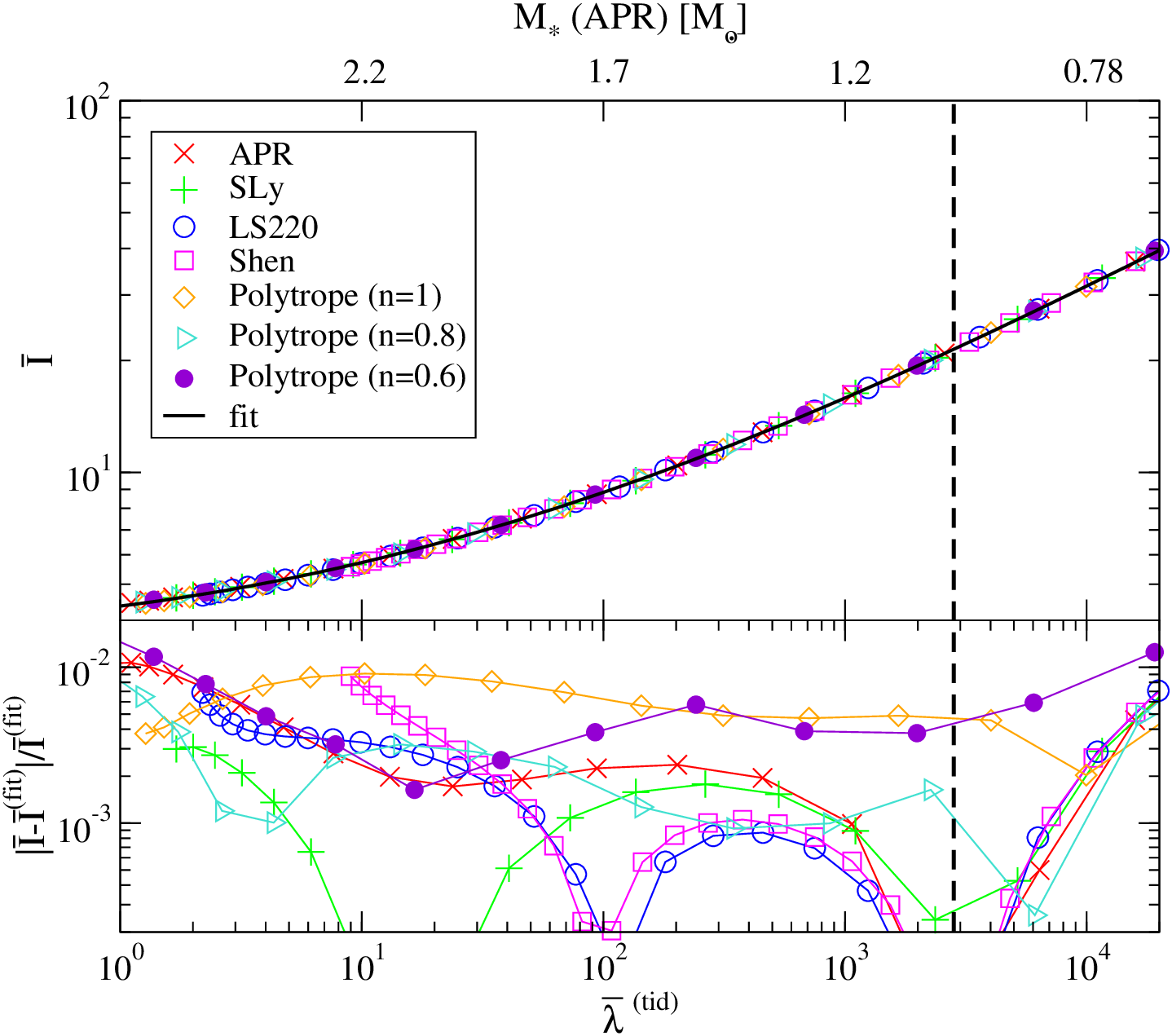}  \qquad
\includegraphics[width=8.5cm,clip=true]{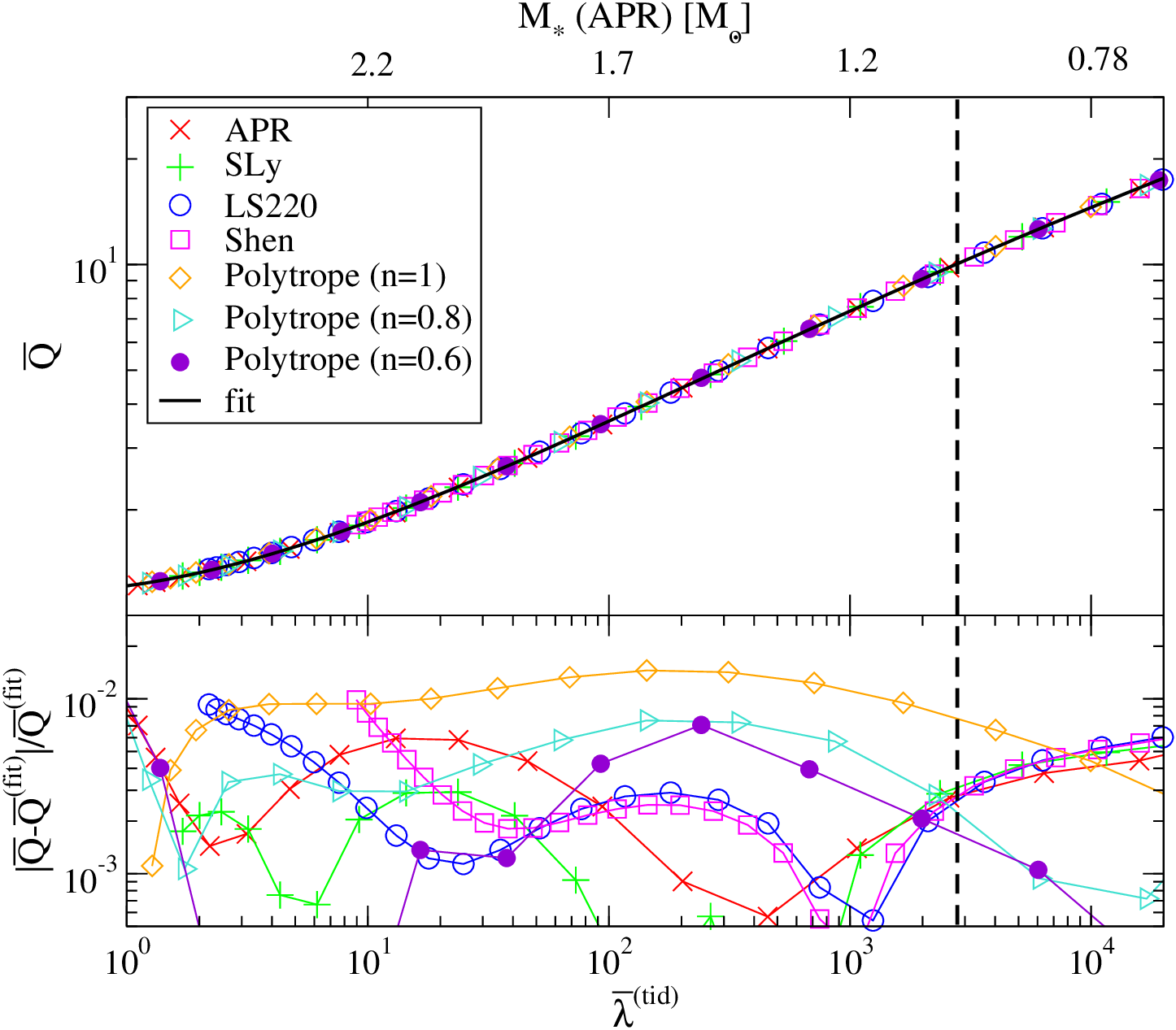}  
\caption{\label{fig:I-Love-fit} (Color Online) (Top) Fitting curves (solid curve), given in Eq.~\eqref{fit}, and numerical results (points) of the universal I-Love (left) and Q-Love (right) relations for various EoSs. These quantities are normalized as follows $\bar{I} = I/M_{\NS}^{3}$, $\bar{\lambda}^\tid = \lambda^\tid/M_{\NS}^{5}$ and $\Qbar = Q^\rot/[M_{\NS}^{3} (S/M_{\NS}^{2})^{2}]$. The parameter varied along each curve is the NS central density, or equivalently the NS compactness, with the latter increasing to the left of the plots. For reference, we also show the corresponding NS mass for the APR EoS on the top axes and a vertical dashed line when $M_{\NS} = 1 M_{\odot}$. (Bottom) Relative fractional errors between the fitting curve and numerical results. Observe that these relations are essentially EoS independent, with loss of universality at the $1\%$ level.}
\end{center}
\end{figure*}
In this paper, we take the first steps toward resolving this problem by discovering new relations between the NS moment of inertia $I$, the NS Love numbers and the (spin-induced) NS  quadrupole moment $Q^\rot$ (I-Love-Q relations) that are essentially EoS independent for slowly-rotating NSs~\cite{Yagi:2013bca}. Physically, the moment of inertia quantifies how fast a NS can spin given a fixed spin angular momentum $S$, the quadrupole moment describes how much a NS is deformed away from sphericity, and the Love number characterizes how easy or difficult it would be to deform a NS. 

The moment of inertia, Love numbers and quadrupole moment can be computed by numerically solving for the interior and exterior gravitational field of a NS in a slow-rotation~\cite{hartle1967} and a small tidal deformation approximation~\cite{hinderer-love}, to quadratic order in the former and to linear order in the latter. The moment of inertia and quadrupole moment can be obtained from the asymptotic behavior of the $(t,\phi)$ and $(t,t)$ components of the metric at spatial infinity respectively, which depend on the interior solution through matching boundary conditions at the NS surface that ensure metric continuity and differentiability. Although the moment of inertia is a first-order in spin quantity, the quadrupole moment is generated by quadratic spin terms. The tidal Love number $\lambda^\mrm{(tid)}$ is defined by the ratio between the tidally-induced quadrupole moment and the tidal field due to a companion NS, which can be calculated in a similar fashion.

One would expect that all of these quantities should depend quite sensitively on the NS EoS; after all, a fluffier star should be more easily deformable than a stiffer star. We find here, however, that, these quantities seem to satisfy almost universal relations when plotted against each other that are essentially independent of the NS EoS. Figure~\ref{fig:I-Love-fit} shows the I-Love (left) and Q-Love (right) relations, where $I$, $\lambda^\mrm{(tid)}$ and $Q^\rot$ are normalized to $M_{\NS}^{3}$, $M_{\NS}^{5}$ and $M_{\NS}^{3} (S/M_{\NS}^{2})^{2}$ respectively, with $M_{\NS}$ the NS mass and $S$ its spin angular momentum. The different curves represent the relations using different EoSs (APR~\cite{APR}, SLy~\cite{SLy}, Lattimer-Swesty (LS220)~\cite{LS}, Shen~\cite{Shen1,Shen2} and polytropic EoSs with indices of $n=0.6$, $0.8$ and $1$). The symbols represent numerical solutions, while the solid curve is a single  fitting function. The bottom of this figure shows the fractional errors between the fitting function and the numerical results. Observe that these relations are EoS independent to within $\mathcal{O}(1)$\%. 

We have found two {\emph{possible}} reasons that could explain such a weak EoS dependence. The first is that the I-Love-Q trio may depend most sensitively on the NS outer layers, far from the core, where all realistic EoSs approach each other. In this interpretation, the I-Love-Q relations do depend on the EoS, but only in a regime where the EoSs contributes similarly to the I-Love-Q trio. The second reason is based on the fact that the I-Love-Q trio for NSs approaches the I-Love-Q relations for a BH, as one increase the NS compactness. For BHs, these relations are clearly independent of the BH internal-structure (or lack thereof) due to the the no-hair theorems~\cite{robinson,israel,israel2,hawking-uniqueness0,hawking-uniqueness,carter-uniqueness}, which lead to well-known expressions for all multipole moments in terms of just the mass and spin~\cite{geroch,hansen}. But for NSs, such expressions do not exist because there is no NS no-hair theorem. In spite of this, we still find a NS universal relation between the moment of inertia (and thus the spin angular momentum) and the quadrupole moment, similar to that which arise for BHs due to the no-hair theorems. 

The universal I-Love-Q relations tell us that there is an {\emph{effacing of internal structure}} in play here, i.e. the expected internal-structure dependence of the I-Love-Q relations effaces away. One might think that such an effect is a consequence of the celebrated effacement principle~\cite{damour-effacement} in GR. However, this is not quite right because the latter states that the motion of compact objects is independent of their internal structure; the effacement principle says nothing about the multipolar-decomposition of the object's gravitational field or of its tidal deformations. Of course, the effacement principle holds in GR for BHs, but it is violated for NSs, with internal-structure corrections to the center of mass acceleration entering first at 5 post-Newtonian (PN) order\footnote{A term of $A$th PN order is suppressed relative to the leading-order term by a factor of ${\cal{O}}(v^{2A}/c^{2A})$, where $v$ is the characteristic velocity of the system and $c$ is the speed of light.} for systems of non-spinning bodies. On the other hand, the I-Love-Q relations interconnect different multipole components of the exterior gravitational field of isolated bodies, saying nothing about their relative motion. 

The I-Love-Q relations have immediate applications to observational astrophysics, GWs and fundamental physics, breaking degeneracies that would otherwise prevent us from taking full advantage of NS observations. On the observational astrophysics front, the measurement of any single member of the I-Love-Q trio would automatically provide information about the other 2 members, even if the latter are not easily accessible from an observational viewpoint. For example, if one could measure the moment of inertia of the primary NS of the double binary pulsar J0737-3039~\cite{burgay,lyne,kramer-double-pulsar}, one could then obtain its quadrupole moment and its tidal Love number through the I-Love-Q relations without any further measurements. This is particularly important because the Love number and the quadrupole moment cannot be easily extracted from binary pulsar observations, since they have a much weaker effect on observables.

On the GW physics front, the I-Love-Q relations can break the degeneracy between the NS quadrupole moment and the NS spins, given a sufficiently large SNR detection of a NS binary inspiral. The first spin-induced modification to the waveform, a spin-orbit coupling, enters at $1.5$ PN order in the waveform phase~\cite{kidder-spin}. Given a large SNR detection, one can then extract this phase term, and thus measure a certain combination of the individual spins. In order to extract both spins, however, one needs to also measure the spin-spin correction to the waveform, which enters at 2PN order. At this same order, however, the quadrupole moment also modifies the waveform phase, leading to a $100\%$ degeneracy between $Q^\rot$ and the individual spins. 

The Q-Love relation can be used to break this degeneracy. One can write the quadrupole moment as a function of the Love number, which enters at 5PN order in the waveform phase~\cite{flanagan-hinderer-love}. This forces a correlation between the quadrupole moment piece of the 2PN term and a 5PN term that is weakly correlated with other binary parameters. Recently,~\cite{lackey,damour-nagar-villain} suggested that second-generation, ground-based detectors could be used to extract the Love number. Therefore, such measurement of the tidal Love number, in combination with the Q-Love relation, determines the NS quadrupole moment, which then allows for a measurement of the averaged spin parameter $\chi_s \equiv (\chi_1 + \chi_2)/2$, where $\chi_{A}$ is the individual (dimensionless) spin parameter of NS A.  

\begin{figure}[t]
\begin{center}
\includegraphics[width=8.5cm,clip=true]{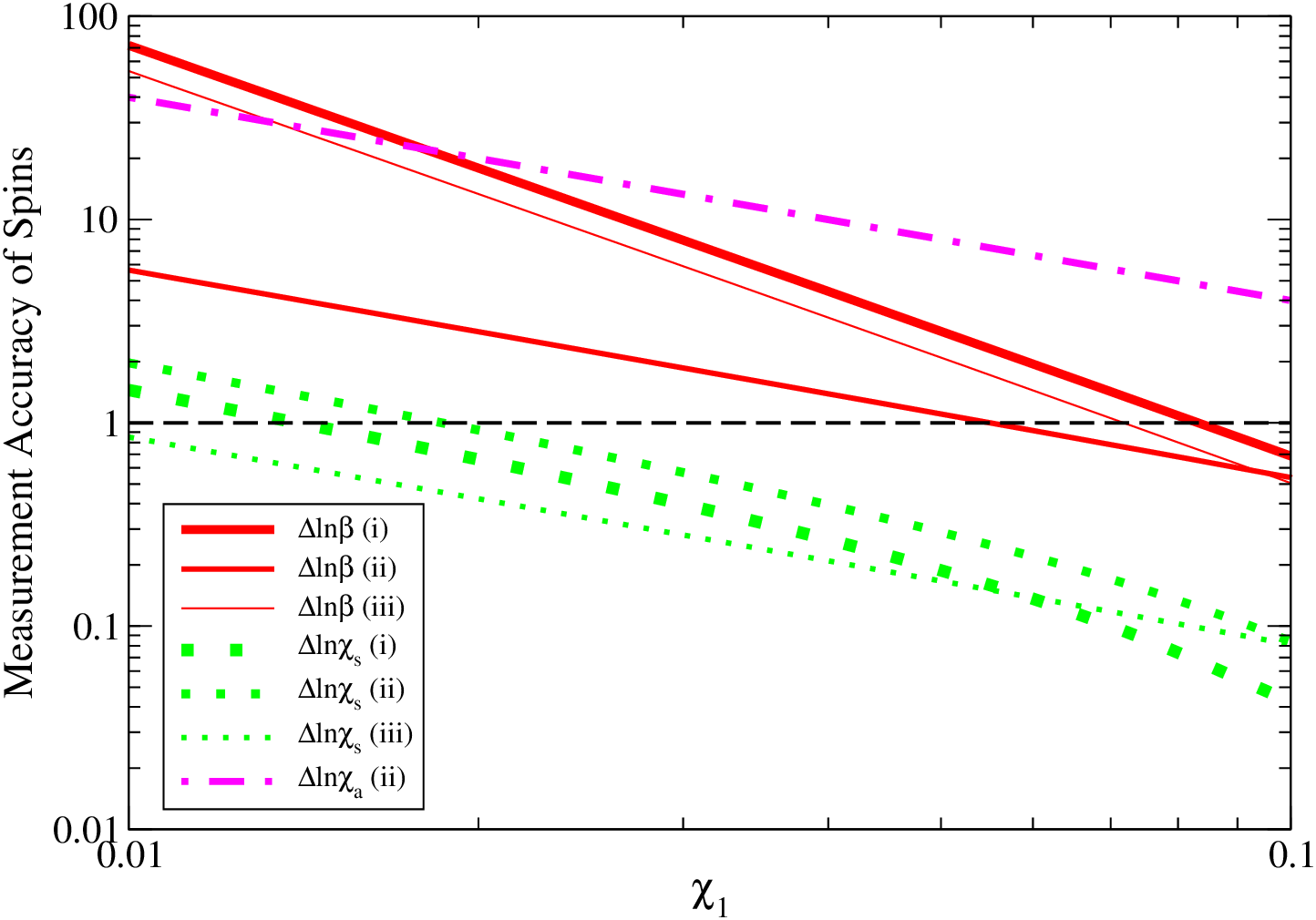}  
\caption{\label{fig:spin} (Color Online) Measurement accuracy of spin parameters $\beta$, $\chi_s$ and $\chi_a$ with Adv.~LIGO given a detection at a luminosity distance of $100$ Mpc with $\mrm{SNR} \approx 30$. We consider three different NS binaries, labeled by (i), (ii) and (iii), as described in the text. 
Observe that $\chi_{s}$ can be measured to approximately $\mathcal{O}(10)\%$ when we use the Q-Love relation to break spin degeneracies.}
\end{center}
\end{figure}

Figure~\ref{fig:spin} shows the projected measurement accuracies of spin parameters $\chi_a \equiv (\chi_1 - \chi_2)/2$, $\chi_{s}$ and the $1.5$PN phase term $\beta$ as functions of $\chi_1$ for 3 different systems: (i) $(m_1,m_2)=(1.45,1.35)M_\odot$, $\chi_1=\chi_2$, (ii) $(m_1,m_2)=(1.45,1.35)M_\odot$, $\chi_1=2 \chi_2$ and (iii) $(m_1,m_2)=(1.4,1.35)M_\odot$, $\chi_1=\chi_2$, where $m_A$ is the NS mass of the $A$ component. We used second-generation, ground-based detectors and luminosity distance of 100Mpc for $\rm{SNRs} \sim 30$. 
One can measure the averaged spin $\chi_s$ to $\mathcal{O}(10)\%$ if one uses the Q-Love relation. Such a measurement accuracy on $\chi_s$ is inaccessible without the Q-Love relation. 

\begin{figure}[t]
\begin{center}
\begin{tabular}{l}
\includegraphics[width=8.0cm,clip=true]{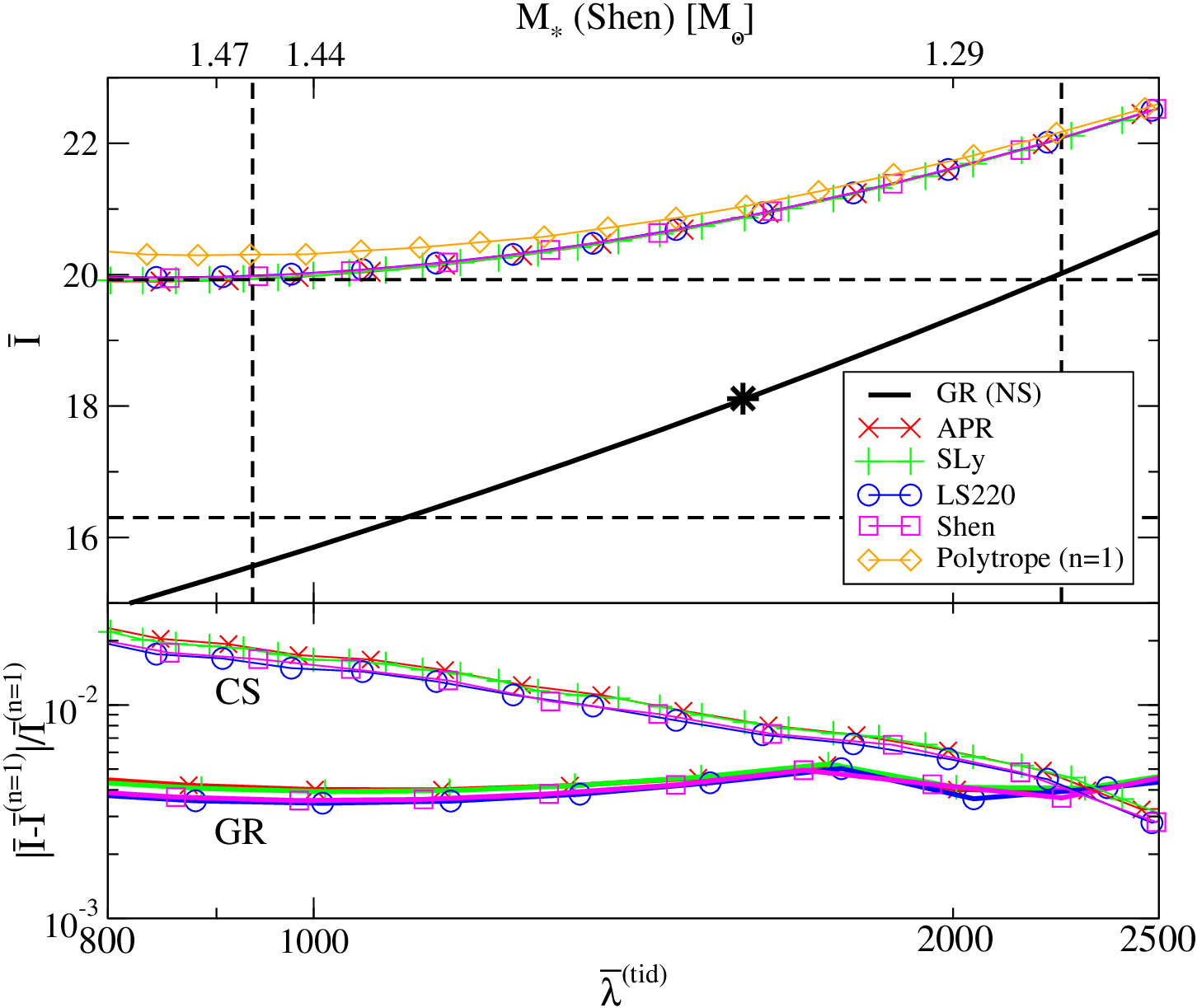}  
\end{tabular}
\caption{\label{fig:CS} (Color Online) (Top) I-Love relation in GR (black thick solid curve), normalized as in Fig.~\ref{fig:I-Love-fit}, with $\Delta\bar{I}$ and $\Delta\bar{\lambda}^\tid$ shown as (black) dashed lines around a fiducial measured value, shown with a cross. $\Delta\bar{I}$ is 10\% of the fiducial value, assuming future double binary pulsar observations~\cite{lattimer-schutz,kramer-wex}. $\Delta\bar{\lambda}^\tid$ is 40\% of the fiducial value, assuming a GW binary NS observation. We also plot the I-Love relation in dynamical CS gravity with a CS parameter $\xi_{\rm{cons}}=1.85 \times 10^4 M_\NS^4$ for various EoSs. This test would constrain $\xi < \xi_{\rm{cons}}$, 6 orders of magnitude more strongly than current Solar System bounds~\cite{alihaimoud-chen}. For reference, $M_\NS$ for the Shen EoS is shown on the top axis. (Bottom) Relative fractional difference of the CS I-Love relations (light solid curves) between the realistic EoSs and the $n=1$ polytrope. For reference, we also show this difference for the GR I-Love relations (thick solid curves). Observe that even in dynamical CS gravity, the universality of the I-Love relation seems to hold to ${\cal{O}}(1) \; \%$.}
\end{center}
\end{figure}

On a fundamental physics front, independents measurement of any two members of the I-Love-Q trio would allow for model-independent and EoS-independent tests of GR. For example, let us assume that one has measured the moment of inertia of the primary NS of the double binary pulsar J0737-3039 to 10\% accuracy\footnote{Notice that this pulsar has a relatively long spin period for a millisecond pulsar, $22.7 \; {\rm{ms}}$, and thus, the slow-rotation approximation is perfectly valid.}~\cite{lattimer-schutz,kramer-wex}. Let us further assume that GW observations have independently measured the NS tidal Love number to roughly 40\% from a detection of an equal-mass NS binary with the same NS mass as the primary in J0737-3039. With these observations, one can then plot a point in the I-Love plane with a measurement error box as shown in the top panel of Fig.~\ref{fig:CS}. Such a figure automatically provides a consistency (null) test of GR: one can test whether GR predicts an I-Love curve that goes through such an error box. Moreover, one can also constrain modified gravity theories by requiring that the I-Love curves in these theories pass through this error box. We will show here that such a test is even possible when the GW binary system has component masses that are up to $10\%$ different from the pulsar ones. 

Such a test, of course, is constraining provided modified gravity theories predict I-Love-Q relations that are not degenerate with the GR ones. Figure~\ref{fig:CS} shows that at least for dynamical Chern-Simons (CS) gravity~\cite{CSreview} this is not the case. Dynamical CS gravity is a parity-violating and quadratic-curvature corrected theory that has been weakly constrained in the Solar System through Gravity Probe B observations~\cite{alihaimoud-chen} and table-top experiments~\cite{kent-CSBH}. The fiducial I-Love measurements of Fig.~\ref{fig:CS} would constrain this theory 6 orders of magnitude more strongly than current tests, down to $\xi < 1.85 \times 10^4 M_\NS^4$, where $\xi^{1/4}$ is the characteristic length scale of the theory. Observe also that the universality of the I-Love relation still holds in dynamical CS gravity within several \% accuracy, although this universality does not hold as well as in GR. Perhaps, this is because NSs in dynamical CS gravity have scalar hair that depends nontrivially on the NS's internal structure~\cite{quadratic}, and thus, the internal-structure effacing in dynamical CS gravity may not be as effective. 

The I-Love-Q relations presented here show universality within the framework (uniform and slow-rotation and small tidal deformations) we work in. Of course, this framework is inappropriate to study newly-born NSs, which are probably differentially rotating and doing so fast. Older NSs that are uniformly rotating usually spin slowly, especially those that serve as a source of GWs for ground-based detectors, as they will have spun down significantly by the time they enter the GW sensitivity band~\cite{bildsten-cutler}. Short-period, millisecond pulsars, on the other hand, spin at a non-negligible rate, and thus, a slow-rotation expansion may not be suitable. In that case, we still expect to find semi-universal I-Love-Q relations that although independent of the EoS will depend on the spin period. Such an analysis requires a full numerical treatment of rapidly rotating NSs~\cite{berti-stergioulas,berti-white,benhar,pappas-apostolatos} and is left for future work.     

The rest of this paper deals with the details of this calculation and it is organized as follows. In Sec.~\ref{sec:decomp}, we explain how the spacetime is decomposed, the approximations used and the stress-energy tensor we used to describe NSs. In Sec.~\ref{sec:zeroth}, we construct non-rotating, isolated NS solutions. Taking these solutions as a background, in Sec.~\ref{sec:linear} we construct slowly-rotating NS solutions to linear order in spin and calculate the NS moment of inertia. In Sec.~\ref{sec:quadratic}, we construct slowly-rotating NS solutions to quadratic order in spin, calculate the NS quadrupole moment and define the rotational Love number.  In Sec.~\ref{sec:tidal}, we define the tidal Love number and construct tidally-corrected NS solutions. In Sec.~\ref{sec:I-Love-Q}, we show how the I-Love-Q relations are essentially independent of the NS EoS. We also study explanations for these relations, by considering analytic relations when using polytropic EoSs in the Newtonian limit. In Sec.~\ref{sec:applications}, we explain possible applications of the I-Love-Q relations to observational astrophysics, GWs and fundamental physics. In Sec.~\ref{sec:conclusions} we conclude and point to future research. 

All throughout the paper, we follow mostly the conventions of Misner, Thorne and Wheeler~\cite{MTW}. We use the Greek letters $(\alpha, \beta, \cdots)$ to denote spacetime indices. The metric is denoted by $g_{\mu \nu}$ and it has signature $(-,+,+,+)$. We use geometric units, with $G=1=c$.

\section{Spacetime Decomposition and Matter Representation}
\label{sec:decomp}

In this paper, we consider uniformly rotating NSs that are slightly deformed either due to rotation or tidal fields. Such solutions can be numerically constructed perturbatively in a slow-rotation and tidal-deformation expansion, taking the non-rotating, isolated solution as a background. In this section, we explain the metric decomposition employed here and the stress-energy tensor we will use to describe NSs.

\subsection{Metric Decomposition}

We choose Boyer-Lindquist type coordinates $(t,r,\theta,\phi)$ and decompose the metric as
\ba
ds^2 &=& -e^{\bar{\nu}(r)} \left[1+2 \epsilon^2 \bar{h}_2 (r) \alpha Y_{2 m}(\theta,\phi) \right] dt^2 \nn \\
& & + e^{\bar{\lambda}(r)} \left[ 1+\frac{2 \epsilon^2 \bar{m}_2 (r) \alpha Y_{2 m}(\theta,\phi)}{r-2\bar{M}(r)} \right] dr^2 \nn \\
& & + r^2 \left[ 1+2 \epsilon^2 \bar{K}_2(r) \alpha Y_{2 m}(\theta,\phi) \right] \nn \\
& & \times \left\{ d \theta^2 + \sin^2 \theta \left[ d\phi - \epsilon [\Omega_\NS - \bar{\omega}_1 (r) P_1' (\cos \theta) ] dt \right]^2 \right\} \nn \\
& & + \mathcal{O}(\epsilon^3)\,, 
\label{metric-ansatz-rth}
\ea
where $\bar{M}(r)$ is defined by
\be
\bar{M}(r) \equiv \frac{\left[ 1-e^{-\bar{\lambda}(r)} \right]r}{2}\,,
\ee
$P_\ell (\cos \theta)$ is the $\ell$-th order Legendre polynomial, $P_{1}' = d P_{1}/d(\cos{\theta})$ and $Y_{\ell m}(\theta, \phi)$ is the spherical harmonic function. The quantity $\epsilon$ here is a book-keeping parameter that we will later set to unity and we only introduce to remind ourselves of the order of the approximation. Terms linear in $\epsilon$ are induced only by linear-order in rotation effects, while tidal-deformation effects enter at ${\cal{O}}(\epsilon^{2})$. We will work here to quadratic order in $\epsilon$.

A slow-rotation expansion is quite appropriate to model old NSs. Recycled millisecond pulsars, the fastest NSs observed to date, have angular velocities in the kHz, but this number is small relative to the NS mass, i.e.~$M_{\NS} \Omega_{\NS} \lesssim 0.01$, where $\Omega_\NS$ is the NS angular velocity. For the fastest millisecond pulsar J1939+2134~\cite{fastest-pulsar}, with period $1.5 \; {\rm{ms}}$, the dimensionless spin parameter, defined via $\chi \equiv S/M_{\NS}^{2} = I \Omega_{\NS}/M_{\NS}^{2}$, is still small $\chi \lesssim 0.3$, using a Newtonian expression for the moment of inertia. Thus, a slow-rotation expansion is well-justified, especially when carried out to second order. This approximation, however, would break down if considering newly-born NSs, which are likely to be differentially rotating, much hotter and with much larger magnetic fields. Notice also that the NSs that will source GWs in the band of ground-based detectors are expected to have significantly smaller spins than that. This is because NSs spin-down~\cite{bildsten-cutler} as they inspiral and ground-based detectors will only be sensitive to the last 17 minutes of the orbit before coalescence.

The free functions in our metric decomposition are $\bar{\nu}$ and $\bar{\lambda}$ at $\mathcal{O}(\epsilon^0)$, $\bar{\omega}_1$ at $\mathcal{O}(\epsilon)$ and $\bar{h}_2$, $\bar{K}_2$ and $\bar{m}_2$ at $\mathcal{O}(\epsilon^2)$. The leading-order correction due to slow rotation enters at $\mathcal{O}(\epsilon)$, while that due to tidal deformations enters at $\mathcal{O}(\epsilon^2)$. For the former, we restrict ourselves to axisymmetric perturbations; at $\mathcal{O}(\epsilon)$ only the $(\ell,m)=(1,0)$ mode survives, while at $\mathcal{O}(\epsilon^2)$ only the $(\ell,m)=(0,0)$ and $(\ell,m)=(2,0)$ modes survive. For the latter, we are only interested in the spin and tidal, {\emph{quadrupolar}} deformations, and thus we only keep $\ell=2$ modes in Eq.~\eqref{metric-ansatz-rth}, but allow for all $m$ modes. Henceforth, we set the constant $\alpha=2 \sqrt{\pi/5}$ so that $\alpha Y_{20}(\theta, \phi) = P_2 (\cos \theta)$. 

As pointed out by Hartle~\cite{hartle1967}, one needs to be careful about choosing coordinates when deriving and solving perturbed equations. A perturbative analysis is valid only if perturbed quantities are much smaller than the unperturbed one. If one were to carry out calculations in $(t,r,\theta,\phi)$ coordinates, such conditions would be violated in certain situations. For example, in the region of spacetime outside the unperturbed star but inside the perturbed star, the ratio of the perturbed pressure (or density) to that of the unperturbed pressure (or density) diverges, which violates our perturbative treatment. 

In order to overcome this problem, we transform the radial coordinate via~\cite{hartle1967} 
\be
r(R,\theta) = R + \epsilon^2 \xi_2(R) \alpha Y_{2 m}(\theta, \phi) + {\cal{O}}(\epsilon^{3})\,,
\ee
where $\xi_{2}(R)$ is such that
\be
\rho [r(R, \theta, \phi)]=\rho(R) = \rho^{(0)}(R)\,.
\ee
In other word, the new radial coordinate $R$ is chosen such that $\rho [r(R, \theta,\phi)]$ is identical to the unperturbed density $\rho^{(0)}(r)$. By construction, the density and pressure in these new coordinates contain only the unperturbed contributions. Notice that $\xi_{2} Y_{2m}$ is well-defined only inside the star and we take it to be constant outside. This means that the exterior metric in $(t,r,\theta,\phi)$ coordinates can be obtained simply by replacing $R \to r$ in the exterior metric in $(t,R,\theta,\phi)$ coordinates. 

The transformed metric in $(t,R,\theta,\phi)$ coordinates can be found in~\cite{kent-CSNS} for the axisymmetric case.  Henceforth, we will relabel the metric coefficients via
\allowdisplaybreaks
\ba
\nu(R) &\equiv& \bar{\nu}(r)  = \bar{\nu}(R + \epsilon^{2} \xi_{2} \alpha Y_{2m})\,, 
\nn \\ 
\lambda(R) &\equiv& \bar{\lambda}(r)  = \bar{\lambda}(R + \epsilon^{2} \xi_{2} \alpha Y_{2m})\,, 
\nn \\ 
\omega_{1}(R) &\equiv& \bar{\omega}_{1}(r)  = \bar{\omega}_{1}(R + \epsilon^{2} \xi_{2} \alpha Y_{2m})\,, 
\nn \\ 
h_{2}(R) &\equiv& \bar{h}_{2}(r)  = \bar{h}_{2}(R + \epsilon^{2} \xi_{2} \alpha Y_{2m})\,, 
\nn \\ 
m_{2}(R) &\equiv& \bar{m}_{2}(r)  = \bar{m}_{2}(R + \epsilon^{2} \xi_{2} \alpha Y_{2m})\,, 
\nn \\ 
K_{2}(R) &\equiv& \bar{K}_{2}(r)  = \bar{K}_{2}(R + \epsilon^{2} \xi_{2} \alpha Y_{2m})\,, 
\nn \\ 
M(R) &\equiv& \bar{M}(r)  = \bar{M}(R + \epsilon^{2} \xi_{2} \alpha Y_{2m})\,.
\ea
%

\subsection{Matter Representation}

We here consider NSs that are uniformly rotating, and thus, we model them as a perfect fluid. Uniform rotation should be a reasonable approximation unless one considers newly-born NSs. The stress energy-momentum tensor of the matter field $T_{\mu\nu}^\MAT$ is then given by
\be
T_{\mu\nu}^\MAT = (\rho + p ) u_\mu u_\nu + p \; g_{\mu\nu}\,,
\ee
where the four-velocity $u^\mu$ is given by
\be
u^\mu = (u^0, 0,0, \epsilon \Omega_\NS u^0)\,,
\ee
and $\Omega_\NS$ is the constant angular velocity of the NS. By using the normalization condition $u_\mu u^\mu = -1$, we obtain the time component of the four-velocity $u^0$ as
\ba
u^0 &=& e^{-\nu/2} +  \epsilon^2\frac{e^{-3\nu/2}}{2} [ \omega_1^2 P_1'{}^2 R^2 \sin^2\theta \nn \\
& & -  e^{\nu} (2 h_2 +\nu' \xi_2) \alpha Y_{2m} ] + \mathcal{O}(\epsilon^4)\,.
\ea

We here consider 4 realistic EoSs: APR~\cite{APR}, SLy~\cite{SLy,shibata-fitting}, Lattimer-Swesty with nuclear incompressibility of 220MeV (LS220)~\cite{LS, ott-EOS} and Shen~\cite{Shen1,Shen2,ott-EOS}, the latter two with temperature of 0.1MeV and an electron fraction determined by the neutrino-less, beta-equilibrium condition. All of the EoS described above are ``realistic'' in that they allow NSs with masses larger than 1.93$M_\odot$, the lower bound of the recently found massive pulsar J1614-2230~\cite{1.97NS}. For comparison purposes, we also consider polytropic EoSs, i.e.~EoSs of the form
\be
p=K \rho^{1+1/n}\,,
\label{polytropic}
\ee
where $K$ is an amplitude constant and $n$ is the constant polytropic index. One can approximate the NS EoS with polytropes in the range $n \approx 0.5-1$~\cite{flanagan-hinderer-love,lattimer-prakash-2001}. No single polytrope, however, is believed to be an accurate representation of a realistic EoS. 

The APR EoS is constructed by using the variational chain summation methods, which is expected to include all leading many-body correlation effects. The APR EoS uses Hamiltonians that include a three-nucleon interaction, which predicts that a transition exists from NS matter to a phase with neutral pion condensation at a baryon number density of $\sim 0.2 \ \mrm{fm^{-3}}$. The SLy EoS is calculated from a non-relativistic mean field theory approach, with a new set of Skyrme-type effective nucleon-nucleon interactions, suitable for describing very neutron rich matter. Unlike the APR EoS that describes only the NS's liquid core, the SLy EoS is a ``unified EoS'' in the sense that it is supposed to describe also the NS crust. The LS220 EoS is constructed from a finite-temperature compressible liquid-droplet model with a Skyrme nuclear force. Such an EoS is derived within the single heavy nucleus approximation and the assumption of nuclear statistical equilibrium. The Shen EoS uses a relativistic mean-field theory model and assumes nuclear statistical equilibrium. Nuclear incompressibility of the Shen EoS occurs at 281MeV.

\section{Slowly Rotating, Isolated Neutron Stars: $\mathcal{O}(\chi^0)$}
\label{sec:zeroth}

In this section, we construct non-rotating, isolated NS solutions, which will later be used as background solutions to construct slowly-rotating, tidally-deformed NS solutions in Secs.~\ref{sec:linear},~\ref{sec:quadratic} and~\ref{sec:tidal}.

\subsection{Einstein Equations and Exterior Solutions}

The $(t,t)$ and $(R,R)$ components of the Einstein Equations yield
\ba
\label{tt-zeroth}
\frac{d M}{dR} &=& 4 \pi R^2 \rho\,, \\
\label{rr-zeroth}
\frac{d \nu}{dR} &=& 2\frac{4 \pi R^3 p + M}{R(R-2M)}\,,
\ea
respectively. Combining the $R$-component of the equation of motion $\nabla^\mu T_{\mu R}^\MAT=0$ and Eq.~\eqref{rr-zeroth}, one obtains the Tolman-Oppenheimer-Volkoff (TOV) equation:
\be
\frac{dp}{dR} = -\frac{(4\pi R^3 p + M) (\rho + p)}{R(R-2M)}\,.
\label{TOV-zeroth}
\ee
Equations~\eqref{tt-zeroth},~\eqref{rr-zeroth} and~\eqref{TOV-zeroth} together with the equation of state $p=p(\rho)$ close the system of differential equations.

The exterior solutions to the above equations can be obtained by setting $\rho=0=p$. One finds~\cite{hartle1967}
\be
\nu^\mrm{ext} (R) = -\lambda^\mrm{ext} (R) = \ln \left(1 - \frac{2 M_\NS}{R} \right)\,.
\label{ext-nu-lambda}
\ee
We use the superscripts ``ext'' to refer to exterior quantities.

\subsection{Interior Solutions}

First, we solve Eqs.~\eqref{tt-zeroth} and~\eqref{TOV-zeroth} together with the equation of state with initial conditions
\allowdisplaybreaks
\ba
\rho (r_\epsilon) &=& \rho_c + \mathcal{O}(r_\epsilon^2)\,, \\
p (r_\epsilon) &=& p_c + \mathcal{O}(r_\epsilon^2)\,, \\
M(r_\epsilon) &=& \frac{4\pi}{3} \rho_c r_\epsilon^3 + \mathcal{O}(r_\epsilon^5)\,, 
\ea
where $\rho_c$ and $p_c$ are the central density and pressure respectively, and $r_\epsilon$ corresponds to the core radius which we take to be $r_\epsilon = 100 \; {\rm{cm}} \ll \RNS$. We have checked that all of our results are independent of the choice of $r_{\epsilon}$ provided this is a very small number relative to the NS radius. We solve Eqs.~\eqref{tt-zeroth} and~\eqref{TOV-zeroth}  outwards from $r=r_\epsilon$ until $p$ vanishes. The NS radius $\RNS$ and the NS mass $M_\NS$ are then defined by $p(\RNS)=0$ and $M_\NS = M(\RNS)$ respectively. For later convenience, we introduce the NS compactness
\be
C \equiv \frac{M_\NS}{\RNS}\,.
\ee
Notice that The central pressure $p_c=p(\rho_{c})$ is determined from the EoS, once $\rho_c$ is chosen. The central density $\rho_c$ is then a free parameter that effectively determines the mass and radius of the NS. 

With this solutions, we can then solve Eq.~\eqref{rr-zeroth}. One approach is to use the boundary condition [see Eq.~\eqref{ext-nu-lambda}]
\be
e^{\nu(\RNS)} = 1- \frac{2 M_\NS}{\RNS} 
\ee
at the NS surface as an initial condition and then integrate inwards toward the core. Another approach is to use the fact that Eq.~\eqref{rr-zeroth} is shift invariant, as done e.g.~in~\cite{kent-CSNS}. All throughout this paper, numerical solutions to the initial value problem are obtained with an adaptive 4th-order Runge-Kutta method~\cite{gsl}.

\begin{figure}[t]
\begin{center}
\includegraphics[width=8.0cm,clip=true]{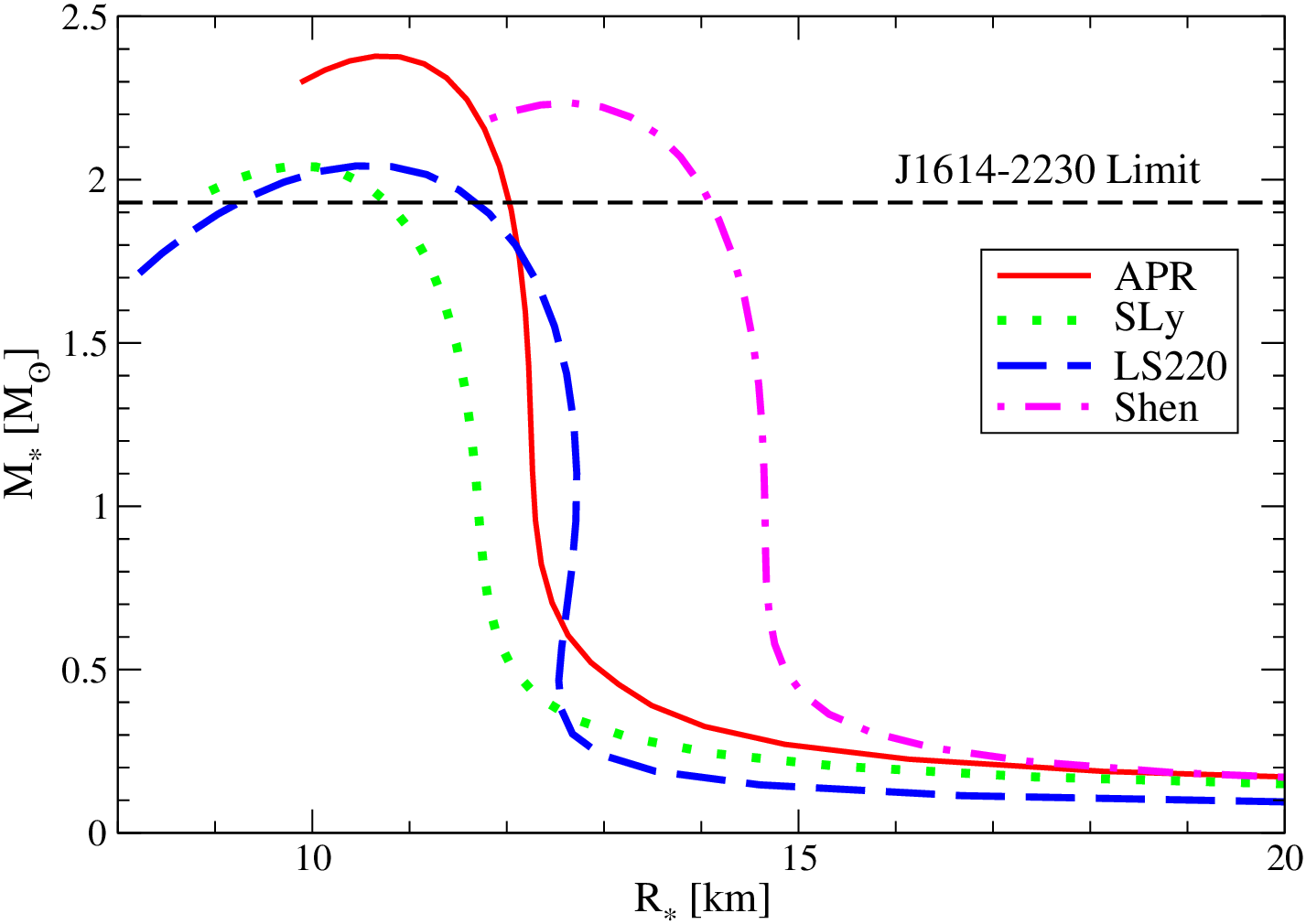}  
\caption{\label{fig:MR} (Color Online) The NS mass-radius relation for a few realistic EoSs. The black, horizontal, dashed line at $1.93M_\odot$ corresponds to the lower bound on the mass of the recently found massive pulsar J1614-2230~\cite{1.97NS}. Observe that all realistic EoSs lead to mass-radius curves that exceed this lower bound.}
\end{center}
\end{figure}

\begin{figure}[htb]
\begin{center}
\includegraphics[width=8.0cm,clip=true]{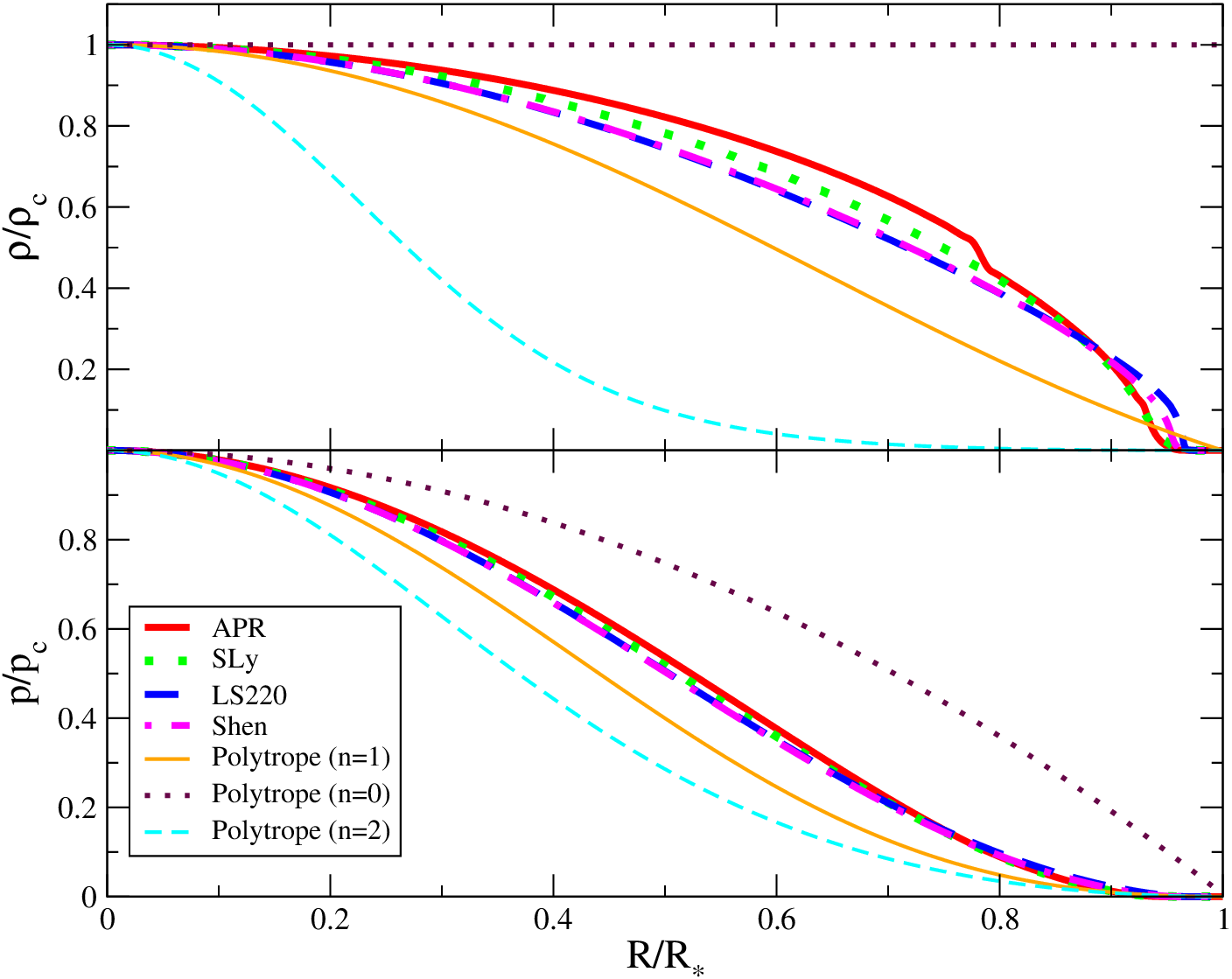}  
\caption{\label{fig:profile} (Color Online) The NS density (top) and pressure (bottom) profile for various EoSs as functions of the radial coordinate. We fix $C=0.17$ which corresponds to $M_\NS = 1.4M\odot$ for the APR EoS. We do not show profiles of $n=2.5$ and $n=3$ polytropic EoSs since the maximum compactness with such EoSs is smaller than 0.17.}
\end{center}
\end{figure}

Figure~\ref{fig:MR} shows the mass-radius relation for various EoSs. We have checked that the mass-radius relation for the SLy EoS agrees with that shown in~\cite{shibata-fitting}. As anticipated, all EoSs lead to NSs with maximum mass larger than $1.93 \; M_\odot$ (the black dashed horizontal line), which is the lower bound for the mass of J1614-2230~\cite{1.97NS}. We do not show the mass-radius relation for the polytropic EoSs because the I-Love-Q relations that we present in Sec.~\ref{sec:I-Love-Q} only depend on the NS compactness and do not depend on the mass-radius relation. Figure~\ref{fig:profile} shows the interior profile of the NS density (top) and pressure (bottom) as functions of the radial coordinate for a compactness of $C=0.17$, which corresponds to a NS with $M_\NS = 1.4 M_\odot$ and $\RNS \approx 12.1 \; {\rm{km}}$ for the APR EoS.

\section{Slowly Rotating, Isolated Neutron Stars: $\mathcal{O}(\chi^1)$}
\label{sec:linear}

Let us now focus on constructing slowly-rotating, isolated NS solutions. In this section, we only consider axisymmetric perturbations and construct NSs to linear order in spin. We will first discuss the differential equation that needs to be solved, and then we will solve them in the exterior region modulo an integration constant. After this, we discuss the asymptotic behavior of the solution at the NS center, which can then be used as an initial condition to solve the equations in the interior region. Finally, we determine the integration constant by matching the interior and exterior solutions at the NS surface.  

\subsection{Einstein Equations and Exterior Solutions}

At linear order in $\epsilon$, the only non-vanishing component of the Einstein Equations is the $(t,\phi)$ one:
\be
\frac{d^2 \omega_1}{d R^2} + 4\frac{1- \pi R^2 (\rho +p) e^{\lambda}}{R}\frac{d \omega_1}{d R}  -16 \pi (\rho +p) e^{\lambda} \omega_1=0\,. 
\label{omega1RR}
\ee
Solving this equation in the exterior (i.e.~setting $p = 0 = \rho$), one finds~\cite{hartle1967}
\be
\omega_1^\ext = \Omega_\NS - \frac{2S}{R^3} = \Omega_\NS \left( 1- \frac{2I}{R^3} \right)\,, 
\label{omega-ext}
\ee
where we have defined the moment of inertia by 
\be
I \equiv \frac{S}{\Omega_\NS}\,.
\label{I}
\ee
This quantity characterizes how fast a body can spin given a fixed spin angular momentum $S$. Notice that the exterior solution depends on two constants $\Omega_{\NS}$ and $S$. The former must be specified {\emph{a priori}}, just like $\rho_{c}$, and it describes how fast the NS is rotating. The latter is determined by matching this exterior solution to an interior solution at the NS surface.

\subsection{Interior Solutions}

Before we can solve for the interior solution, we first need initial conditions at the NS center. Taylor-expanding Eq.~\eqref{omega1RR} about the NS center, we find that the interior solution must asymptotically behave as
\be
\label{omega10}
\omega_1(R) = \omega_c + \frac{8\pi}{5} (\rho_c + p_c) \omega_c R^2 + \mathcal{O}(R^3) \quad (R \to 0^+)\,.
\ee
This solution contains a single constant, $\omega_c$, because we have eliminated another constant by requiring regularity of the solution at the NS center. The constant $\omega_{c}$ determines the NS spin angular momentum $S$, or equivalently, the NS moment of inertia $I$; in particular, $I$ increases as $\omega_c$ increases.

\begin{figure*}[htb]
\begin{center}
\includegraphics[width=8.0cm,clip=true]{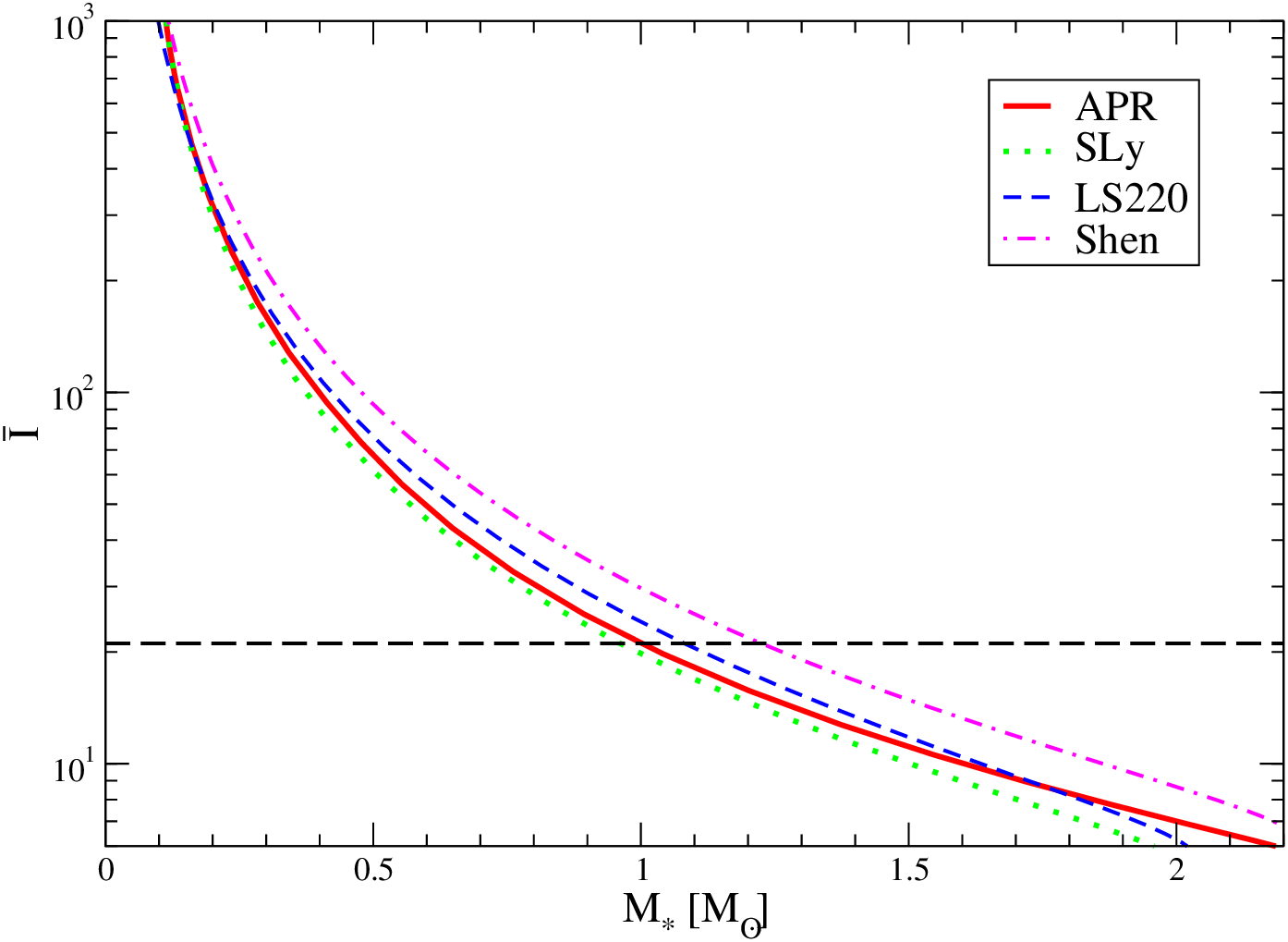}  
\includegraphics[width=8.0cm,clip=true]{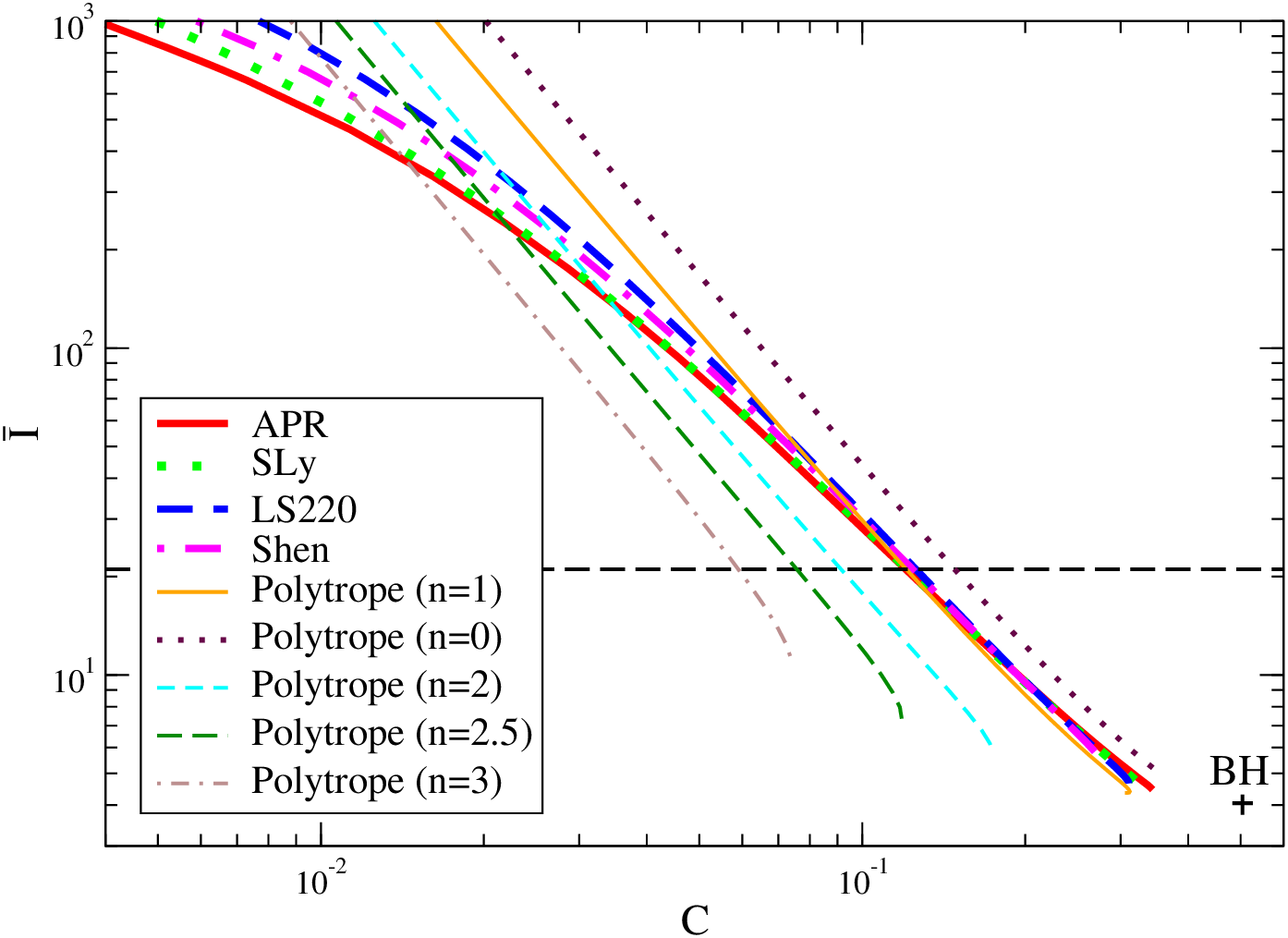}  
\caption{\label{fig:IbarMC} (Color Online) Dimensionless moment of inertia $\bar{I}$, defined in Eq.~\eqref{bar-I}, as functions of $M_\NS$ (left) and $C$ (right) for various EoSs. The horizontal dashed lines at $\Ibar = 21.1$ correspond to $M_\NS = 1M_\odot$ for the APR EoS; NSs below this line have higher $M_{\NS}$ and $C$. The solid cross indicates the value of $\Ibar$ for a BH. Observe that the $\Ibar$ curves for realistic EoSs approach each other as $C$ increases, and moreover they approach the BH limit $\Ibar_{\BH} = 4$ as $C \to 0.5$.}
\end{center}
\end{figure*}

We numerically solve Eq.~\eqref{omega1RR} with the initial condition in Eq.~\eqref{omega10} via an adaptive 4th-order Runge-Kutta method~\cite{gsl}. In solving this equation, one can take advantage of its homogeneity, its scale-invariance, as done e.g.~in~\cite{yunespsaltis,kent-CSNS}. Once the interior solution has been found, we match it to the exterior one in Eq.~\eqref{omega-ext} at the NS surface $R=\RNS$. The matching ensures that the solution is continuous and differentiable at the NS surface: 
\be
\omega_1^{\inter}(\RNS) = \omega_1^{\ext}(\RNS), \quad  \omega_1'{}^{\inter}(\RNS) = \omega_1'{}^{\ext}(\RNS)\,,
\label{BC-omega}
\ee
where we use the superscript ``int'' to refer to interior quantities. Through these conditions, we determine $S$ (or equivalently $I$)  and $\omega_c$ as a function of $\Omega_{*}$. In practice, due to the scale invariance of Eq.~\eqref{omega1RR}, the exterior solution can be divided by $\Omega_{*}$ and thus it only depends on the single constant $I$. Similarly, the interior solution can be obtained for $\omega_{1}^{\inter}/\Omega_{*}$ as a function of a single constant $\bar{\omega}_{c} = \omega_{c}/\Omega_{*}$. Therefore, the conditions in Eq.~\eqref{BC-omega} uniquely determine $I$ and $\omega_{c}$. This then determines the full solution, and thus also $S$, up to the overall constant of proportionality $\Omega_{*}$. 

The moment of inertia can be expressed entirely as a function of the interior solution. From Eqs.~\eqref{tt-zeroth}--\eqref{omega1RR},~\eqref{omega-ext} and~\eqref{BC-omega}, $I$ takes the form~\cite{hartle1967,kalogera-psaltis}
\be
I = \frac{8\pi}{3} \frac{1}{\Omega_\NS} \int_0^{\RNS} \frac{e^{-(\nu^\inter + \lambda^\inter )/2} R^5 (\rho + p) \omega_1^{\inter}}{R-2M(R)} dR\,.
\label{I}
\ee 
In the Newtonian limit (superscript ``N''), Eq.~\eqref{I} reduces to~\cite{hartle1967}
\be
I^\N = \frac{8\pi}{3} \int^{\RNS}_0 R^4 \rho (R)dR\,,
\label{I-Newton}
\ee
For later convenience, we define the dimensionless moment of inertia $\bar{I}$
\be
\label{bar-I}
\bar{I} \equiv \frac{I}{M_\NS^3}\,.
\ee

Figure~\ref{fig:IbarMC} shows $\bar{I}$ as a function of the NS mass $M_\NS$ and compactness $C$. We have verified that the moment of inertia $I$ obtained here agrees exactly with previous results in the literature~\cite{pani-NS-EDGB}. Observe that the different $\bar{I}$ curves for realistic EoSs approach each other as $C$ increases. Moreover, observe that all these curves approach the value of $\bar{I}$ for a BH as $C \to 0.5$, shown with a solid cross in Fig.~\ref{fig:IbarMC}. Of course, none of the NS sequences considered here will ever lead to a BH solution for any finite choice of central density.

\section{Slowly Rotating, Isolated Neutron Stars: $\mathcal{O}(\chi^2)$}
\label{sec:quadratic}

Let us now look at slowly-rotating NS solutions at quadratic order in spin. Following Sec.~\ref{sec:linear}, we first discuss the differential equations that describe the solution and then we solve them in the exterior region. We then discuss the asymptotic behaviors of the solutions at the NS center, obtain the interior solutions numerically, and match it to the exterior solution at the NS surface.

\subsection{Einstein Equations and Exterior Solutions}

At quadratic order in spin, the $\theta$-component of the equation of motion $\nabla^{\mu} T_{\mu\theta}^\MAT=0$, valid only inside the star, yields
\be
\xi_2 = -\frac{R^2 e^{-\lambda} (3 h_2+e^{-\nu} R^2 \omega_1^2)}{3(M+4\pi p R^3)}\,.
\label{xi2}
\ee
\bw
The $(\theta,\theta)-(\phi,\phi)$, $(R,\theta)$ and $(R,R)$ components of the Einstein Equations give respectively,
\ba
\label{m2}
m_2 &=& -R e^{-\lambda} h_2  +\frac{1}{6} R^4 e^{-(\nu+\lambda)} \left[ R e^{-\lambda} \left(\frac{d \omega_1}{dR} \right)^2
+ 16 \pi R \omega_1^2 (\rho +p) \right]\,, \\
\label{k2R}
\frac{dK_2}{dR} &=& -\frac{dh_2}{dR} + \frac{R-3M-4\pi pR^3}{R^2} e^{\lambda} h_2 
+ \frac{R-M+4\pi p R^3}{R^3} e^{2\lambda} m_2\,, \\
\label{h2R}
\frac{dh_2}{dR} &=& -\frac{R-M+4 \pi p R^3}{R}e^{\lambda} \frac{dK_2}{dR} +\frac{3-4\pi (\rho +p) R^2}{R} e^{\lambda} h_2 +\frac{2}{R} e^{\lambda} K_2 +\frac{1+8\pi p R^2}{R^2} e^{2\lambda} m_2 +\frac{R^3}{12}e^{-\nu} \left( \frac{d\omega_1}{dR} \right)^2 \nn \\
& & -\frac{4 \pi (\rho +p) R^4 \omega_1^2}{3R}e^{-\nu+\lambda}\,.
\ea
By imposing asymptotic flatness at spatial infinity, one finds the exterior solutions ~\cite{hartle1967}
\ba
\label{h2-ext}
h_2^\ext &=& \frac{1}{M_\NS R^3} \left( 1 + \frac{M_\NS}{R} \right) S^2  + AQ_2^2 \left( \frac{R}{M_\NS}-1 \right) \nn \\
             &=&  \frac{1}{M_\NS R^3} \left( 1 + \frac{M_\NS}{R} \right) S^2 -\frac{3A R^2}{M_\NS (R-2 M_\NS)} \left[ 1- 3 \frac{M_\NS}{R} + \frac{4}{3} \frac{M_\NS^2}{R^2} + \frac{2}{3} \frac{M_\NS^3}{R^3} + \frac{R}{2 M_\NS} f(R)^2 \ln f(R) \right]\,, \\
\label{k2-ext}
K_2^\ext &=& - \frac{1}{M_\NS R^3} \left( 1+\frac{2M_\NS}{R} \right) S^2 + \frac{2AM_\NS}{\sqrt{R (R-2 M_\NS)}}Q_2^1\left( \frac{R}{M_\NS}-1 \right) - AQ_2^2 \left( \frac{R}{M_\NS}-1 \right) \nn \\
              &=& - \frac{1}{M_\NS R^3} \left( 1+\frac{2M_\NS}{R} \right) S^2 + \frac{3AR}{M_\NS} \left[1+ \frac{M_\NS}{R} - \frac{2}{3} \frac{M_\NS^2}{R^2} + \frac{R}{2M_\NS} \left( 1- \frac{2 M_\NS^2}{R^2} \right) \ln f(R)  \right]\,, \\
\label{m2-ext}
m_2^\ext &=&  - \frac{1}{M_\NS R^2} \left( 1 -7\frac{M_\NS}{R} + 10 \frac{M_\NS^2}{R^2} \right)S^2 +\frac{3 A R^2}{M_\NS} \left[ 1 - 3 \frac{M_\NS}{R} + \frac{4}{3} \frac{M_\NS^2}{R^2} + \frac{2}{3} \frac{M_\NS^3}{R^3} + \frac{R}{2 M_\NS} f(R)^2 \ln f(R)  \right]\,,
\ea
\ew
with $f(R) \equiv 1- 2M_\NS/R$, $Q_2^2$ and $Q_2^1$ the associated Legendre functions of the second kind and $A$ an integration constant that is to be determined by matching with the interior solution at the NS surface. 

The spin-induced quadrupole moment $Q^\rot$ can be read off from the coefficient of the $P_2(\cos\theta)/R^3$ term in the Newtonian potential~\cite{hartlethorne}:
\be
Q^\rot = - \frac{S^2}{M_\NS} - \frac{8}{5} A M_\NS^3\,.
\label{quadrupole}
\ee
Notice that the quadrupole moment depends both on the magnitude of the spin angular momentum $S$ and the integration constant $A$, determined after matching the interior and exterior linear- and quadratic-order in spin solutions at the NS surface. The quadrupole moment represents the quadrupolar deformation of a body away from sphericity, with $Q^\rot<0$ corresponding to an oblate deformation. Notice also that the first term of Eq.~\eqref{quadrupole} is identical to the relation one obtains for BH, which means that $A\to0$ in the GR test-particle limit.

\begin{figure*}[htb]
\begin{center}
\includegraphics[width=8.0cm,clip=true]{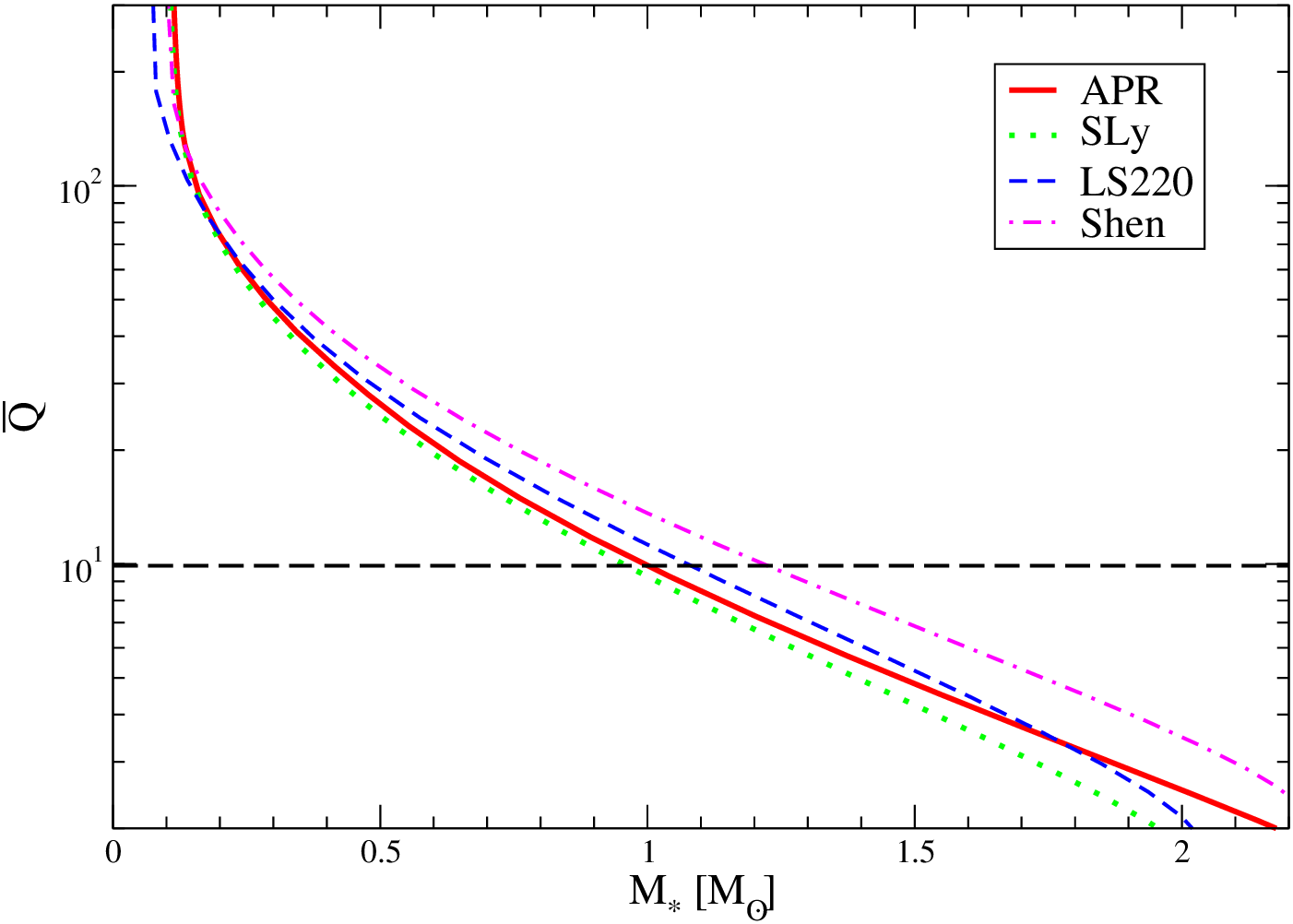}  
\includegraphics[width=8.0cm,clip=true]{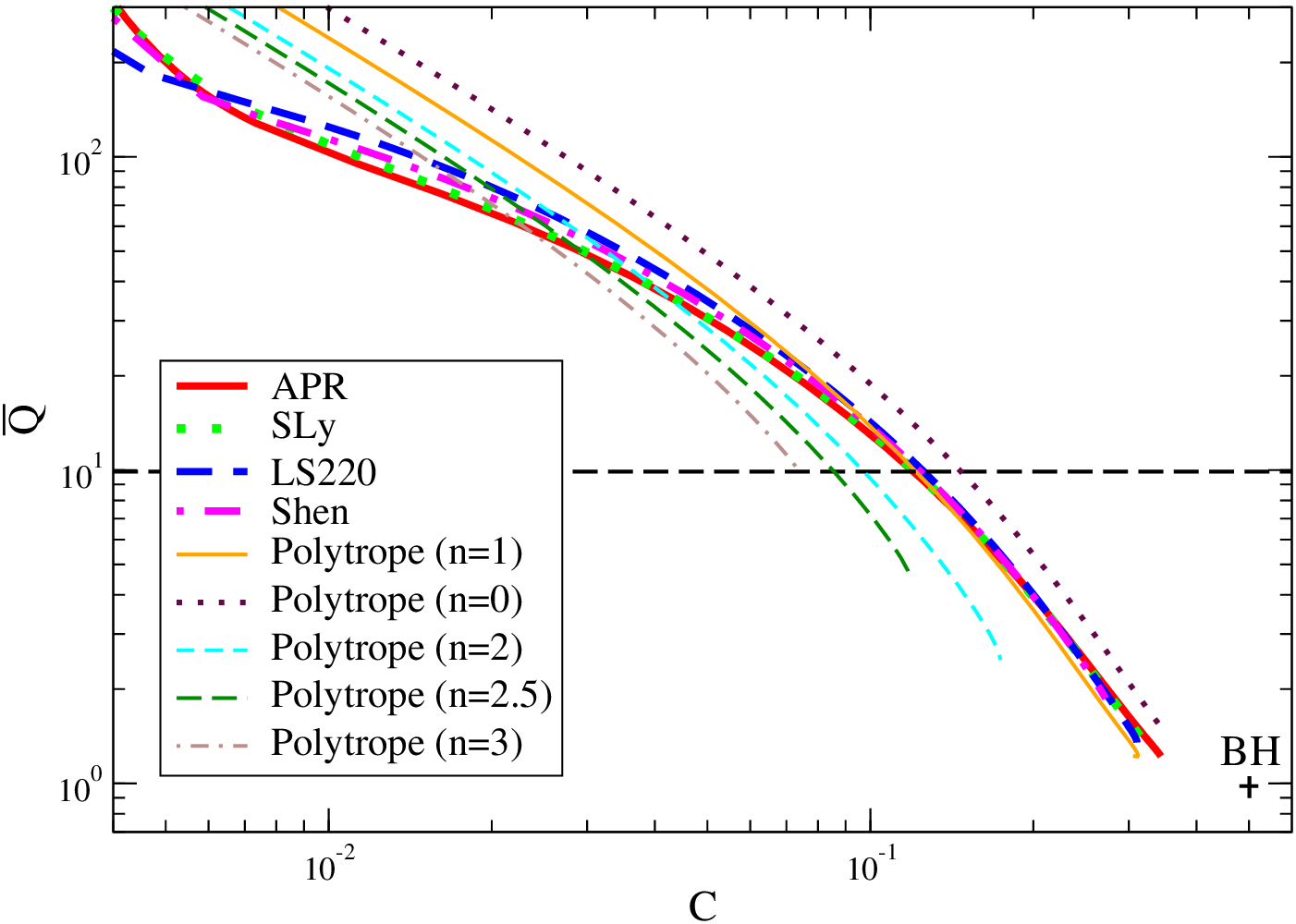}  
\caption{\label{fig:QbarMC} (Color Online) Dimensionless quadrupole moment $\bar{Q}$, defined in Eq.~\eqref{bar-Q}, as functions of $M_\NS$ (left) and $C$ (right) for various EoSs. The horizontal dashed lines at $\Qbar = 9.89$ correspond to a NS with $M_{\NS} = 1 M_{\odot}$; curves below this line have higher $M_{\NS}$ and $C$. Observe that the $\Qbar$ curves for realistic EoSs approach each other as $C$ increases, and moreover, they approach the BH limit $\Qbar_{\BH} = 1$ as $C \to 0.5$.}
\end{center}
\end{figure*}

\subsection{Interior Solutions}

Let us begin by Taylor-expanding Eqs.~\eqref{xi2}--\eqref{h2R} about the NS center and solving the expanded equations to obtain
\allowdisplaybreaks
\ba
\label{h20}
h_2 (R) &=&  B R^2 + \mathcal{O}(R^4)\,, \\
\label{k20}
K_2(R) &=& - B R^2 + \mathcal{O}(R^4)\,, \\
\label{m20}
m_2(R) &=& - B R^3 + \mathcal{O}(R^5)\,, \\
\label{xi20}
\xi_2(R) &=& - \frac{3 B + e^{-\nu_c} \omega_c^2}{4\pi (\rho_c +3p_c)}  R + \mathcal{O}(R^3), \quad (R \to 0^+)\,, \nn \\
\ea
where $B$ is a constant that determines the NS quadrupole moment. As before, the constant $\nu_c$ is defined as $\nu_c \equiv \nu(r_\epsilon)$. 

We numerically solve the evolution Eqs.~\eqref{k2R} and~\eqref{h2R} with the initial conditions of Eqs.~\eqref{h20} and~\eqref{k20}, using an adaptive 4th-order Runge-Kutta algorithm~\cite{gsl}. As before, when solving these equations we must impose the following boundary conditions, such that $h_2$ and $K_2$ are continuous at the NS surface:
\be
h_2^\inter (\RNS) = h_2^\ext (\RNS), \quad K_2^\inter (\RNS) = K_2^\ext (\RNS)\,.
\ee
These matching conditions determine the constants $A$ in Eqs.~\eqref{h2-ext} and~\eqref{k2-ext} and $B$ in Eqs.~\eqref{h20} and~\eqref{k20}. 

In practice, we follow~\cite{hartle1967,kent-CSNS} and first solve the interior solution as a sum of a particular solution, with some test-value for $B$, and the product of an undetermined constant and the homogeneous solution. We then fix this undetermined constant, together with $A$, by requiring that the interior and exterior solutions match at the NS surface. We have checked the results obtained through this method by solving the equations using the Riccati method~\cite{dieci1,dieci2,takata}.

Figure~\ref{fig:QbarMC} shows the dimensionless rotationally-induced quadrupole moment $\bar{Q}$ as functions of $M_\NS$ and $C$, where $\bar{Q}$ is defined by
\be
\label{bar-Q}
\bar{Q} \equiv -\frac{Q^\rot}{M_\NS^3 \chi^2}\,,
\ee
where we recall that the dimensionless spin parameter $\chi$ is defined by $\chi \equiv S/M_\NS^2$. This $\bar{Q}$ is the same as the dimensionless quadrupole moment $a$ in~\cite{poisson-quadrupole}. As in the $\bar{I}$ case, the $\bar{Q}$ curves for realistic EoSs approach each other as $C$ increases. Moreover, these curves also approach the $\Qbar$ value for a BH as the compactness approaches $0.5$. As before, however, the NS sequence does not go to a BH solution for any finite choice of central density.

\subsection{Rotational Love Number}

With the quadratic isolated NS solutions at hand, we can now introduce the rotational Love number~\cite{mora-will}. In general, Love numbers represent the deformability of a NS away from sphericity. The rotational Love number, in particular, refers to the deformability of a NS due to its spin. 

Love numbers are defined in a {\emph{buffer zone}}, the region ${\cal{R}} \gg R \gg \RNS$, where ${\cal{R}}$ is the radius of curvature of the source of the perturbation. For example, the $(t,t)$ component of the metric can be expanded in the buffer zone as~\cite{mora-will,Yunes:2005nn,Yunes:2006iw,JohnsonMcDaniel:2009dq,hinderer-love,Chatziioannou:2012gq} 
\ba
\label{tidal-asymp}
\frac{1-g_{tt}}{2} &=& - \frac{M_\NS}{R} - \frac{4\pi}{5} \frac{Q^\rot}{R^3} \sum_m Y_{2m} ( \hat{\Omega} ) Y_{2m}^* ( \hat{n} ) + \mathcal{O} \left( \frac{\RNS^{4}}{R^4} \right) \nn \\
&& + \frac{4\pi}{15} \mathcal{E}^\rot R^2 \sum_m Y_{2m} ( \hat{\Omega} ) Y_{2m}^* (\hat{n} )  + \mathcal{O}\left(\frac{R^{3}}{{\cal{R}}^3}\right) \nn \\
&=& - \frac{M_\NS}{R} - \frac{Q^\rot}{R^3} P_2 (\hat{\Omega} \cdot \hat{n}) + \mathcal{O} \left( \frac{\RNS^{4}}{R^4} \right) \nn \\
&& + \frac{1}{3} \mathcal{E}^\rot R^2 P_2 (\hat{\Omega} \cdot \hat{n})  + \mathcal{O}\left(\frac{R^{3}}{{\cal{R}}^3}\right)\,.
\ea
The quantity $\mathcal{E}^\rot$ is related to the trace of the rotationally-induced, electric, quadrupole tidal tensor, i.e.~the quadrupolar contribution of the centrifugal potential. In the Newtonian limit, this quantity reduces to $\mathcal{E}^\rot = \Omega_\NS^2$~\cite{mora-will}. As usual, $Y_{2m} ( \hat{\Omega} )$ are the $\ell =2$ spherical harmonics in the $\hat{\Omega}$ direction, where $\hat{n}$ is the principal axis of the perturbation, which in this case corresponds to the unit vector of the spin angular momentum $\hat{S}$.

The $\ell=2$ rotational Love number $\lambda^\rot$ is defined by~\cite{mora-will,berti-iyer-will}
\be
\lambda^\rot \equiv - \frac{Q^\rot}{\mathcal{E}^\rot} = -\frac{Q^\rot}{\Omega_{\NS}^2}\,,
\ee
where the second equality uses the Newtonian expression for ${\cal{E}}^{\rot}$. As defined here, $\lambda^\rot$ has unit of (mass)$^5$ or (length)$^5$ (recall that we use geometric units throughout this paper, where $c=1=G$), and thus, there are 2 natural ways of normalizing it~\cite{mora-will,berti-iyer-will};
\ba
k_2^\rot & \equiv & \frac{3}{2} \frac{\lambda^\rot}{\RNS^5}\,,  \\
\bar{\lambda}^\rot &\equiv& \frac{\lambda^\rot}{M_\NS^5} = \frac{2}{3} k_2^\rot C^{-5}\,.
\ea
By using Eqs.~\eqref{I},~\eqref{bar-I} and~\eqref{bar-Q}, one can rewrite $\bar{\lambda}^\rot$ as
\be
\bar{\lambda}^\rot = \Ibar^2 \Qbar\,.
\ee
In this paper, we refer to $k_2^\rot$ as the $\ell=2$ rotational apsidal constant, while we refer to $\bar{\lambda}^\rot$ as the $\ell=2$ dimensionless rotational Love number.

\section{Tidally-Deformed NS Solutions}
\label{sec:tidal}

Up until now we have concentrated on isolated NSs in the slow-rotation approximation. We will now switch gears and consider NSs in a binary system. We focus on one of the binary components, the primary, and study how it is tidally deformed by its companion, assuming the primary is not spinning. One can construct tidally-deformed NS solutions in a manner similar to the construction of slowly-rotating solutions. In both cases, the deformation (either due to rotation or tidal effects) is treated as a small deformation away from sphericity. 

\subsection{Einstein Equations and Exterior Solutions}

The leading-order effect of tidal perturbations enters at $\mathcal{O}(\epsilon^2)$. This is because this effect is generated by an electric tidal perturbation, which must be of even parity. Moreover, in this section we are interested in non-rotating tidally deformed NSs, so we can set $\omega_1=0$ in Eqs.~\eqref{m2}--\eqref{h2R}. By eliminating $m_2$ and $K_2$ from these 3 equations, one obtains a master equation for $h_2$~\cite{hinderer-love}:
\begin{align}
\label{h2RR}
0=& \frac{d^2 h_2}{dR^2} + \left\{ \frac{2}{R} + \left[ \frac{2M}{R^2} + 4 \pi R (p-\rho) \right] e^{\lambda} \right\} \frac{dh_2}{dR} \nn \\
-& \left\{ \frac{6 e^{\lambda}}{R^2} - 4\pi \left[ 5 \rho + 9p + (\rho+p) \frac{d\rho}{dp} \right] e^{\lambda} + \left( \frac{d\nu}{dR} \right)^2 \right\} h_2\,.
\end{align}

The observable related to tidally-deformed NS will eventually be a tidal Love number, and thus, we will need to asymptotically expand the exterior solution in the buffer zone. This time, however, the radius of curvature that defines the buffer zone is related to the tidal field generated by the companion. This radius is approximately equal to the orbital separation of the binary. Therefore, when solving Eq.~\eqref{h2RR} in the exterior region, one cannot impose asymptotic flatness to eliminate one of the constants of integration. Keeping this in mind, the solution to the above equation is~\cite{hinderer-love}
\ba
h_2^\ext &=& c_1 \left( \frac{R}{M_\NS} \right)^2 \left( 1-\frac{2M_\NS}{R} \right) \nn \\
& & \times \left[ - \frac{2M_\NS (R-M_\NS) (3 R^2 - 6 M_\NS R - 2 M_\NS^2)}{R^2 (R-2M_\NS)^2} \right. \nn \\
& & \left. + 3 \ln \left( \frac{R}{R-2M_\NS} \right) \right] +c_2 \left( \frac{R}{M_\NS} \right)^2 \left( 1-\frac{2M_\NS}{R} \right)\,, \nn \\
\label{h2-ext-tides}
\ea
where $c_1$ and $c_2$ are integration constants. 

\subsection{Interior Solutions and the Tidal Love Number}

The interior solution to Eq.~\eqref{h2RR} can be obtained by solving this equation numerically with the initial condition in Eq.~\eqref{h20} and its derivative. We obtain this numerical solution in the same way as we obtained $h_{2}$ for slowly-rotating NSs. As before, the interior solution will depend on the integration constant $B$, which, in principle, is determined by matching this solution to the exterior solution in Eq.~\eqref{h2-ext-tides} at the NS surface:
\be
h_2^\inter (\RNS) = h_2^\ext (\RNS), \quad h_2'{}^\inter (\RNS) = h_2'{}^\ext (\RNS)\,.
\label{h2-tidal-matching}
\ee
Notice that by using Eq.~\eqref{h2-tidal-matching}, we can re-express $c_1$ and $c_2$ in terms of $h_2(\RNS)$, $h_2'(\RNS)$ and the NS compactness $C$.

With the interior solution in hand, let us now define the tidal Love number. As in the case of the rotational Love number, the tidal one characterizes the deformability of a NS away from sphericity, but this time due to the presence of a tidal field induced by a companion. In the buffer zone, the $(t,t)$ component of the metric takes the form of Eq.~\eqref{tidal-asymp}, but with $Q^\rot \to Q^{\tid}$ and $\mathcal{E}^\rot \to {\cal{E}}^{\tid}$, where $Q^\tid$ and ${\cal{E}}^{\tid}$ correspond to the tidally-induced quadrupole moment and the tidal potential, induced by the companion. We then define the tidal Love number $\lambda^\tid$ by
\ba
\lambda^\tid & \equiv & - \frac{Q^\tid}{\mathcal{E}^\tid}\,,
\label{tid-Love-def-ori}
\ea
and its dimensionless versions $k_2^\tid$ and $\lambdabartid$ by 
\ba
k_2^\tid &\equiv & \frac{3}{2} \frac{\lambda^\tid}{\RNS^5}\,, \\
\bar{\lambda}^\tid &\equiv & \frac{\lambda^\tid}{M_\NS^5} = \frac{2}{3} k_2^\tid C^{-5}\,.
\label{tid-Love-def}
\ea
Following~\cite{flanagan-hinderer-love}, we here refer to $\lambda^\tid$ as the tidal Love number\footnote{In some references, $\lambda^\tid$ is called the tidal deformability and the word ``tidal Love number'' is reserved for $k_2^\tid$.} and we refer to $k_2^\tid$ and $\lambdabartid$ as the tidal apsidal constant and dimensionless tidal Love number, respectively.

\begin{figure*}[htb]
\begin{center}
\includegraphics[width=8.0cm,clip=true]{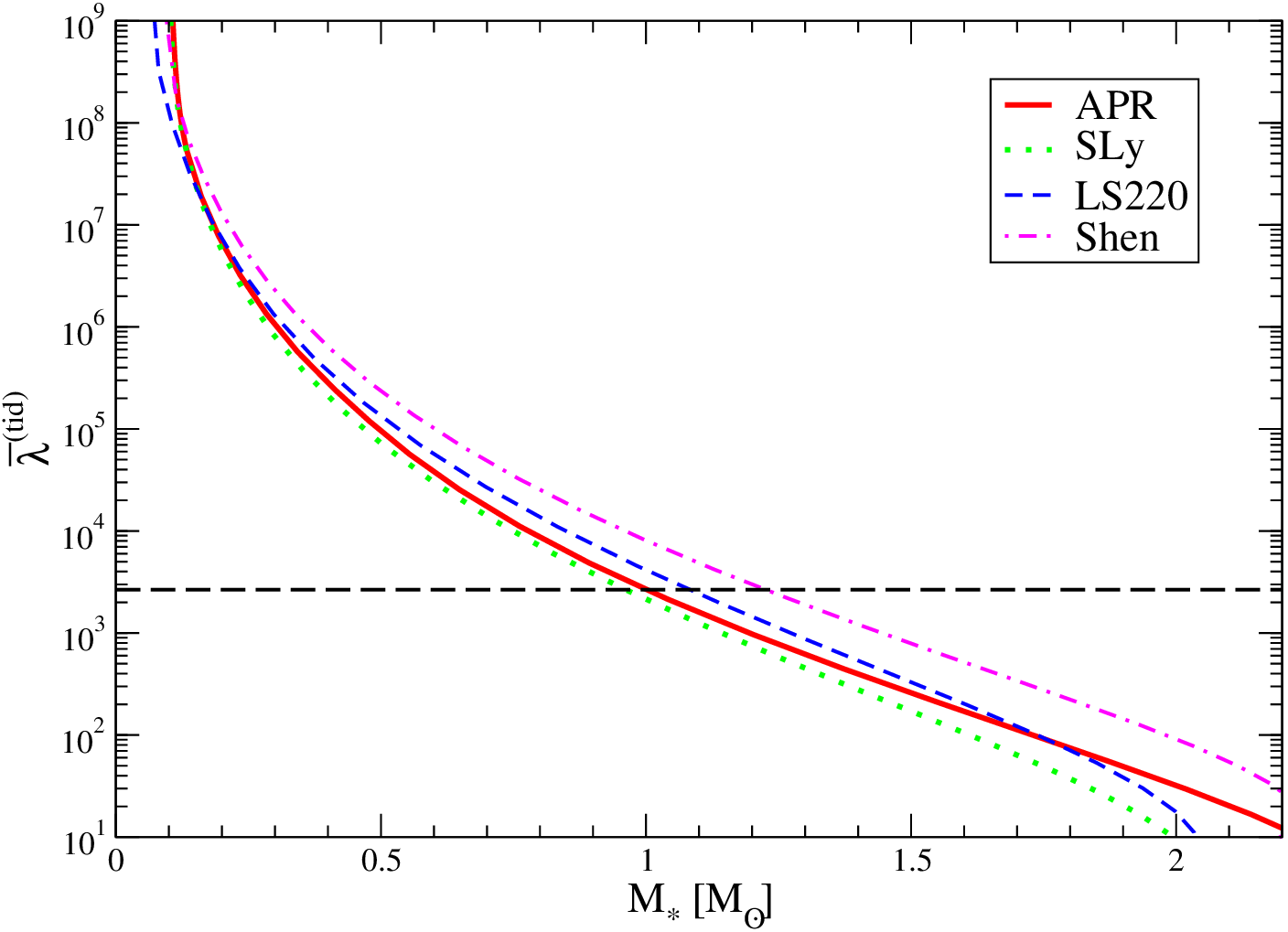}  
\includegraphics[width=8.1cm,clip=true]{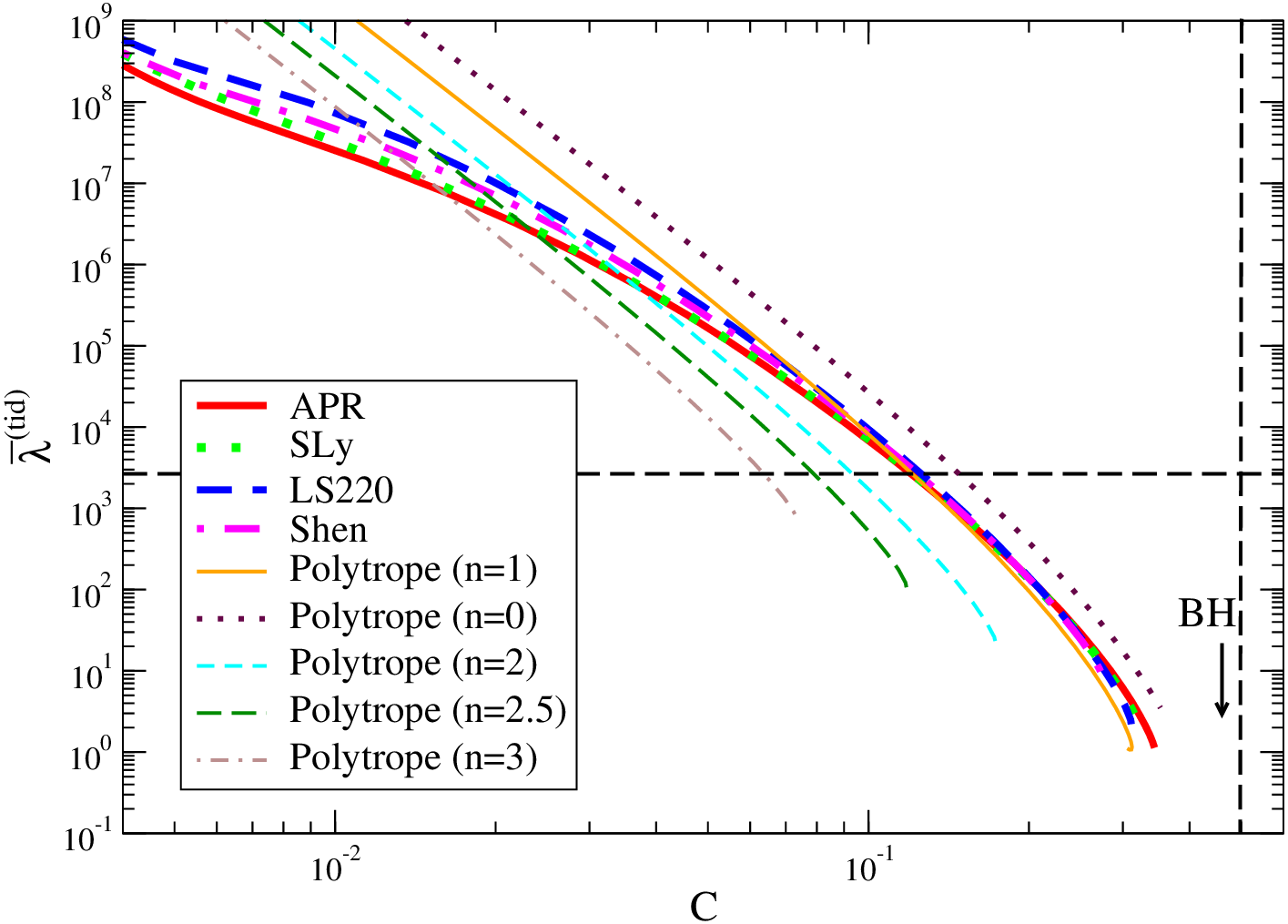}  
\caption{\label{fig:lambdaMC} (Color Online) Dimensionless tidal Love number $\bar{\lambda}^\tid$ as functions of $M_\NS$ (left) and $C$ (right) for various EoSs. The horizontal dashed lines at $\lambdabartid = 2.66 \times 10^3$ corresponds to a star with $M_{\NS} = 1 \; M_{\odot}$; the region below this line corresponds to stars with larger mass and compactness. Observe that the $\lambdabartid$ curves for realistic EoSs approach each other as $C$ increases, and moreover, approaches the BH limit $\lambdabartid_{\BH} = 0$ as $C \to 0.5$.}
\end{center}
\end{figure*}
The prescription of the tidal Love number is completed by finding the value of ${\cal{E}}^{\tid}$, which is determined by the asymptotic behavior of $h_{2}$ in the buffer zone. Taylor-expanding this quantity in the buffer zone, one finds~\cite{hinderer-love}
\be
h_2^\ext = \frac{16}{5} c_1 \frac{M_\NS^3}{R^3} + c_2 \frac{R^2}{M_\NS^2} + \mathcal{O}\left( \frac{M_\NS^4}{R^4}, \frac{R}{M_\NS} \right)\,.
\ee
As shown in Eq.~\eqref{tidal-asymp}, the term in the asymptotic expansion of $g_{tt}$ (or $h_{2}$) in the buffer zone that is proportional to $R^{-3}$ gives us the tidal quadrupole moment, while the term proportional to $R^{2}$ gives us the tidally-induced electric quadrupole tidal tensor. Thus, we find that $c_1$ is related to $Q^{\tid}$, while $c_2$ is related to ${\cal{E}}^{\tid}$. 

The tidal apsidal constant can then be found by taking the ratio of $c_1$ and $c_2$~\cite{hinderer-love}:
\begin{align}
\label{k2}
k_2^\tid &= \frac{8}{5} C^5 \frac{c_{1}}{c_{2}} = \frac{8}{5} C^5 (1-2 C)^2[2+2C(y-1)-y] \nn \\
&\times  \left\{ 2C[6-3y+3C(5y-8)] \right. \nn \\
&\left. +4C^3[13-11y+C(3y-2)+2C^2(1+y)] \right. \nn \\
& \left. +3(1-2C)^2[2-y+2C(y-1)] \ln(1-2C)  \right\}^{-1}\,,
\end{align}
with $y\equiv \RNS h_2'(\RNS)/h_2(\RNS)$. In the second equality, we have rewritten $c_{1,2}$ in terms of $h_{2}$, its derivative and the NS compactness.

We see then that the tidal apsidal constant only depends on $y$, which simplifies the way one must solve Eq.~\eqref{h2RR}. First, we notice that Eq.~\eqref{h2RR} is a homogeneous equation for $h_{2}$, and thus, the integration constant $B$ in Eq.~\eqref{h20} only changes the solution $h_2$ by a constant factor. Since $y \propto h_{2}'/h_{2}$ does not depend on this overall factor, it suffices to solve Eq.~\eqref{h2RR} with an arbitrary test value for $B$, if one is only interested in the tidal apsidal constant. We have calculated the tidal apsidal constant, as well as the tidal Love number for a sequence of stars with varying $M_{\NS}$ and $C$. We have found that our results agree exactly with Figs.~1 and~2 of~\cite{hinderer-lackey-lang-read}. 

Figure~\ref{fig:lambdaMC} shows the dimensionless Love number $\bar{\lambda}^\tid$ as  functions of $M_\NS$ and $C$. Observe that the $\bar{\lambda}^\tid$ curves for realistic EoSs approach each other as $C$ increases, and moreover, they approach the BH limit as $C \to 0.5$. Once more, as before, the BH limit cannot be taken from the sequence of NS considered, as there is no finite central density that would lead to BH formation.

\section{I-Love-Q Relations}
\label{sec:I-Love-Q}

Now that the moment of inertia, quadrupole moment and Love numbers have been calculated, let us present the universal I-Love-Q relations. We first show numerical results and a fitting curve through these. Then, we obtain analytic I-Love-Q relations for the $n=0$ and 1 polytropic EoSs in the Newtonian limit.

\subsection{Numerical Results}

\begin{figure*}[htb]
\begin{center}
\includegraphics[width=8.0cm,clip=true]{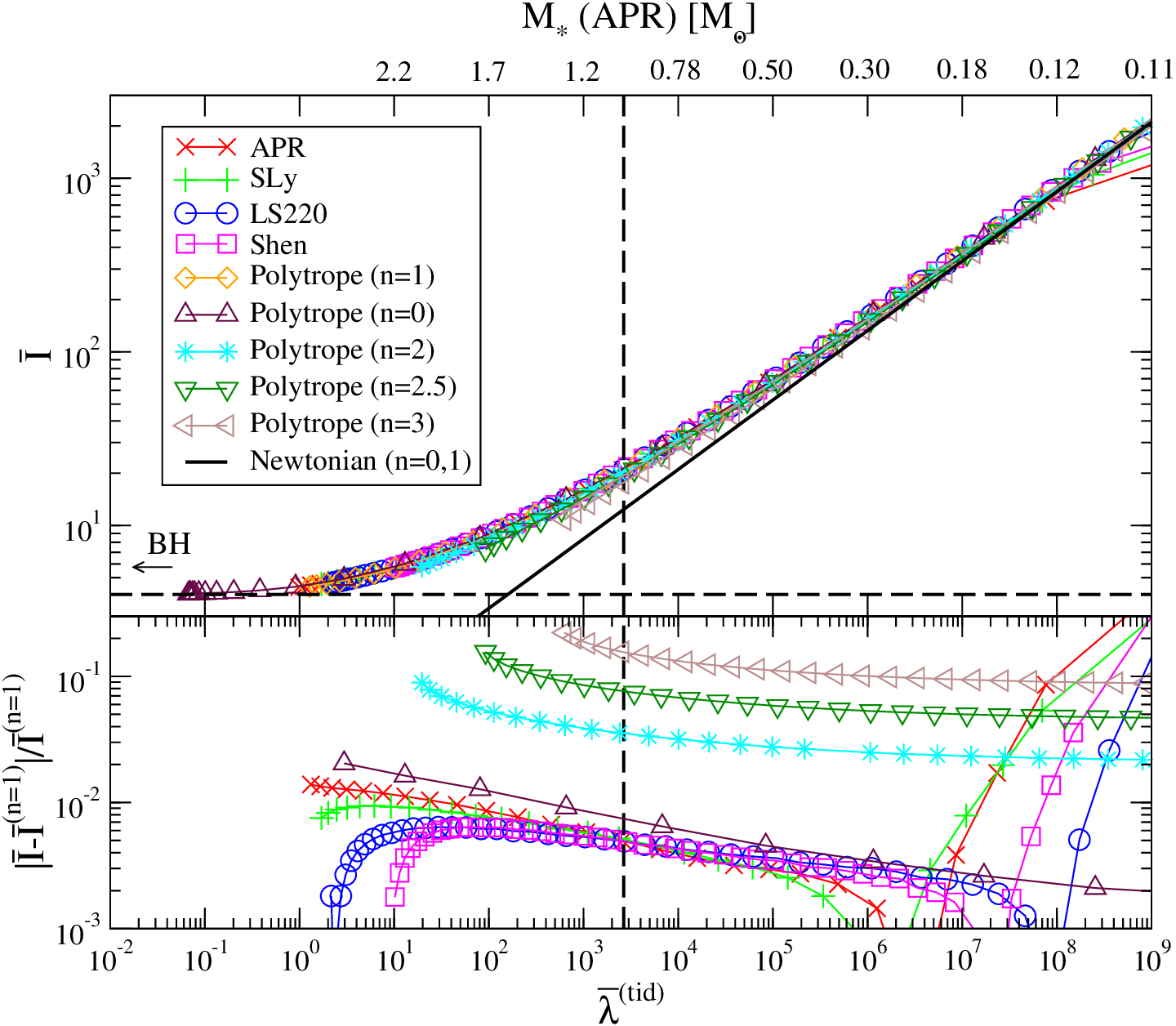}  
\hspace{0.5cm} 
\includegraphics[width=8.0cm,clip=true]{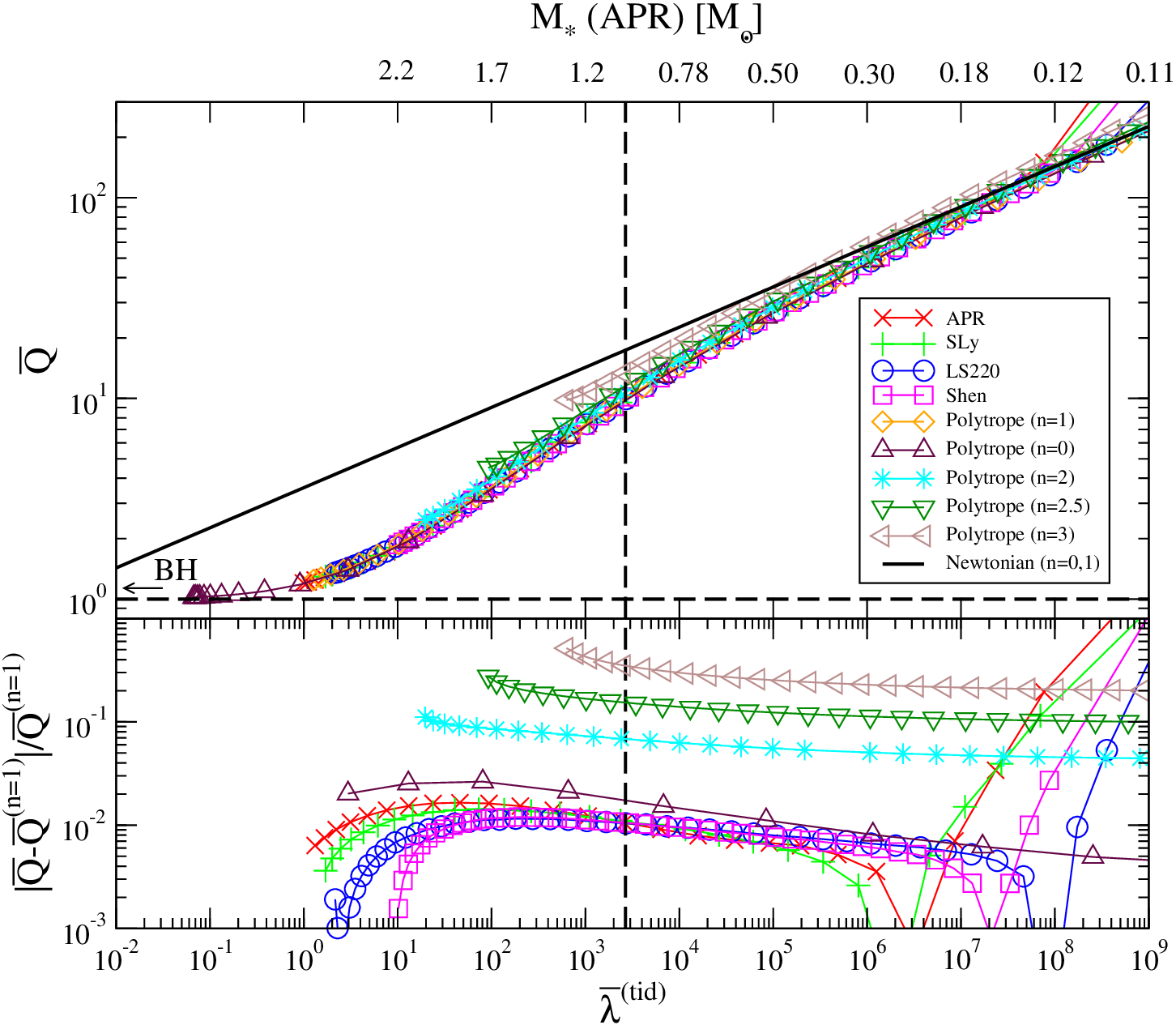}  \\
\vspace{0.5cm}
\includegraphics[width=8.0cm,clip=true]{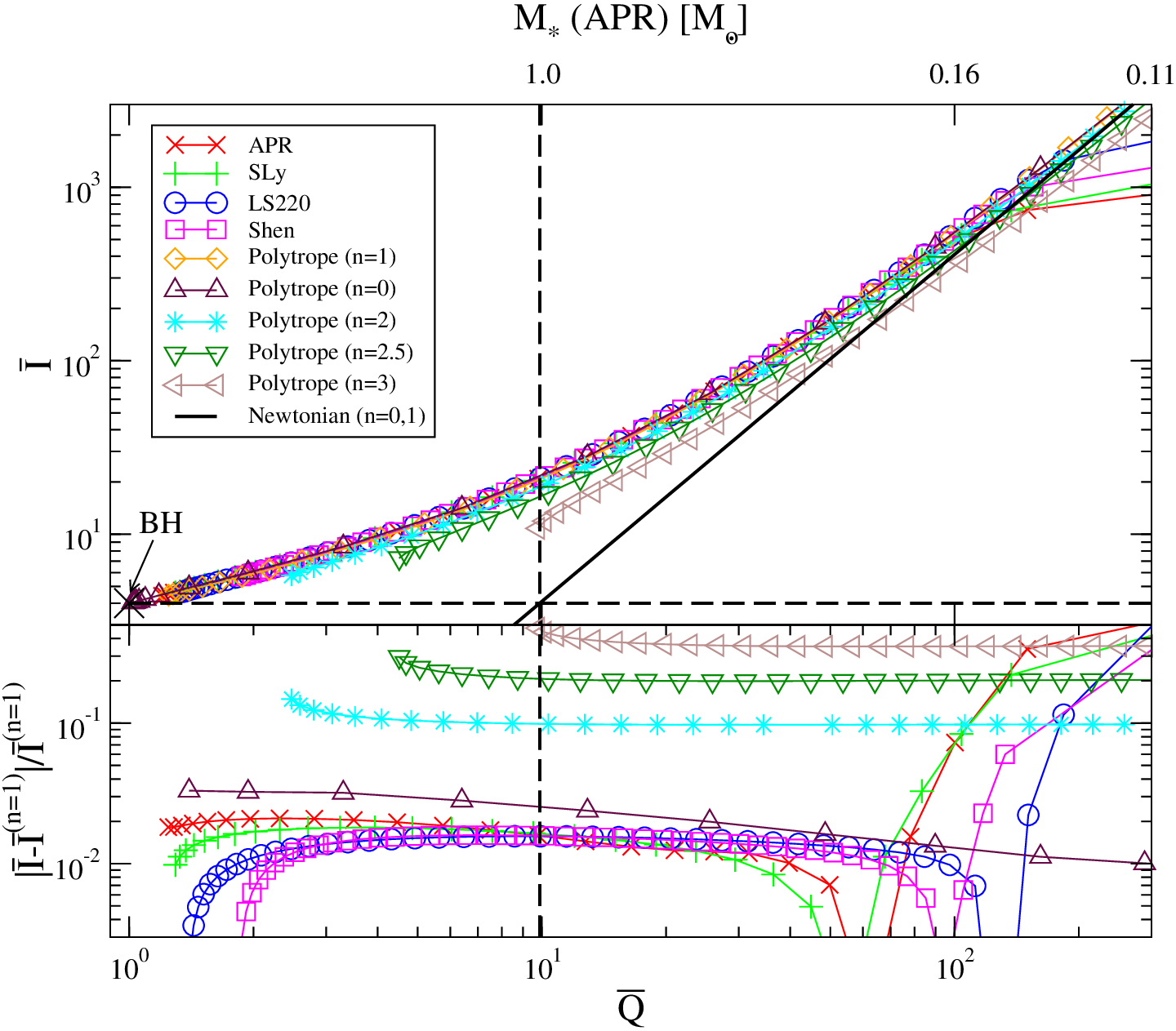}  
\hspace{0.5cm} 
\includegraphics[width=8.0cm,clip=true]{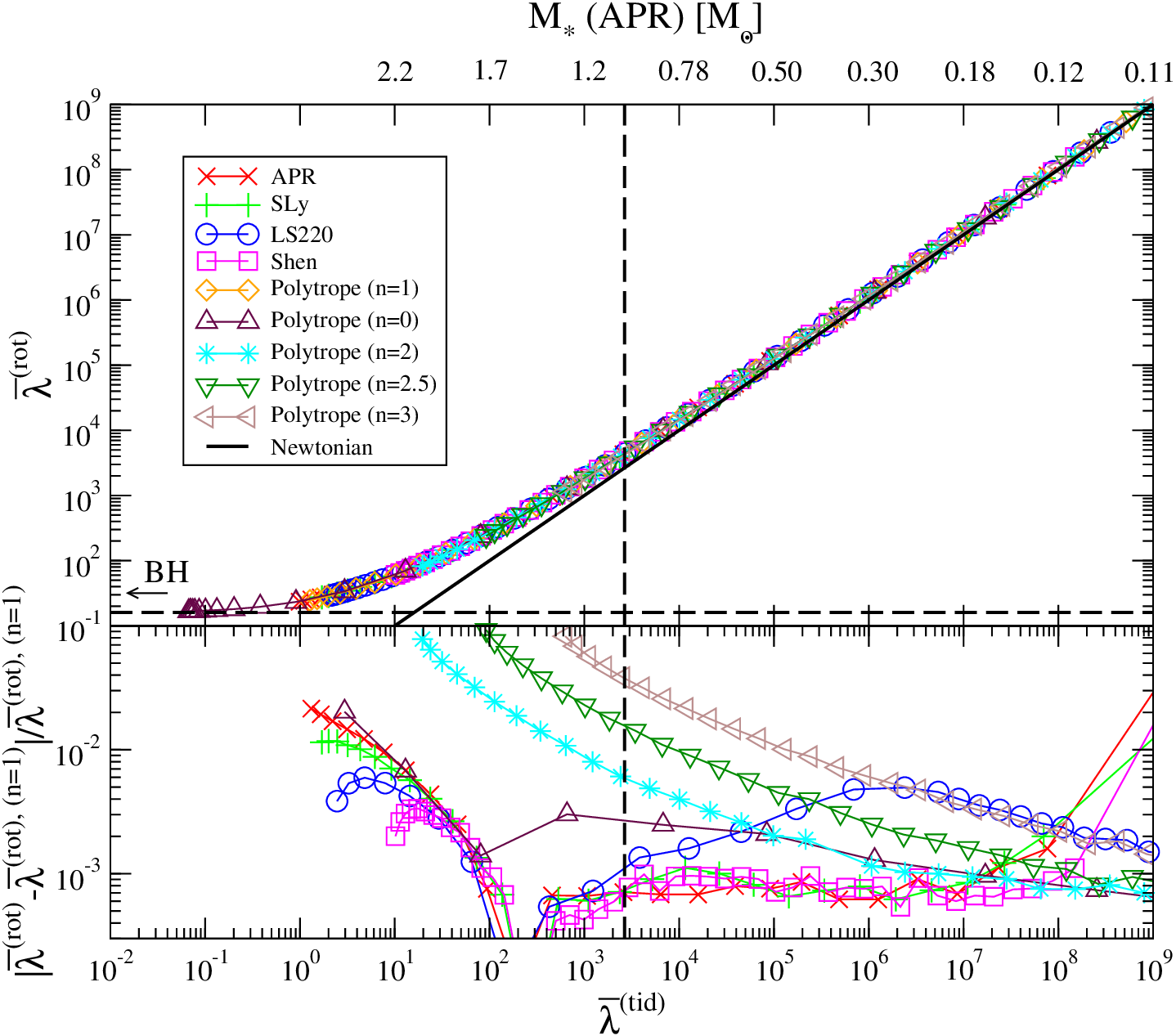}  
\caption{\label{fig:I-Love-Q} (Color Online) I-Love (top left), Q-Love (top right), I-Q (bottom left) and Love-Love (bottom right) relations for various EoSs. Black thin solid lines represent the Newtonian limit for the polytropic EoSs with $n=0$ and $n=1$. The horizontal lines at $\Ibar = 4$, $\Qbar = 1$, $\lambdabartid = 0$ and $\lambdabarrot=16$ correspond to the (non-rotating) BH limiting values. Observe that as one increases the NS compactness (toward the left of each panel), the I-Love-Q relations approach the BH limit. The (barred) quantities $\Ibar$, $\Qbar$, $\lambdabartid$ and $\lambdabarrot$ do not depend on the NS spin to second-order in the slow-rotation approximation. The parameter varied along each curve is the NS central density, or equivalently the NS compactness, both increasing to the left of the plots. For reference, the vertical dashed lines correspond to $M=1M_\odot$ for the APR EoS. The top axis of each panel shows the corresponding NS mass for the APR EoS. The bottom part of each panel shows the relative fractional differences between the relations, using the $n=1$ polytropic curve as a reference.}
\end{center}
\end{figure*}
Figure~\ref{fig:I-Love-Q} shows universal relations between dimensionless quantities, $\bar{I}$, $\Qbar$, $\bar{\lambda}^\tid$ and $\bar{\lambda}^\rot$ for various EoSs. Notice that these (barred) 4 quantities are essentially {\emph{independent of the NS spin}}, to second order in the slow-rotation approximation\footnote{Formally, the barred quantities depend on the mass $M_{*}$, which is not the observed mass. The two are related by $M_{\rm ons} = M_{*} \left(1 + \chi^{2} \delta M\right)$. Rapidly rotating NS calculations indicate that $\delta M = {\cal{O}}(0.3)$, and thus, for $\chi < 0.1$, such spin dependence introduces corrections of ${\cal{O}}(0.003)$.}.  The parameter varied along each curve is the NS central density, or equivalently the NS compactness. Therefore, for the polytropic EoS the I-Love-Q relations are independent of the polytropic amplitude coefficient $K$ in Eq.~\eqref{polytropic}. The bottom part of each panel shows the relative fractional difference between each of the curves and curve corresponding to the $n=1$ polytropic EoS. For reference, the top axes show the NS mass with the APR EoS. The vertical dashed lines correspond to $M_\NS=1M_\odot$ for the APR EoS and points to the left of these lines correspond to more massive NSs with higher compactness. Observe that, for realistic EoSs with $M_\NS>1M_\odot$, the fractional relative differences are $\mathcal{O}(1)\%$. Observe also that the polytropic I-Love-Q relations deviate from those with realistic EoSs as one increases $n$, ie.~as the NS becomes more centrally-concentrated. We see this as evidence that the I-Love-Q trio is most sensitive to the NS outer layers, where realistic EoSs mostly agree with each other. Curiously, the fractional relative difference between the $n=1$ and $n=0$ polytrope (constant density NS star) is also of ${\cal{O}}(1)\%$, but this case will be studied analytically in Sec.~\ref{analytic-exp}.

Observe that, in general, the dependence of the I-Love-Q relations on any EoS becomes weaker as the NS compactness $C$ increases, ie.~from right to left in any of the panels of Fig.~\ref{fig:I-Love-Q}. This may be evidence that at least part of the universality observed is due to the NS sequence approaching a BH as $C \to 0.5$, where the latter has no internal-structure dependence by the no-hair theorems. As $C \to 0.5$, the asymptotic values of $\bar{I}$, $\bar{\lambda}^\tid$ and $\Qbar$ for a BH are $\bar{I} \to 4$~\cite{membrane}, $\bar{\lambda}^\tid \to 0$~\cite{binnington-poisson} [see also Eq.~\eqref{k2}] and $\Qbar \to 1$~\cite{geroch,hansen,poisson-quadrupole} respectively. In each panel of Fig.~\ref{fig:I-Love-Q}, we show the BH limit of either $\Ibar=4$, $\Qbar =1$ or $\lambdabarrot = 16$ as a horizontal dashed line. Observe that the I-Love-Q relations asymptote to such BH values as $C$ increases. However, one cannot quite reach this limit, as one can never construct a BH solution by increasing the central density of a NS solution by a finite amount. We think that this is why the relative fractional differences shown in the bottom panels of Fig.~\ref{fig:I-Love-Q} do not decrease to zero as one increases the central density. 

Unlike the other relations, the $\bar{\lambda}^\rot$--$\bar{\lambda}^\tid$ relation (bottom, right panel of Fig.~\ref{fig:I-Love-Q}) depends very weakly on the EoS even when the NS compactness is relatively small (as one approaches the Newtonian limit). In fact, one can show that the relation $\bar{\lambda}^\rot=\bar{\lambda}^\tid$ holds exactly in the Newtonian limit for any EoS, as we will discuss in Sec.~\ref{sec:Love-Newton}.

{\renewcommand{\arraystretch}{1.2}
\begin{table}
\begin{centering}
\begin{tabular}{cccccccc}
\hline
\hline
\noalign{\smallskip}
$y_i$ & $x_i$ &&  \multicolumn{1}{c}{$a_i$} &  \multicolumn{1}{c}{$b_i$}
&  \multicolumn{1}{c}{$c_i$} &  \multicolumn{1}{c}{$d_i$} &  \multicolumn{1}{c}{$e_i$}  \\
\hline
\noalign{\smallskip}
$\bar{I}$ & $\bar{\lambda}^\tid$ && 1.47 & 0.0817  & 0.0149 & $2.87\times 10^{-4}$ & $-3.64\times 10^{-5}$\\
$\bar{I}$ & $\Qbar$ && 1.35  & 0.697 & -0.143  & $9.94\times 10^{-2}$ & $-1.24\times 10^{-2}$\\
$\Qbar$ & $\bar{\lambda}^\tid$ && 0.194  & 0.0936 & 0.0474  & $-4.21\times 10^{-3}$ & $1.23\times 10^{-4}$\\
\noalign{\smallskip}
\hline
\hline
\end{tabular}
\end{centering}
\caption{Estimated numerical coefficients for the fitting formula of the I-Love, I-Q and Q-Love relations given in Eq.~\eqref{fit}.}
\label{table:coeff}
\end{table}
}

\begin{figure}[htb]
\begin{center}
\includegraphics[width=8.0cm,clip=true]{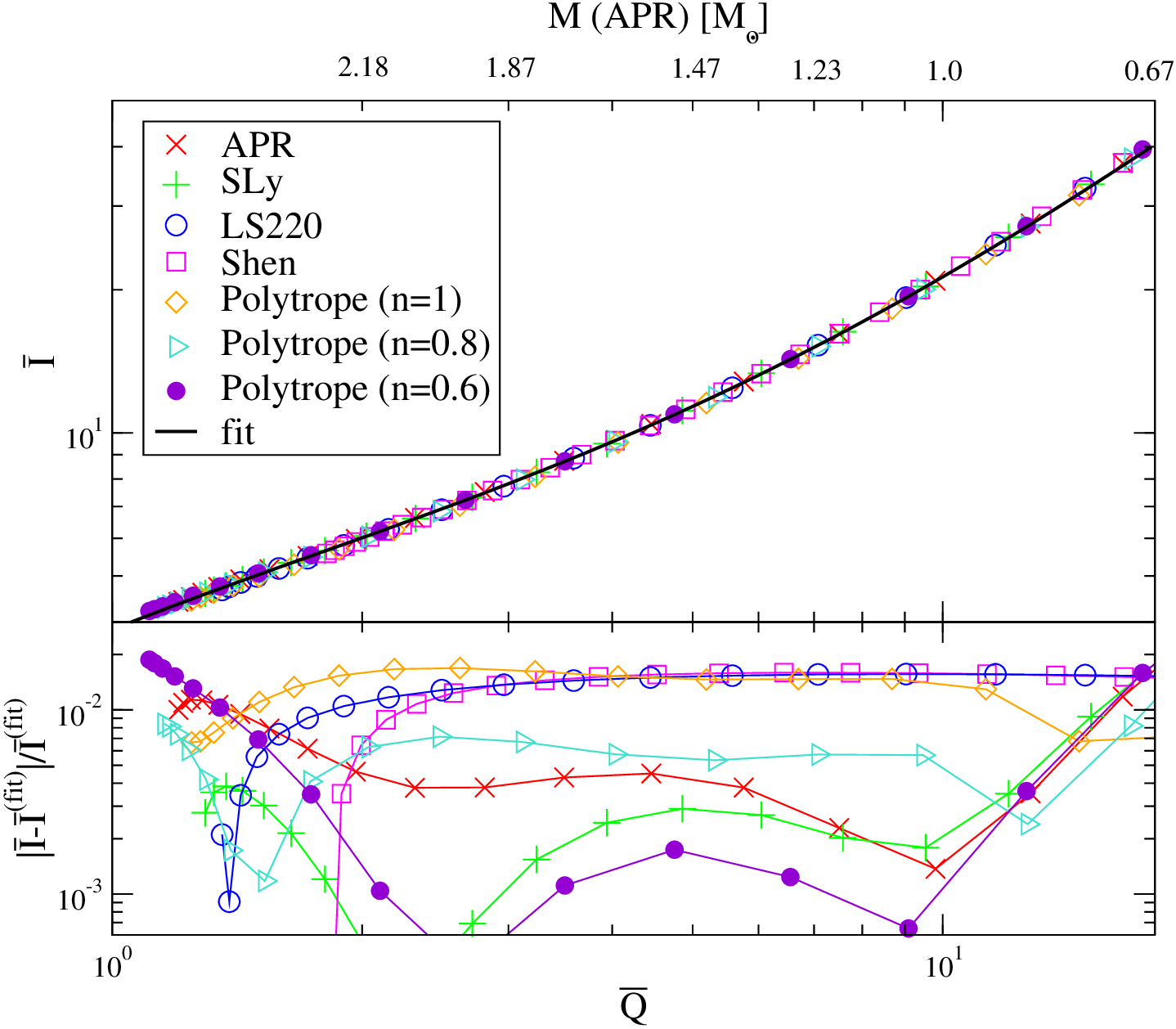}  
\caption{\label{fig:I-Q-fit} (Color Online) (Top) Fitting curve (solid curve) and numerical results (points) of the I-Q relation with various EoSs. (Bottom) Fractional errors between the fitting curve and numerical results.
}
\end{center}
\end{figure}

Given the universality of the I-Love-Q relations, one can fit them all with a single curve:
\be
\ln y_i = a_i + b_i \ln x_i + c_i (\ln x_i)^2 + d_i (\ln x_i)^3+ e_i (\ln x_i)^4\,,
\label{fit}
\ee
where the coefficients are summarized in Table~\ref{table:coeff}.
Figures~\ref{fig:I-Love-fit}  and~\ref{fig:I-Q-fit} show the fitting curves for the I-Love, Q-Love and I-Q relations, together with the relative fractional difference between the fitting curves and all other EoS curves. For the polytropic EoSs, we do not show the results when $n=0$, 2, 2.5 and $3$ since such EoSs do not model NSs well; instead, we add results when $n=0.6$ and 0.8. As one can see, the fitting curves are accurate to within $\mathcal{O}(1)\%$ accuracy.

\subsection{Analytical Explanations}
\label{analytic-exp}

The universal I-Love-Q relations presented in the previous subsection are quite intriguing, and thus, they beg for an analytic explanation. We will attempt one here, by investigating these relations for certain EoSs that allow for an analytical treatment. In particular, we will study the $n=0$ and $n=1$ polytropic EoSs in the Newtonian limit, for which the moment of inertia, the quadrupole moment and the Love numbers can be computed fully analytically. We can then derive the I-Love-Q relations analytically as well to try to obtain an analytical explanation for these relations. 

\subsubsection{Newtonian $\lambda$ for Generic EoSs}
\label{sec:Love-Newton}

In Newtonian theory, the curl of the equation of hydrostatic equilibrium, $\nabla p = \rho \nabla \Psi$ where $\Psi$ is the total gravitational potential, with the Newtonian force vanishes, ie.~$\nabla \rho \times \nabla \Psi =0$. Thus, surfaces of constant $\rho$ and $\Psi$ coincide. One can express such surfaces in terms of a radial parameter $a$ as~\cite{mora-will}
\allowdisplaybreaks
\ba
r(a, \theta, \phi) &=& a \left[ 1 + \sum_{\ell, m} f_{\ell} Y_{\ell m} (\hat{\Omega}) Y_{\ell m} (\hat{n}) \right]\,, \nn \\
&=& a \left[ 1 + \frac{5}{4\pi} \sum_{\ell} f_{\ell} P_\ell (\hat{\Omega} \cdot \hat{n}) \right]\,, 
\ea
where $f_\ell$ is the dimensionless distortion function of the constant $a$. This function is related to the $\ell =2$ tidal apsidal constant $k_2^\tid{}^{,\N}$ by
\be
k_2^\tid{}^{,\N} = \frac{3-\eta_2(a_*)}{2[2 + \eta_2(a_*)]}\,,
\label{k2N1}
\ee
where $a=a_*$ denotes the surface of the star and
\be
\eta_2 (a) \equiv \frac{d \ln f_2}{d \ln a}\,.
\ee
This function can be obtained by solving the Clairaut-Radau equation~\cite{brooker-olle,mora-will}
\be
a\frac{d\eta_2}{d a} + 6 \mathcal{D} (\eta_2 +1) + \eta_2 (\eta_2 -1) -6 =0\,,
\label{Clairaut-Radau}
\ee
with the boundary condition $\eta_2 (0) =0$, where $\mathcal{D}(a) \equiv \rho(a)/\bar{\rho}(a)$ with $\bar{\rho}$ representing the mean density of the star. 

In the Newtonian limit, both rotational and tidal apsidal constants can be calculated from Eq.~\eqref{Clairaut-Radau} with the same boundary condition~\cite{mora-will}, and hence, 
\be
\bar{\lambda}^\tid{}^{,\N} = \bar{\lambda}^\rot{}^{,\N}\,.
\label{Love-Love-Newton}
\ee
This shows that the rotationally-induced and tidally-induced NS deformabilities are exactly the same in the Newtonian limit. In GR, non-linear effects modify this relation and break the equality. Equation~\eqref{Love-Love-Newton} is shown as a black, thin, solid line in the bottom, right panel of Fig.~\ref{fig:I-Love-Q}. Notice that all the curves approach this Newtonian result as one decreases the compactness, as expected. 

One can also calculate the $\ell=2$ tidal apsidal constant $k_2^\tid{}^{,\N}$ by taking the Newtonian limit ($R \gg M(R)$, $\rho \gg p$) of Eq.~\eqref{k2}~\cite{hinderer-love}, where Eq.~\eqref{h2RR} in the Newtonian limit is given by
\be
\frac{d^2 h_2}{dR^2} + \frac{2}{R} \frac{d h_2}{dR} + \left( 4 \pi \rho \frac{d\rho}{dp} - \frac{6}{R^2} \right)h_2 =0\,.
\label{h2RR-Newton}
\ee

\subsubsection{Polytropic I-Love-Q relations: $n=0$}

Now, let us investigate the I-Love-Q relations in the Newtonian limit for specific EoSs. First, we focus on the polytropic EoS with $n=0$, which corresponds to the incompressible EoS with 
\be
\rho = \rho_c \; \Theta(R_{\NS} - R)= \frac{3}{4\pi} \frac{M_\NS}{\RNS^3} \Theta(R_{\NS} - R)\,, 
\label{rho-n0}
\ee
where $\Theta(R _{\NS}- R)$ is the Heaviside function, which is unity inside the star
and zero outside. By substituting Eq.~\eqref{rho-n0} into Eq.~\eqref{I-Newton}, one obtains
\be
I^\N = \frac{2}{5} M_\NS \RNS^2\,, \quad \bar{I}^\N = \frac{2}{5} \frac{1}{C^2}\,.
\label{I-bar-n0}
\ee
Not surprisingly, this is the Newtonian moment of inertia for a sphere of constant density. 

Next, we solve Eq.~\eqref{h2RR-Newton} to obtain $\lambda^\tid{}^{,\N}$. As pointed out in~\cite{damour-nagar}, one must be careful when solving Eq.~\eqref{h2RR-Newton} for the incompressible EoS. This is because $\rho$ can be expressed as a step-function with a discontinuity at the NS surface. Thus, $d\rho/dp$ in Eq.~\eqref{h2RR-Newton} gives a delta-function centered at the NS surface. To be more precise, by using $d\rho/dR = - \rho_c \delta (R_\NS-R)$, $p(\RNS)=0$, $M_\NS=(4\pi/3) \RNS^3 \rho_c$ and the hydrostatic equilibrium equation [or the Newtonian limit of Eq.~\eqref{TOV-zeroth}] at the NS surface, $dp(\RNS)/dR = -M_\NS \rho(\RNS)/\RNS^2$, the coefficient of $h_2$ in Eq.~\eqref{h2RR-Newton} that is proportional to $d\rho/dp$ becomes $4\pi \rho d\rho/dp = 4\pi \rho (d\rho/dR) (dp/dR)^{-1} = (3/\RNS) \delta (R_{\NS}-R)$ near the surface. By taking such term into account, one can obtain the correct $k_2^\tid{}^{,\N}$ by first solving Eq.~\eqref{h2RR-Newton} with $d\rho/dp=0$ and then shifting $y^\N$ by -3~\cite{damour-nagar}. By taking all of this into account, one obtains $y^\N = -1$, and thus, $k_2^\tid{}^{,\N} = 3/4$~\cite{damour-nagar}. This agrees with the classic result in~\cite{brooker-olle}. By using Eq.~\eqref{tid-Love-def}, $\bar{\lambda}^\tid{}^{,\N}$ becomes
\be
\bar{\lambda}^\tid{}^{,\N} = \frac{1}{2}\frac{1}{C^5}\,.
\label{lambda-bar-n0}
\ee

Let us now move on to the rotationally-induced quadruple moment. Rotating configurations of constant density stars can be described by Maclaurin spheroids~\cite{shapiro-teukolsky}. The quadrupole moment in Newtonian theory is given by~\cite{laarakkers}
\be
Q^\rot{}^{,\N} = 2\pi \int_0^{\pi} \int_0^{r_* (\theta)} \rho(r,\theta) r^4 P_2 (\cos \theta) \sin \theta \ dr \ d\theta\,. 
\label{Q-integ}
\ee
The surface of the star $r=r_* (\theta)$ is in turn given by
\be
r_* (\theta) = \left( \frac{\sin^2\theta}{b^2} + \frac{\cos^2\theta}{c^2} \right)^{-1/2}\,,
\label{RNS-n0}
\ee
where $b$ and $c$ are the semi-major and semi-minor axes, respectively. By substituting Eq.~\eqref{RNS-n0} and $\rho=\rho_c$ into Eq.~\eqref{Q-integ}, one obtains
\be
Q^\rot{}^{,\N} = -\frac{4\pi}{15} \rho_c b^2 c (b^2-c^2)\,.
\label{Q-n0}
\ee

From the equation of hydrostatic equilibrium,
\be
\frac{d\bm{v}}{dt} = - \frac{1}{\rho} \bm{\nabla}p - \bm{\nabla} \Psi\,,
\ee
where boldfaced quantities refer to three-dimensional Euclidean vectors, 
with $\bm{v} = \bm{\Omega} \times \bm{r}$ and $\bm{\Omega}$ the NS angular velocity vector, one obtains~\cite{shapiro-teukolsky}
\ba
\Omega_\NS &=& \left\{ 2\pi \rho_c \left[ \frac{\sqrt{1-e^2}(3-2e^2)}{e^3} \sin^{-1}e - \frac{3(1-e^2)}{e^2} \right] \right\}^{1/2} \nn \\
&=& \sqrt{\frac{8\pi}{15}\rho_c} \ e + \mathcal{O}(e^3)\,,
\label{Omega-n0}
\ea
where $\Omega_\NS = |\bm{\Omega}|$  and $e$ is the eccentricity defined by
\be
e \equiv \sqrt{1- \frac{c^2}{b^2}}\,.
\label{e-n0}
\ee
Using Eqs.~\eqref{Omega-n0} and~\eqref{e-n0}, we can eliminate $c$ from Eq.~\eqref{Q-n0} and substitute $b= \RNS + \mathcal{O}(\Omega_\NS^2)$ to obtain
\be
Q^\rot{}^{,\N} = - \frac{1}{2} \RNS^5 \Omega_\NS^2 + \mathcal{O}(\Omega_\NS^4)\,. 
\ee
Keeping only the leading term, one obtains~\cite{laarakkers}
\be
\Qbar^{\N} = \frac{25}{8} \frac{1}{C}\,.
\label{Q-bar-n0}
\ee

The dimensionless rotational Love number $\bar{\lambda}^\rot{}^{,\N}$ can be calculated as
\be
\bar{\lambda}^\rot{}^{,\N} = (\bar{I}^\N)^2 \Qbar^{\N} = \frac{1}{2} \frac{1}{C^5}\,,
\ee
which agrees with $\bar{\lambda}^\tid{}^{,\N}$ given in Eq.~\eqref{lambda-bar-n0}. This then verifies Eq.~\eqref{Love-Love-Newton}.

From Eqs.~\eqref{I-bar-n0},~\eqref{lambda-bar-n0} and~\eqref{Q-bar-n0}, one obtains the \emph{I-Love-Q} relations in the Newtonian limit for the incompressible EoS:
\ba
\bar{I}^\N &=& C_{\bar{I} \bar{\lambda}}^{(n=0)} \left[ \bar{\lambda}^\rot{}^{,\N} \right]^{2/5}\,, \quad \bar{I}^\N = C_{\bar{I} \bar{Q}}^{(n=0)} \left[ \Qbar^{\N} \right]^{2}\,, \nn \\ 
\Qbar^{\N} &=& C_{\bar{Q} \bar{\lambda}}^{(n=0)} \left[ \bar{\lambda}^\rot{}^{,\N} \right]^{1/5}\,,
\label{I-Love-Q-n0}
\ea
with
\ba
\label{coeff-I-Love-n0}
C_{\bar{I} \bar{\lambda}}^{(n=0)} &=& \frac{2^{7/5}}{5} \approx 0.528\,, \\
C_{\bar{I} \bar{Q}}^{(n=0)} &=& \frac{128}{3125} \approx 0.0410\,, \\
\label{coeff-Q-Love-n0}
C_{\bar{Q} \bar{\lambda}}^{(n=0)} &=& \frac{25}{2^{14/5}} \approx 3.59\,. 
\ea
Of course, universality would be established if the constants $C_{A}$ are independent of the EoS, with $A$ any pair in the I-Love-Q trio. We will compute the same relations for the $n=1$ polytrope next, and thus, we will verify the degree of universality quantitatively.  

\subsubsection{Polytropic I-Love-Q relations: $n=1$}

Let us now concentrate on the $n=1$ polytrope and look first at the moment of inertia. 
Equation~\eqref{TOV-zeroth} in the Newtonian limit gives the equation of hydrostatic equilibrium:
\be
\frac{dp}{dR} = - \frac{\rho M}{R^2}\,.
\label{TOV-zeroth-Newton}
\ee
From Eqs.~\eqref{tt-zeroth},~\eqref{TOV-zeroth-Newton} and $p=K \rho^2$, one can solve this equation to obtain 
\be
\rho = \frac{1}{4} \frac{M_\NS}{\RNS^2} \frac{1}{R} \sin \left( \frac{\pi R}{\RNS} \right)\,.
\ee
One can then calculate $I^\N$ (and $\bar{I}^\N$) by substituting the above equation in Eq.~\eqref{I-Newton} to find
\be
I^\N = \frac{2 (\pi^2-6)}{3\pi^2} M_\NS \RNS^2\,, \quad \bar{I}^\N = \frac{2 (\pi^2-6)}{3\pi^2} \frac{1}{C^2}\,.
\label{I-bar-n1}
\ee

Let us now consider the tidal apsidal constant. The solution to Eq.~\eqref{h2RR-Newton} can be written in terms of Bessel functions, 
as $h_2^\N \propto (R/\RNS)^{-1/2} J_{5/2} (\pi R/\RNS)$~\cite{hinderer-love,damour-nagar}. With this, we find that the apsidal constant is
\be
k_2^\tid{}^{,\N} = -\frac{1}{2} + \frac{15}{2\pi^2}\,.
\ee
This constant agrees with the numerical results of~\cite{brooker-olle}. The dimensionless tidal Love number, $\bar{\lambda}^\tid{}^{,\N}$,
is then
\be
\bar{\lambda}^\tid{}^{,\N} = \frac{15-\pi^2}{3\pi^2} \frac{1}{C^5}\,.
\label{lambda-bar-n1}
\ee

Finally, let us look at the rotationally-induced quadrupole moment. From Eq.~\eqref{Love-Love-Newton}, one easily finds that
\be
\Qbar^{\N} = \frac{\bar{\lambda}^\tid{}^{,\N}}{(\bar{I}^\N)^2} = \frac{3\pi^2 (15-\pi^2)}{4(\pi^2-6)^2} \frac{1}{C}\,.
\label{Q-bar-n1}
\ee

We now have all the necessary ingredients to compute the I-Love-Q relations in the Newtonian limit for an
$n=1$ polytrope. From Eqs.~\eqref{I-bar-n1},~\eqref{lambda-bar-n1} and~\eqref{Q-bar-n1}, one finds 
\ba
\bar{I}^\N &=& C_{\bar{I} \bar{\lambda}}^{(n=1)} \left[ \bar{\lambda}^\tid{}^{,\N} \right]^{2/5}\,, \quad \bar{I}^\N = C_{\bar{I} \bar{Q}}^{(n=1)} \left[\Qbar^{\N} \right]^{2}\,, \nn \\ 
\Qbar^{\N} &=& C_{\bar{Q} \bar{\lambda}}^{(n=1)} \left[ \bar{\lambda}^\tid{}^{,\N} \right]^{1/5}\,,
\label{I-Love-Q-n1}
\ea
with
\ba
\label{coeff-I-Love-n1}
C_{\bar{I} \bar{\lambda}}^{(n=1)} &=& \frac{ 2(\pi^2 -6)}{3^{3/5} \pi^{6/5} (15-\pi^2)^{2/5}} \approx 0.527\,, \\
C_{\bar{I} \bar{Q}}^{(n=1)} &=& \frac{32 (\pi^2-6)^5}{27 \pi^6 (\pi^2 -15)^2} \approx 0.0406\,, \\
\label{coeff-Q-Love-n1}
C_{\bar{Q} \bar{\lambda}}^{(n=1)} &=& \frac{3^{6/5} \pi^{12/5} (15-\pi^2)^{4/5}}{4(\pi^2-6)^2} \approx 3.60\,. 
\ea
Observe that the numbers shown in Eqs.~\eqref{coeff-I-Love-n1}--\eqref{coeff-Q-Love-n1} are almost identical to those in Eqs.~\eqref{coeff-I-Love-n0}--\eqref{coeff-Q-Love-n0}.

The I-Love-Q relations for the $n=0$ and $n=1$ polytropic EoS in the Newtonian limit are shown as black thin solid lines in Fig.~\ref{fig:I-Love-Q}. Notice that the relations for the $n=0$ and $n=1$ polytropic EoSs in GR approach the Newtonian ones as the compactness decreases. Notice, however, that we have here analytically shown that the I-Love-Q relations are very similar for the $n=0$ and $n=1$ polytropic EoSs \emph{only}. This does not mean that the dependence of the I-Love-Q relations on the EoSs is weak in the Newtonian limit for all EoSs. Indeed, Fig.~\ref{fig:I-Love-Q} shows that these relations for some EoSs, such as APR and SLy,  do not approach the Newtonian limit of the $n=0$ and 1 polytrope. 

\subsubsection{Analytical Reasoning}

The universality of the I-Love-Q relations rests on two ingredients. The first ingredient is that the functional form of the relations must
be the same for different EoSs. For the $n=0$ and $n=1$ polytropes in the Newtonian limit, this is verified by comparing Eqs.~\eqref{I-Love-Q-n0} to~\eqref{I-Love-Q-n1}, and noting that regardless of the EoS, $\bar{I}^\N \propto  \left[ \bar{\lambda}^\tid{}^{,\N} \right]^{2/5}$, $\bar{I}^\N \propto \left[ \Qbar^{\N} \right]^{2}$ and $\Qbar^{\N} \propto \left[ \bar{\lambda}^\tid{}^{,\N} \right]^{1/5}$. This fact is perhaps expected, since all multipole moments must be proportional to the product of a dimensionless constant and the compactness to some power. The power is determined by the Newtonian dimensional structure of the particular multipole moment, e.g.~$I \propto \RNS^{2}$ and thus $\bar{I} \propto C^{-2}$. This power will be the same regardless of the EoS, and thus the power exponent in the I-Love-Q relations will also be EoS independent. 

The second ingredient, and perhaps the most difficult to understand, is the requirement that the constants of proportional  (ie.~the $C_{A}$'s) be the same regardless of the EoS. For the $n=0$ and $n=1$ polytropes in the Newtonian limit, this is again verified by noting that the coefficients in Eqs.~\eqref{coeff-I-Love-n0}--\eqref{coeff-Q-Love-n0} are almost identical to those in Eqs.~\eqref{coeff-I-Love-n1}--\eqref{coeff-Q-Love-n1}; their ratios are
\ba
\frac{C_{\bar{I} \bar{\lambda}}^{(n=0)}}{C_{\bar{I} \bar{\lambda}}^{(n=1)}} &=& \frac{2^{2/5} 3^{3/5} \pi^{6/5} (15-\pi^2)^{2/5}}{5 \pi^2-30} \approx 1.002\,, \\
\frac{C_{\bar{I} \bar{Q}}^{(n=0)}}{C_{\bar{I} \bar{Q}}^{(n=1)}} &=& \frac{108 \pi^6 (\pi^2-15)^2}{3125 (\pi^2-6)^5} \approx 1.008\,, \\
\frac{C_{\bar{Q} \bar{\lambda}}^{(n=0)}}{C_{\bar{Q} \bar{\lambda}}^{(n=1)}} &=& \frac{25 (\pi^2-6)^2}{ 2^{4/5} 3^{6/5} \pi^{12/5} (15-\pi^2)^{4/5}} \approx 0.997\,.
\ea

In principle, there is no reason to expect that these coefficients should be equal to each other regardless of the EoS. Rather, one expects them to depend on the NS internal structure. One possible explanation is to argue that these coefficients depend on integrals of the energy density that are more heavily weighted toward the NS's outer layers, i.e.~the structure of the NS in its outer layers is what is mostly determining these coefficients. But it is precisely in the outer layers that nuclear physics uncertainties are lowest. Therefore, the EoSs in this regime are more similar to each other than in the core, thus leading to some degree of universality. 

We have found some evidence to support this interpretation, shown in Fig.~\ref{fig:I-Love-Q}. Focus on the I-Love-Q relations for the polytropic EoSs with $n=2$, $2.5$ and $3$, which greatly modify the internal structure in the NS's outer layers, far from the core. In fact, these polytropes lead to essentially no energy density near the NS surface, with most of it concentrated near the core. We see that, indeed, when we choose an EoS that affects the NS structure far from the core, we significantly lose universality in the I-Love-Q relations, as one can see from the $n=2$, $2.5$ and $3$ curves in the bottom part of each panel in Fig.~\ref{fig:I-Love-Q}. 

\begin{figure}[htb]
\begin{center}
\includegraphics[width=8.0cm,clip=true]{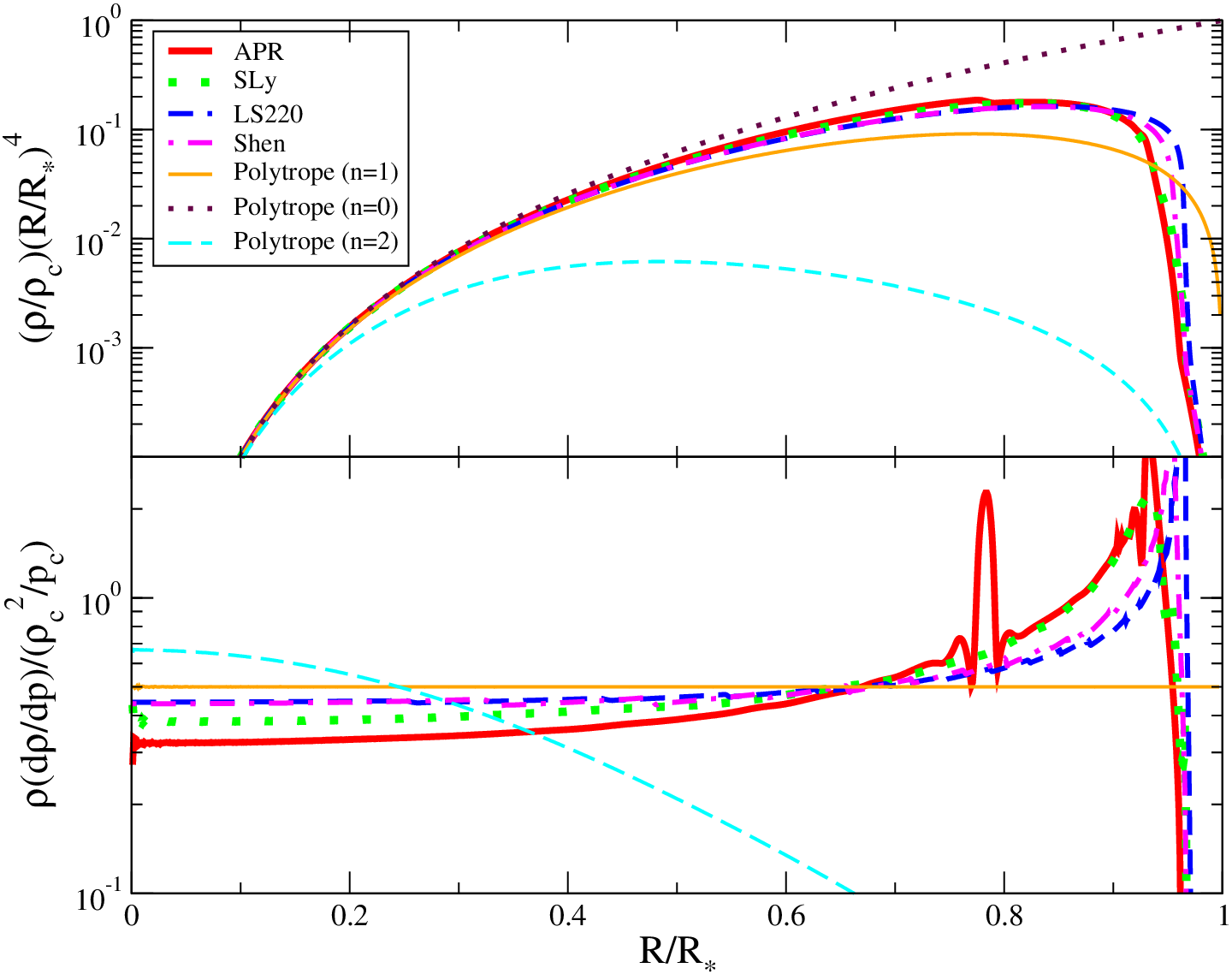}  
\caption{\label{fig:drhodp} (Color Online) $(\rho/\rho_c) (R/\RNS)^4$ (top) and $\rho (d\rho/dp)/(\rho_c^2 /p_c)$ (bottom) as functions of $R/\RNS$ for $C=0.17$ with various EoSs. The former corresponds to the integrand (modulo normalization constants) of $I^\N$ [Eq.~\eqref{I}] and $Q^\rot{}^{,\N}$ [Eq.~\eqref{Q-integ}] in the Newtonian limit, while the latter corresponds to the EoS-dependent coefficient in Eq.~\eqref{h2RR-Newton}, which gives $\lambda^\tid{}^{,\N}$. Sudden changes in the bottom panel correspond to nuclear phase transitions. Observe that the dominant contribution of  $(\rho/\rho_c) (R/\RNS)^4$ and $\rho (d\rho/dp)/(\rho_c^2 /p_c)$ for realistic EoSs come from the NS outer layer.}
\end{center}
\end{figure}

Further evidence can be found by investigating a few of the terms that control the behavior of the moment of inertia, the quadrupole moment and the tidal Love number as a function of $C$. First, in the Newtonian limit, both $I^\N$ [Eq.~\eqref{I}] and $Q^\rot{}^{,\N}$ [Eq.~\eqref{Q-integ}] can be written in integral form, where the radial dependence of the integrand is proportional to $\rho(R) R^4$. The top panel of Fig.~\ref{fig:drhodp} shows  $(\rho/\rho_c) (R/\RNS)^4$ as a function of  $R/\RNS$ for $C=0.17$ with various EoSs. $I$ and $Q^\rot$ are proportional to the area under the curves in this panel. Observe that the curves are similar, even for different realistic EoSs, and the dominant contribution comes from the NS outer layer, somewhere between $R/\RNS \approx 0.7-0.9$. This may explain the similar behavior of the I-C and Q-C curves with different realistic EoSs in the high-compactness regime (see Figs.~\ref{fig:IbarMC} and~\ref{fig:QbarMC}), as well as the universal I-Q behavior. Observe also that as one increase the polytropic index $n$, the NS becomes more centrally-condensed, and $I$ and $Q$ are not dominated just by the NS outer layers. 

Similarly, we can study the behavior of the structure-dependent term that determines the tidal Love number in the Newtonian limit. Equation~\eqref{h2RR-Newton} shows that there is only one such term and it is proportional to $\rho d\rho/dp = \rho (d\rho/dR) (dp/dR)^{-1}$. The bottom panel of Fig.~\ref{fig:drhodp} plots $\rho (d\rho/dp)/(\rho_c^2 /p_c)$ as a function of $R/\RNS$ for $C=0.17$ with various EoSs. Sudden changes in the slope corresponds to nuclear phase transitions. Similar to the top panel, the behavior of  $\rho d\rho/dp$ is similar among realistic EoSs and the dominant contribution comes from the NS outer layers. This partially explains the similar behavior observed in the Love-C curves with different realistic EoSs in the high-compactness regime (see Fig.~\ref{fig:lambdaMC}), as well as the universal Q-Love and I-Love relations.

Another possible explanation for the universality of the I-Love-Q relations involves the behavior of this trio as one approaches the BH limit. The no-hair theorems~\cite{robinson,israel,israel2,hawking-uniqueness0,hawking-uniqueness,carter-uniqueness} of GR state that the exterior multipolar structure of an isolated, stationary, axisymmetric BH solution in GR is completely determined by its mass and its spin angular momentum. Therefore, the quadrupole moment, for example, is completely determined by the spin angular momentum through Eq.~\eqref{quadrupole} with $A = 0$~\cite{geroch,hansen}. For NSs, such a result does not exist, but one might still expect the I-Q relation to become less structure dependent as one approaches the BH limit ($C \to 0.5$). Indeed, Fig.~\ref{fig:I-Love-Q} shows that the loss of universality (the relative fractional difference shown in the bottom part of each panel) decreases logarithmically as the mass increases from $0.3 M_{\odot}$ to $1.4 M_{\odot}$.

Of course, one can never increase the compactness enough by a finite amount to turn a NS into a BH, ie.~the NS sequence of varying compactness does not terminate in a BH for finite central density.  Still, it is interesting to see that as $C$ increases, the I-Love-Q relations are indeed approaching the BH limit, as explicitly shown in Fig.~\ref{fig:I-Love-Q}. Moreover, universality in the I-Q curve suggests a universal relation between the NS spin and the NS quadrupole moment that is almost independent of the internal structure. Such a relation is similar to the no-hair relations for BHs~\cite{geroch,hansen}.

Before proceeding, let us point out that this {\emph{effacing of internal structure}} is not the same as what is discussed in the effacement principle~\cite{damour-effacement} in GR. The latter states that the equations of motion of compact objects of any size and structure depend only on integral parameters, like the mass and spin, and it is independent of its actual shape and internal structure. Of course, this principle holds in GR but only for BHs because of the no-hair theorems. The effacement principle is violated for NSs, but the violation is small, with corrections to the acceleration entering at 5PN order for a binary of non-spinning compact objects. Since the effacement principle deals with the motion of a body only, and not on the multipolar structure of its exterior gravitational field, the effacement we find here is not a consequence of the standard effacement principle. 

\section{Applications}
\label{sec:applications}

The I-Love-Q relations have 3 immediate applications to observational astrophysics, GWs and fundamental physics. Let us look at each application in turn.

\subsection{Observational Astrophysics}

On an astrophysical front, a measurement of any single member of the I-Love-Q trio automatically provides information about the other two, even when measuring the other two directly might not be possible with current observations. For example, one might be able to measure $\Ibar$ within 10\% accuracy by measuring the orbits of binary pulsars sufficiently accurately, so as to extract the spin-orbit coupling effect in the advance rate of the periastron of the double binary pulsar J0737-3039~\cite{lattimer-schutz,kramer-wex}. If such measurement is accomplished, one can then automatically obtain the quadrupole moment and tidal Love number of the primary pulsar by using the I-Love-Q relations. Similarly, if an equal-mass NS binary within 300Mpc is about to coalesce, one might be able to determine the $\lambdabartid$ of the constituents with second-generation, ground-based GW interferometers~\cite{flanagan-hinderer-love,read-love,hinderer-lackey-lang-read,lackey,damour-nagar-villain}. Then, from the I-Love-Q relations, one may obtain the moment of inertia and quadrupole moment of the binary constituents, which again, would be very difficult to measure with GWs.  

Let us stress that the I-Love-Q relations cannot be used to measure the equation of state, but rather to infer two members in the I-Love-Q trio when the third is measured. The main result in this paper is, in fact, that the I-Love-Q relations seem to be rather insensitive to the EOS. Inferring the quadrupole moment and the Love number would provide important information about the properties of NSs. The quadrupole moment would tell us how much a NS can be quadrupolarly deformed (squeezed at the poles), while the Love number would tell us how much it can be deformed, for example, in the presence of a companion. 

A small caveat should be presented here. The I-Love-Q relations hold for the dimensionless (barred) moment of inertia, quadrupole moment and Love number, which are normalized by the NS mass and spin. In particular, the observed NS mass differs from the mass used to normalize the I-Love-Q relations by factors of ${\cal{O}}(\chi^{2})$, ie.~$M_{\rm obs} =  M_{*} \left(1 + \chi^{2} \delta M\right)$, where $\delta M = {\cal{O}}(0.3)$~\cite{Berti:2004ny,berti-iyer-will}. For stars spinning with $\chi \lesssim 0.1$, this induces differences between $M_{\rm obs}$ and $M_{*}$ of ${\cal{O}}(10^{-3})$, which would spoil the I-Love-Q universality. However, this non-universality is much smaller than the accuracy to which $M_{*}$ can be observationally determined, and thus, it does not spoil the use of the I-Love-Q relations in observational astrophysics.  

Of course, these application assumes that the universality of the I-Love-Q relations holds, which rests on the assumptions of uniform and slow-rotation, small tidal perturbations and that GR is the correct theory. Therefore, this technique cannot be applied to, for example, newly-born NSs that are differentially and rapidly rotating. NSs that source GWs in the sensitivity band of ground based detectors, however, are expected to be old, and thus uniformly rotating with large spin periods (they should have spun down by the time they are visible by GW detectors~\cite{bildsten-cutler}), so that the slow-rotation approximation is well-justified. The primary NS in the double binary pulsar has a period of $22 \; {\rm{ms}}$, which implies a $\chi \sim 0.018$, small enough that the slow-rotation approximation is again well-justified. 

Millisecond binary pulsars with short-periods, ie.~periods below $1 \; {\rm{ms}}$, would be spinning too fast for the above relations to be directly applicable. However, we expect I-Love-Q type universality with respect to the EoS to still hold in this case, except that now the coefficients in Table~\ref{table:coeff} will also depend somewhat on the spin angular frequency (or the spin period). One can correct the universal I-Love-Q relations for non-negligible spins by considering rapidly rotating NSs~\cite{berti-stergioulas,berti-white,benhar,pappas-apostolatos}, but we leave this to future work.

\subsection{Gravitational Wave Astrophysics}

Another application of the I-Love-Q relations is to GW astrophysics, as a means to break the degeneracy between individual spins and the quadrupole moments of NSs in the GWs emitted during binary NS inspirals. Let us first discuss gravitational waveforms of spinning, tidally-deformed NS binaries, and then, carry out a back-of-the-envelope parameter estimation study using Fisher theory. The latter will allow us to determine the degree to which degeneracies are broken through the I-Love-Q relations and the projected accuracy to which individual NS spins could be measured given a GW detection. 

\subsubsection{Waveforms}

The sky-averaged gravitational waveform (in the Fourier domain) generated by a compact NS binary in a quasi-circular orbit with masses $m_1$ and $m_2$ and at distance $D_L$ is given by~\cite{cutlerflanagan} $\tilde{h}(f) = A(f) \exp [i \Psi(f)]$, with\footnote{$A(f)$ needs to be multiplied by $\sqrt{3}/2$ when calculating the Fisher matrix for LISA and DECIGO/BBO.}
\ba
A(f) &=& \frac{1}{\sqrt{30} \pi^{2/3}} \frac{\mathcal{M}^{5/6}}{D_L} f^{-7/6}\,, \\
\Psi(f) &=& \Psi_\mrm{tp}(f) + \Psi_{\bar{Q}} (f) +\Psi_{\bar{\lambda}} (f)\,.
\ea
Here, $f$ is the GW frequency, $\mathcal{M} =  m \eta^{3/5}$ is the chirp mass, $\eta = {m_1 m_2}/{m^2}$ is the symmetric mass ratio and $m =  m_1 + m_2$ is the total mass. The quantity $\Psi_\mrm{tp}(f)$ is the gravitational waveform phase in the test-particle limit while $\Psi_{\Qbar}$ and $\Psi_{\bar{\lambda}}$ represent terms that deviate from this limit, where the former corresponds to a quadrupole moment deformation, while the latter depends on the tidal Love number. 

The test-particle term, to 3.5 PN order, is given by~\cite{arun35PN,arunbuonanno,blanchet3PN} 
\allowdisplaybreaks
\bw
\ba
\Psi_\mrm{tp} (f) &=& 2\pi f t_c - \phi_c - \frac{\pi}{4} +\frac{3}{128} (\pi \mathcal{M} f)^{-5/3} \Bigg \{ 1 + \left( \frac{3715}{756} + \frac{55}{9} \eta \right) x - (16\pi -4 \beta ) x^{3/2} \nn \\
& & + \left( \frac{15293365}{508032} + \frac{27145}{504} \eta + \frac{3085}{72} \eta^2 - 10\sigma \right) x^2 + \left( \frac{38645}{756} \pi - \frac{65}{9}\pi \eta - \gamma \right) (1+3\log{v}) x^{5/2} \nn \\
& & +\Bigg[ \frac{11583231236531}{4694215680} - \frac{640 \pi^2}{3} - \frac{6848}{21} \gamma_{\E}  - \left(\frac{15737765635}{3048192} - \frac{2255}{12} \pi^2 \right) \eta + \frac{76055}{1728} \eta^2 - \frac{127825}{1296} \eta^3  \nn \\
& & - \frac{6848}{21} \log (4v) + \alpha \Bigg] x^3 + \left( \frac{77096675}{254016} + \frac{1014115}{3024} \eta - \frac{36865}{378} \eta^2 \right) \pi x^{7/2} \Bigg \}\,,
\label{phase}
\ea
\ew
where $x\equiv v^2 = (\pi m f)^{2/3}$ and $(t_c,\phi_c)$ correspond to the time and phase at coalescence, respectively, with $\gamma_{\E}$ the Euler constant. The spin parameters $\beta$ and $\sigma$~\cite{kidder-spin,vasuth-spinspin,arunbuonanno}, $\gamma$~\cite{arunbuonanno} and $\alpha$~\cite{blanchet3PN} are given by\footnote{$\sigma$ includes the quadrupole-monopole interaction in the test-particle limit.}
\ba
\label{beta}
\beta &=&  \left( \frac{113}{12} - \frac{19}{3}\eta \right)   \left(\hat{\bm{L}} \cdot \bm{\chi}_s \right) + \frac{113}{12} \delta_m (\bm{\chi}_s \cdot \bm{\chi}_a) \,, \\
\sigma &=&\frac{719}{48} \delta_m   \left(\hat{\bm{L}} \cdot \bm{\chi}_s \right)  \left(\hat{\bm{L}} \cdot \bm{\chi}_a \right) -\frac{233}{48} \delta_m (\bm{\chi}_s \cdot \bm{\chi}_a) \nn \\
& & + \left( \frac{719}{96}+\frac{1}{24} \eta \right) \left(\hat{\bm{L}} \cdot \bm{\chi}_s \right) ^2 + \left(\frac{719}{96} -30 \eta \right)  \left(\hat{\bm{L}} \cdot \bm{\chi}_a \right) ^2 \nn \\
& & - \left( \frac{233}{96}+\frac{7}{24} \eta \right) \chi_s^2 - \left( \frac{233}{96}-10 \eta  \right) \chi_a^2\,,  \\
\gamma &=& \left(\frac{732985}{2268}-\frac{24260}{81} \eta-\frac{340}{9} \eta^2 \right) \left( \hat{\bm{L}} \cdot \bm{\chi}_s \right)  \nn \\
& & + \left( \frac{732985}{2268}+\frac{140}{9} \eta \right) \delta_m \left( \hat{\bm{L}} \cdot \bm{\chi}_a \right)\,, \\
\alpha &=& \frac{2270 \pi}{3} \left[ \left(1-\frac{227}{156} \eta \right) \left( \hat{\bm{L}} \cdot \bm{\chi}_s \right)+ \delta_m \left(  \hat{\bm{L}} \cdot \bm{\chi}_a \right) \right]\,, \nn \\
\ea
where $\hat{\bm{L}}$ is the unit orbital angular momentum, $\delta_{m} \equiv (m_{1} - m_{2})/m$ is the dimensionless mass difference, $\bm{\chi}_s \equiv (\bm{\chi}_1 + \bm{\chi}_2)/2$ and $\bm{\chi}_a \equiv (\bm{\chi}_1 - \bm{\chi}_2)/2$ with $\bm{\chi}_i \equiv \bm{S}_i/m_i^2$ denoting the dimensionless spin vector of the $i$-th body. Notice that we are here referring to the individual NS spin vectors by $\bm{S}_{i}$

The quadrupole moment contribution correction to the test-particle limit in the GW phase enters at 2PN order and it is given by~\cite{poisson-quadrupole,vasuth-spinspin}
\ba
\Psi_{\bar{Q}} (f) &=& \frac{3}{128} \frac{x^{-5/2}}{\eta} \left\{ -50  \left[  \left( \frac{m_1^2}{m^2} \chi_1^2 + \frac{m_2^2}{m^2} \chi_2^2 \right) (\bar{Q}_s -1) \right. \right. \nn \\
&& \left. \left. +\left( \frac{m_1^2}{m^2} \chi_1^2 - \frac{m_2^2}{m^2} \chi_2^2 \right) \bar{Q}_a \right] x^2 \right\}\,, 
\ea
where
\be
\bar{Q}_s \equiv \frac{\bar{Q}_1 + \bar{Q}_2}{2}\,, \quad \bar{Q}_a \equiv \frac{\bar{Q}_1 - \bar{Q}_2}{2}\,.
\ee
$\Qbar_s$ is strongly correlated with $\sigma$, which enters at the same PN order as $\Qbar_s$.

The leading-order contribution of $\Psi_{\bar{\lambda}} (f)$ to the GW phase enters at 5PN order through~\cite{flanagan-hinderer-love}
\ba
\Psi_{\bar{\lambda}}^\mrm{5PN} (f) &=& -\frac{3}{128} \frac{x^{-5/2}}{\eta}24 \left[ (1+7\eta -31 \eta^2 ) \bar{\lambda}_s \right.  \nn \\
& & \left. + (1+9 \eta -11 \eta^2 ) \bar{\lambda}_a \delta_m \right] x^{5}\,, 
\ea
where 
\be
\bar{\lambda}_s \equiv \frac{\lambdabartid_1+\lambdabartid_2}{2}\,, \quad \bar{\lambda}_a \equiv \frac{\lambdabartid_1-\lambdabartid_2}{2}\,.
\ee
Higher PN contributions to $\Psi_{\bar{\lambda}} (f)$ can be found in~\cite{vines1,vines2,damour-nagar-villain}. When carrying out parameter estimation studies, as explained below, we will use $\Psi_{\bar{\lambda}} (f)$ as given in~\cite{damour-nagar-villain}, which includes up to 2.5PN order corrections relative to $\Psi_{\bar{\lambda}}^\mrm{5PN}$\footnote{Tidal effects on the gravitational waveform phase have been calculated to 1.5PN order relative to the leading 5PN contribution, in addition to tail effects at 2.5PN order. Ref.~\cite{damour-nagar-villain} estimated that currently unknown terms should be subdominant, at least for an equal-mass binary.}.

\subsubsection{Parameter Estimation}
\label{sec:parameter-estimation}

For stationary and Gaussian detector noise, the measurement accuracy of parameters $\theta^a$ can be estimated as
\be
\Delta \theta^a = \sqrt{\frac{(\Gamma^{-1}){}^{aa}}{N}}\,,
\ee
where $N$ is the number of effective interferometers and
\be
\Gamma_{ab} \equiv 4 \; \mrm{Re} \int^{f_\mrm{max}}_{f_\mrm{min}} \frac{\partial_a \tilde{h}(f) \partial_b \tilde{h}(f)}{S_n(f)} df
\ee
is the Fisher matrix, where the partial derivatives are with respect to the parameters $\theta^a$. The noise spectral density $S_n(f)$ is given in Refs.~\cite{cornish-PPE,mishra,yagi:brane} for Adv.~LIGO, ET and DECIGO/BBO, respectively. We take the lower cutoff frequencies $f_\mrm{min}$ to be 10Hz for Adv.~LIGO, 1Hz for ET and the frequency 1yr before coalescence for DECIGO/BBO. For Adv.~LIGO and ET, we take the higher cutoff frequency to be that of the innermost stable circular orbit (ISCO), $f_\mrm{max}=f_\mrm{ISCO}=1/(6^{3/2} \pi m)$, while we set $f_\mrm{max}=100$Hz for DECIGO/BBO. $N$ is the number of effective interferometers, which we take to be 5 for 2nd-generation ground-based detectors (corresponding to 2 Adv.~LIGO, Adv.~VIRGO, KAGRA and INLIGO), 2 for ET (like LISA~\cite{cutler1998}) and 8 for DECIGO/BBO~\cite{yagi:brane}.

We focus here on GWs emitted during the quasi-circular inspiral of NSs with aligned spins, since this is a realistic astrophysical scenario~\cite{kesden-berti}. Given that the NS masses are expected to be approximately the same, we will not include $\bar{Q}_a$ and $\bar{\lambda}_a$ in the parameter vector, as this must be close to zero. We will consider the case of slightly unequal NS masses. We choose two parameterizations of the waveform. Parameterization $A$ uses the parameter set~\cite{damour-nagar-villain}
\be
\{ \theta_A^i \} =  (\ln \mathcal{M}, \ln\eta, \beta, D_L, t_c, \phi_c, \bar{\lambda}_s)
\label{parameter-A}
\ee
with the priors $|\eta| < 0.25$ and $|\beta| < 0.8$. We do not include $\sigma$ in this set because the NS spins at the time of coalescence are expected to be small~\cite{bildsten-cutler}\footnote{Damour \textit{et al.}~\cite{damour-nagar-villain} estimated that at the time of coalescence, $|\beta| < 0.2$ and $|\sigma | < 10^{-4}$. We use a conservative prior $|\beta| < 0.8$ which corresponds to $|\chi|<0.1$.}.
Parametrization $B$ uses the parameter set
\be
\{ \theta_B^i \} =  (\ln \mathcal{M}, \delta_m, \chi_s, \chi_a, D_L, t_c, \phi_c, \bar{Q}_s(\bar{\lambda}_s), \bar{\lambda}_s)\,,
\label{parameter-B}
\ee
with the priors $|\delta_m| < 1/3$, $|\chi_s| < 0.1$ and $|\chi_a| < 0.1$. Moreover, we use the Q-Love relation to express $\Qbar_s$ in terms of $\bar{\lambda}_s$, and thus, partially break the degeneracy between $\Qbar_s$ and $\chi_s$.

Figure~\ref{fig:spin} shows the measurement accuracies of spin parameters using second-generation, ground-based detectors. We assume that the detected GW was emitted by a source at $D_L=100$Mpc with ${\rm{SNR}} \sim 30$. We consider 3 different systems:  (i) $(m_1,m_2)=(1.45,1.35)M_\odot$, $\chi_1=\chi_2$, (ii) $(m_1,m_2)=(1.45,1.35)M_\odot$, $\chi_1=2 \chi_2$ and (iii) $(m_1,m_2)=(1.4,1.35)M_\odot$, $\chi_1=\chi_2$. 
Observe, that the averaged spin $\chi_s$ can be measured to $\mathcal{O}(10)\%$. Such an accuracy on $\chi_s$ is inaccessible without the Q-Love relation. 

We can understand this enhanced accuracy in the extraction of $\chi_{s}$ as follows. First, notice that, for an equal-mass and spin-aligned binary, $\beta \sim \mathcal{O}(10) \chi_s$ [see e.g.~Eq.~\eqref{beta}]. Given that the measurement accuracy of $\beta$ is $\Delta \beta = \mathcal{O}(0.1)$, this implies a measurement accuracy of $\chi_{s}$ of $\Delta \chi_s \approx 0.01$, which corresponds to $\Delta \ln \chi_s \approx 0.1$ for $\chi_s \approx 0.1$.  The measurement accuracy of $\chi_s$ for system (ii) increases as $\chi_1 \to 0.1$, as shown in Fig.~\ref{fig:spin}. This is because $\partial \tilde{h}/\partial \delta_m \approx 0 \approx \partial \tilde{h}/\partial \chi_a$ as $\chi_1 \approx 0.1$ and the priors lead to $\delta_m$, $\chi_a$ and other parameters being effectively uncorrelated. In such a case, however, the assumptions that underlie the Fisher approximation may be violated~\cite{vallisneri-fisher}, and hence, one requires a Bayesian analysis~\cite{cornish-PPE} to confirm these results. Finally, we have checked that the Q-Love relation does not improve the measurement accuracy of $\bar{\lambda}_s$.

\begin{figure}[htb]
\begin{center}
\includegraphics[width=8.0cm,clip=true]{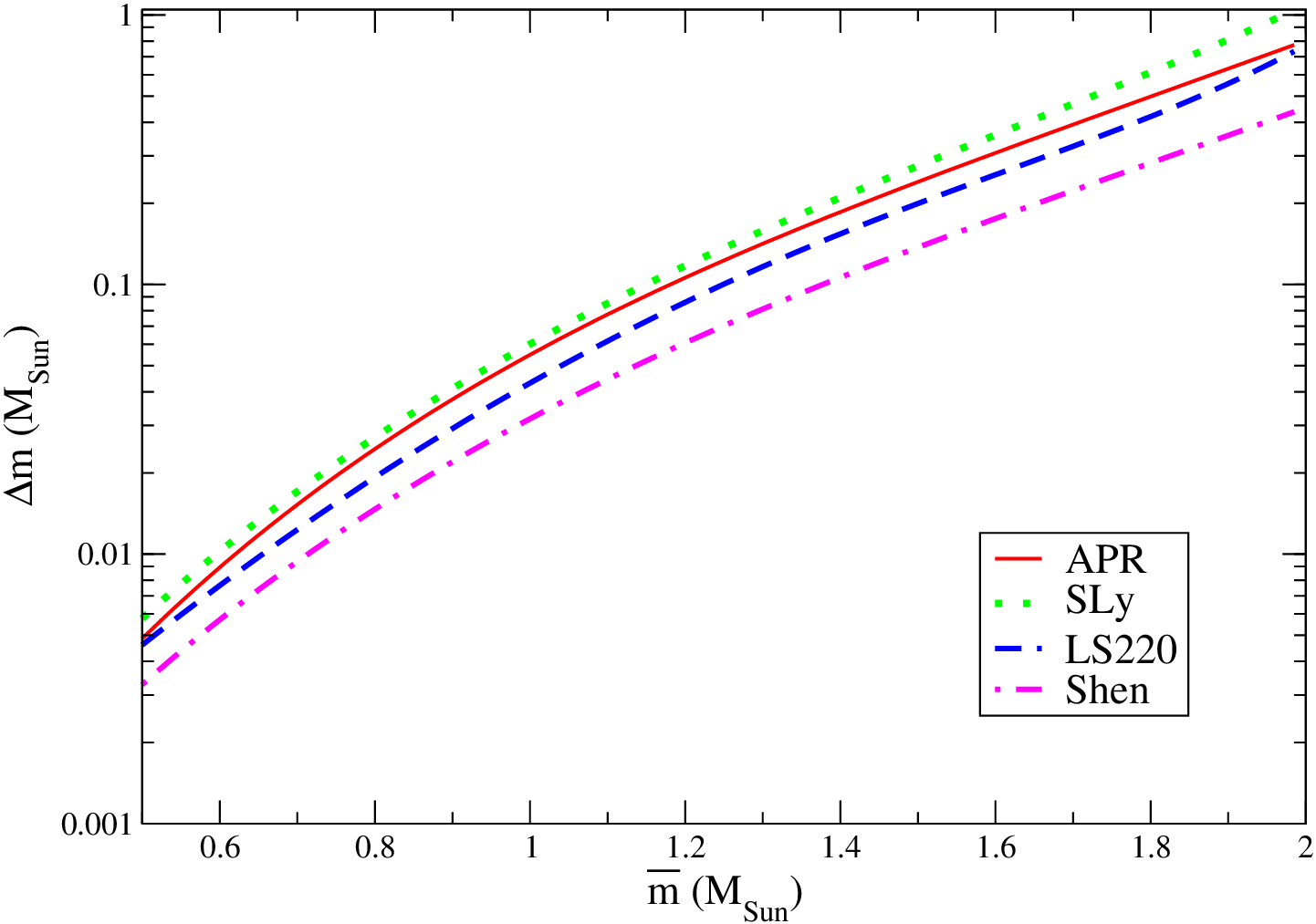}  
\caption{\label{fig:deltam} (Color Online) Separatrix between systems with $\Psi_{\bar{\lambda}_a}>1$ (above the curves) and $\Psi_{\bar{\lambda}_a}<1$ (below the curves) as a function of $\Delta m \equiv |m_1 - m_2|$ and $\bar{m} \equiv (m_1 + m_2)/2$, and for different realistic EoSs. For systems below the curves, we can safely neglect $\bar{\lambda}_a$ in parameter estimation, provided the $\mrm{SNR} \approx {\cal{O}}(10)$. $\Psi_{\bar{Q}_a}$ is smaller than $\Psi_{\bar{\lambda}_a}$ for $|\chi|<0.1$.}
\end{center}
\end{figure}

Up until now, we have considered equal-mass NS binaries, but realistic systems may not have identical masses. If the NS masses are not equal, one must then take into account the parameters $\bar{Q}_a$ and $\bar{\lambda}_a$, whose inclusion could in principle degrade the accuracy to which other parameters are extracted. Let us then study the range of masses for which neglecting $\bar{Q}_a$ and $\bar{\lambda}_a$ is a good approximation. A rough estimate of this range can be obtained by investigating the systems for which the accumulated GW phase induced by terms proportional to $\bar{Q}_a$ and $\bar{\lambda}_a$ is less than one radian. Let us then define the latter by $\Psi_{\bar{Q}_a}$ and $\Psi_{\bar{\lambda}_a}$ respectively, where 
\ba
\Psi_{\bar{Q}_a} (f) &=& - \frac{75}{64} \frac{1}{\eta} \left[ \left( \frac{m_1^2}{m^2} \chi_1^2 - \frac{m_2^2}{m^2} \chi_2^2 \right) \bar{Q}_a \right] x^{-1/2}\,, 
\nonumber \\
\Psi_{\bar{\lambda}_a}^\mrm{5PN} (f) &=&  -\frac{9}{16} \frac{1}{\eta} (1+9 \eta -11 \eta^2 ) \bar{\lambda}_a \delta_m  x^{5/2}\,,
\ea
to leading PN order. 

Figure~\ref{fig:deltam} shows the range of masses for which $\Psi_{\bar{\lambda}_a} = 1$ for various realistic EoSs. Systems above these lines would lead to  $\Psi_{\bar{\lambda}_a} > 1$, while those below this line lead to  $\Psi_{\bar{\lambda}_a} < 1$. In particular, for systems that satisfy the latter inequality, we can in principle neglect $\bar{\lambda}_a$ if the SNR is $\mathcal{O}(10)$. This figure implies that for second-generation, ground-based detectors, the parameter estimation study presented above is probably valid, even for unequal-mass systems provided, for example, that $\bar{m}=1.4 M_\odot$ and $\Delta m \lesssim \mathcal{O}(0.1)M_\odot$, where $\bar{m} \equiv (m_1 + m_2)/2$ is the averaged mass of the binary and $\Delta m \equiv |m_1 - m_2|$ is the mass difference. This same conclusion also applies to neglecting $\bar{Q}_{a}$, provided $|\chi| < 0.1$.

\subsection{Fundamental Physics}

The independent measurement of any 2 members of the I-Love-Q trio would allow us to perform model-independent and EoS-independent tests of GR.  For example, if one can measure $\Ibar$ and $\lambdabartid$ independently, one can plot a point in the I-Love plane with an error box. If the I-Love relation in GR crosses the error box, then GR is consistent with the observations. Otherwise, one would have found model-independent evidence for some type of departure from GR. Moreover, one can constrain non-GR theories by requiring that the I-Love relation in that theory crosses the error box. 

The accuracy of such a test depends, of course, in how accurately two elements in the I-Love-Q set can be measured. One way to measure $\bar{I}$ would be to look for a spin-orbit correction to the rate of advance of the periastron of a binary system. Future double binary pulsar observations may measure $\bar{I}$ with an accuracy of roughly 10\%~\cite{lattimer-schutz,kramer-wex}. Probably, the best way to measure $\bar{\lambda}^\tid$ and $\Qbar$ would be to use GW observations. 

In what follows, we first discuss the possibility of measuring $\Qbar$ and $\lambdabartid$ simultaneously, given GW observations. Then, we discuss how well GR tests can be carried out by combining GW observations with binary pulsar observations. For concreteness, we apply all of this to a specific modified gravity theory (dynamical CS gravity~\cite{CSreview}).

\subsubsection{Redundancy Tests with GW Observations Only}

{\renewcommand{\arraystretch}{1.2}
\begin{table}
\begin{centering}
\begin{tabular}{r|cccccc|ccccc}
\hline
\hline
\noalign{\smallskip}
& & \multicolumn{4}{c}{$\bar{\lambda}^\tid=400.0$}  & & \multicolumn{3}{c}{$M_\NS=1.3382M_\odot$}   \\
 EoS &&  \multicolumn{1}{c}{$M_\NS$} &  \multicolumn{1}{c}{$\RNS$}
&  \multicolumn{1}{c}{$\Qbar$}  &   \multicolumn{1}{c}{$f_\mrm{spin}$}  & & \multicolumn{1}{c}{$\RNS$}
&  \multicolumn{1}{c}{$\bar{I}$}  &   \multicolumn{1}{c}{$\bar{\lambda}^\tid$} 
  \\
  & & ($M_\odot$) &  (km) & &  (Hz)  & &  (km) & &  \\
\hline
APR && 1.40  & 12.2 & 5.52 & 194  & &12.2 & 13.3 & 520  \\
 SLy && 1.32  & 11.6 & 5.54 & 206  & & 11.6 & 12.2 & 375 \\
 LS220 && 1.38 & 13.5 & 5.56 & 198  & & 13.6 & 13.1 & 506  \\
 Shen && 1.55 & 14.6 & 5.54 & 176 & & 15.0 & 15.9 & 1012 \\
\noalign{\smallskip}
\hline
\hline
\end{tabular}
\end{centering}
\caption{NS parameters for $\bar{\lambda}^\tid=400.0$ and $M_\NS=1.3382M_\odot$. $f_\mrm{spin}$ corresponds to the NS spin frequency with $\chi=0.1$.}
\label{table:parameters}
\end{table}
}
Let us estimate how accurately future ground-based detectors may determine $\lambdabartid$ and $\Qbar$ \emph{simultaneously}.  The measurability of $\bar{\lambda}^\tid$ has been discussed extensively in~\cite{flanagan-hinderer-love,read-love,hinderer-lackey-lang-read,kiuchi,kyutoku,hotokezaka,vines1,vines2,lackey,damour-nagar-villain}, while the effect of the quadrupole moment on compact binary GWs has been estimated in~\cite{bildsten-cutler,lai-rasio-shapiro,poisson-quadrupole}, but so far no study has been performed to study their simultaneous extraction. Let us consider an equal-mass NS binary with $\lambdabartid = 400$, which corresponds to NSs with $M_\NS=1.4M_\odot$ with the APR EoS (other parameters are shown in Table~\ref{table:parameters}).

\begin{figure}[htb]
\begin{center}
\includegraphics[width=8.5cm,clip=true]{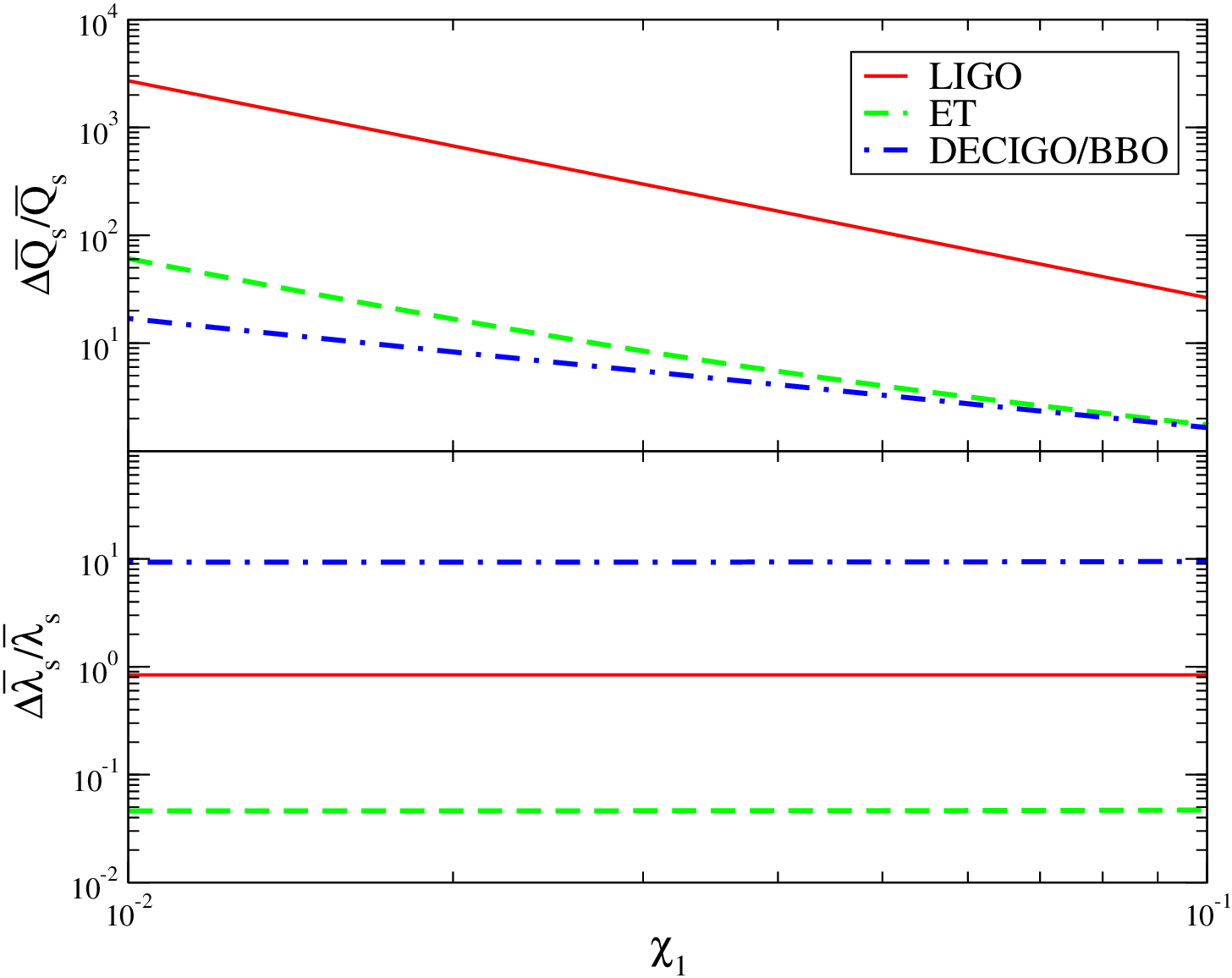}  
\caption{\label{fig:error} (Color Online) Measurement accuracies for $\bar{Q}_s$ (top) and $\bar{\lambda}_s$ (bottom) given a GW detection, emitted by a $(1.4,1.4) \; M_\odot$, spin-aligned NS/NS binary system with Adv.~LIGO, ET and DECIGO/BBO. We assume $\bar{\lambda}_s=400$, $\chi_2=0$, $D_L=100$Mpc and the APR EoS, with priors $|\delta|\leq 1/3$, $|\chi_1|\leq 0.1$ and $|\chi_2|\leq 0.1$. Although it may be difficult to measure $\Qbar_s$, it should be possible to measure $\bar{\lambda}_s$ with ground-based detectors.}
\end{center}
\end{figure}
Figure~\ref{fig:error} shows the measurement accuracy of $\Qbar_s$ and $\bar{\lambda}_s$ with second-generation, ground-based detectors, ET and DECIGO/BBO. We assume an equal-mass, spin-aligned NS/NS binary with $(1.4, 1.4)M_\odot$, $\chi_1 < 0.1$ and $\chi_2 =0$ at $D_L=100$Mpc, and we also assume that the APR EoS is the correct one. The measurement accuracy of $\Delta \Qbar_s$ with second-generation, ground-based detectors is $\Delta \ln \Qbar_s \approx 30$, which would increase to  $\Delta \ln \Qbar_s \approx 2 $ using future detectors, such as ET or DECIGO/BBO. These results imply that it may be difficult to measure $\Delta \Qbar_s$ due to its strong degeneracies with spin parameters. Notice, however, that even though one may not be able to detect $\Qbar_s$, one can still place an upper and lower bound on $\Qbar_s$. Such a bound would be sufficient to perform model-independent GR tests. The measurement accuracy of $\bar{\lambda}_s$ with second-generation, ground-based detectors is $\Delta\bar{\lambda}_s \approx 0.8$, which would increase by roughly an order of magnitude using ET. Although the error bars are large with current detectors, it may be possible to measure $\bar{\lambda}_s$ with future GW observations.

\begin{figure}[htb]
\begin{center}
\includegraphics[width=8.5cm,clip=true]{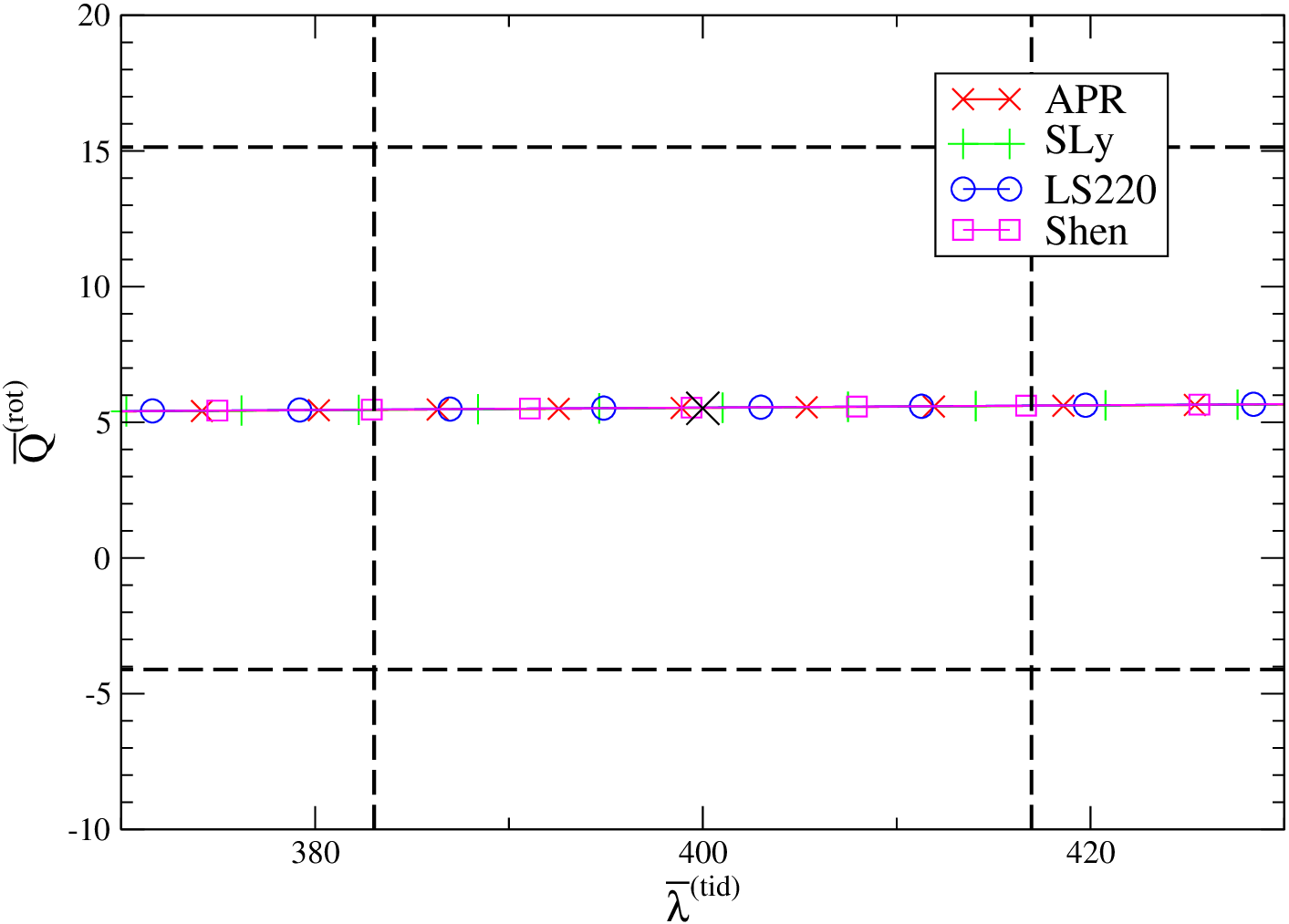}  
\caption{\label{fig:QLoveerror} (Color Online) The Q-Love relations for realistic EoSs with proposed measurement errors. We assume that we determine $\Qbar$ and $\lambdabartid$ simultaneously by detecting GWs from an equal-mass, spin-aligned NS/NS binary with $\chi_1=0.1$, $\chi_2=0$ and $\lambdabartid = 400$ at $D_L=100$Mpc with ET. The fiducial values of $(\lambdabartid, \Qbar)$ are shown as a big black cross. Although the measurement error of $\Qbar$ is greater than its fiducial value, one can still perform model-independent and EoS-independent tests of gravity by requiring that the Q-Love curve must pass through the error box.}
\end{center}
\end{figure}

Given the above measurement errors, we can now simulate a GR test. Figure~\ref{fig:QLoveerror} presents the Q-Love relation for realistic EoS, together with a fiducial GW measurement of the pair $(\Qbar,\lambdabartid)$ and its estimated errors. Notice that the error in $\Qbar$ is larger than the value about which the error is centered. This implies that a GW measurement would not be able to measure $\Qbar$, but it would be able to say the region of allowed $\Qbar$ that is consistent with the GW detection. Therefore, such a GW detection would automatically constitute a model-independent test of GR; for GR to be consistent with these measurements, the GR Q-Love curve must cross the GW error box in $(\Qbar,\lambdabartid)$.     

The above test is quite robust. First, although we employ a uniform and slow-rotation approximation, the NS/NS binaries that ground-based detectors will observe will have spun down by the time they enter the detector's sensitivity band, and thus, the slow-rotation approximation should be excellent. Second, the error box of Fig.~\ref{fig:QLoveerror} depends on how accurately $(\bar{\lambda}_s, \Qbar_s)$ can be measured, which in turn depends on whether $(\bar{\lambda}_{a}, \Qbar_{a})$ need to be included in the parameter set. This would be the case if the binary system detected were not an equal-mass one. As shown in Fig.~\ref{fig:deltam}, however, there is a wide range of mass ratios for which these parameters can be neglected, even outside of the equal-mass point; thus, the discussion presented above should be robust. 

\subsubsection{Joint Tests with GW and Electromagnetic Observations}

Since $\Qbar$ is a quantity that is difficult to measure with GW observations, let us consider model-independent and EoS-independent tests of GR with the I-Love relation that uses a combination of GW and double binary pulsar observations. Let us then assume that $\Ibar$ has been measured to $10\%$ by future double binary pulsar (J0737-3039) observations~\cite{lattimer-schutz,kramer-wex}, and that $\lambdabartid$ has been measured to $40\%$ with future GW observations. The latter assumes an ET detection of an equal-mass, non-spinning NS/NS binary at 3Gpc, with the individual NS masses exactly equal to that of the primary pulsar in J0737-3039, $M_\NS=1.3382M_\odot$, assuming the Shen EoS\footnote{Adv.~LIGO is expected to detect NS/NS binaries out to $D_L \approx 300$Mpc with the detection rate of $\mathcal{O}(10)$/yr~\cite{abadie}. Therefore, if we consider ET detecting GW signals from a NS/NS binary at 3Gpc, the expected detection rate would be $\mathcal{O}(10)/\mrm{yr}\times 10^3 \sim \mathcal{O}(10^4)$/yr. With this detection rate, we may detect an equal-mass NS/NS binary, with the individual masses very close to that of the primary pulsar of J0737-3039, $M_\NS=1.3382M_\odot$}. All of this is shown in Fig.~\ref{fig:CS}, together with the fiducial measurement of $(\Ibar,\lambdabartid)$ as a big black cross. As shown in that figure, one can constrain modified theories of gravity, such as dynamical CS gravity, by requiring that the I-Love curve crosses the error box.

The test described above has one major problem: the NS mass $m_\mrm{pulsar}$ of the primary pulsar in J0737-3039 and the individual NS masses in the binary system that generated the detected GW will all in principle be different from each other. As explained in Sec.~\ref{sec:parameter-estimation}, the accuracy to which $\bar{\lambda}_{s}$ can be measured assumed that $\bar{Q}_{a}$ and $\bar{\lambda}_{a}$ could be neglected, which holds for a certain range of mass differences $\Delta m$, shown in Fig.~\ref{fig:deltam}. In general, the typical maximum mass difference would need to be $\Delta m = \mathcal{O}(0.1)M_\odot$, assuming observations with $\mrm{SNR} \approx 10$. Given current event rate estimates, one expects to detect NS/NS binaries with such similar masses, and thus, this is not in principle a problem. The test above, however, also requires that $m_\mrm{pulsar} \approx \bar{m}$, since after all the I-Love-Q relations assume one is investigating NSs with the same mass. 

\begin{figure}[htb]
\begin{center}
\includegraphics[width=8.0cm,clip=true]{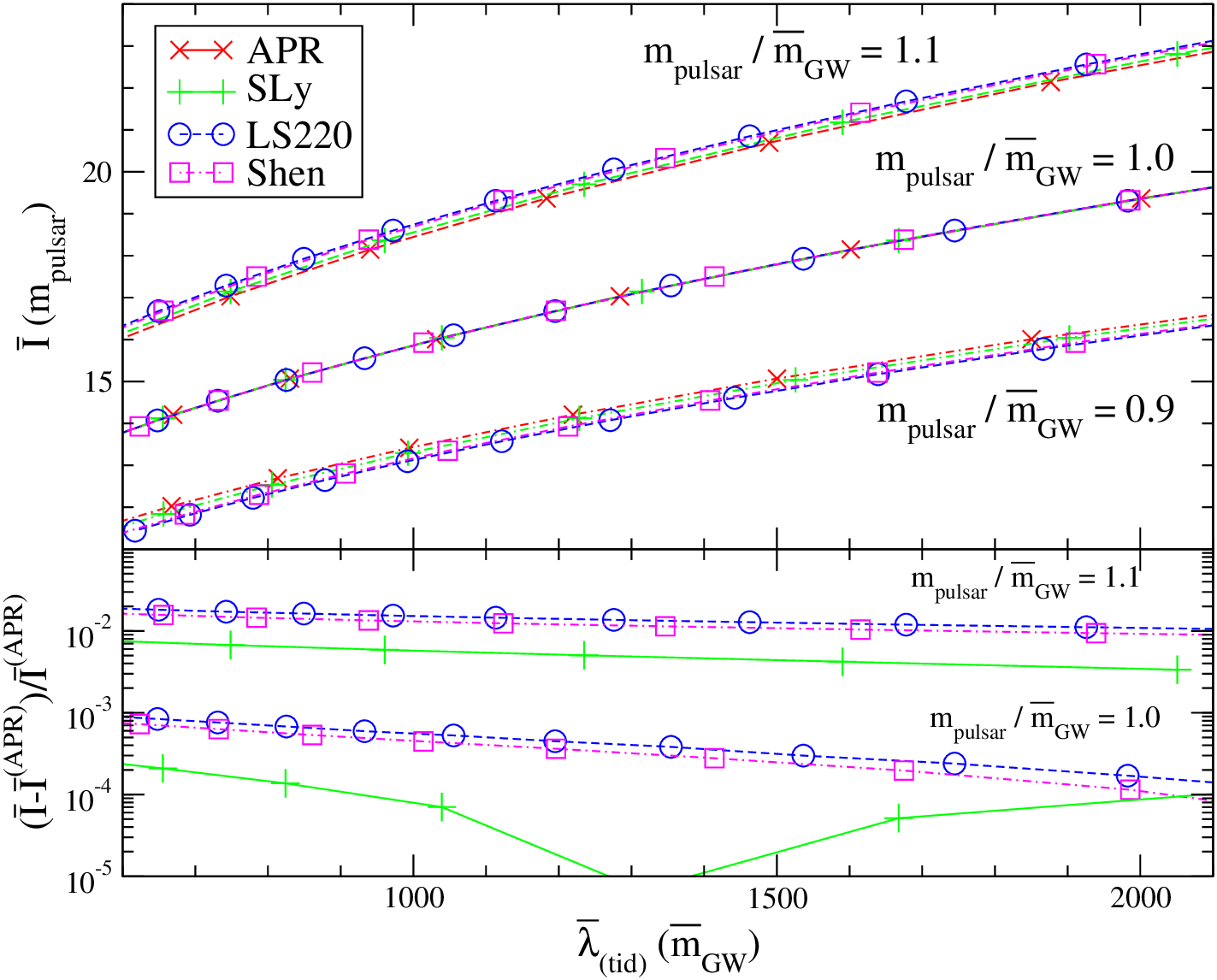}  
\caption{\label{fig:I-Love-massratio} (Color Online) (Top) I-Love relation of NSs with mass ratios of $m_\mrm{pulsar}/\bar{m}_\mrm{GW} = 1$ (solid), 1.1 (dashed) and 0.9 (dotted-dashed)  for realistic EoSs. (Bottom) Relative fractional difference with the APR curve as the reference. Observe that the loss of universality when $m_\mrm{pulsar}/\bar{m}_\mrm{GW} = 0.9$ is similar to that when $m_\mrm{pulsar}/\bar{m}_\mrm{GW} = 1.1$. Observe also that the loss of universality is small relative to the observational error in measuring the moment of inertia or the tidal Love number.}
\end{center}
\end{figure}

Let us then estimate how much the I-Love-Q relations would change if $m_\mrm{pulsar} \neq \bar{m}_\GW$. The top panel of Fig.~\ref{fig:I-Love-massratio} shows the I-Love relation with $m_\mrm{pulsar}/\bar{m}_\GW=0.9$, $1.0$ and $1.1$ for realistic EoSs, while the bottom panel shows the relative fractional difference between the I-Love curves for different EoSs and the APR EoS as a reference. The relative fractional difference when $m_\mrm{pulsar}/\bar{m}_\GW=0.9$ (not shown in this figure) is similar to that of $m_\mrm{pulsar}/\bar{m}_\GW=1.1$. One sees that the dependence on the EoS becomes stronger as the mass difference $m_\mrm{pulsar}$ and $\bar{m}_\GW$ increases. However, this dependence is still weak if the mass difference is sufficiently small (or order $0.1 M_{\odot}$), and most importantly, the loss of universality (the difference between curves with different EoS) is much, much smaller than the observational error in measuring either the moment of inertia or the Love number. Therefore, one can perform the GR test described above, even when $m_\mrm{pulsar} \neq \bar{m}_\GW$.

Of course, the test described here assumes that the uniform and slow-rotation approximation used to derive the I-Love-Q relation holds for binary pulsars. This is indeed the case, provided the period is sufficiently long, such that each binary component is spinning slowly. However, the approximation might break down for (currently unobserved) sub-millisecond pulsars, ie.~those with periods shorter than $1 \; {\rm{ms}}$. For such systems, the I-Love-Q relations will also now depend on the spin frequency. A cursory analysis, however, suggests that the spin-frequency effect breaks universality at the $10\%$ level~\cite{berti-stergioulas,berti-white,benhar,pappas-apostolatos}. Therefore, the difference in I-Love-Q relations for different EoSs will be rather small, and in particular smaller than the errors in the first binary pulsar measurements of the moment of inertia.   

Binary pulsar observations may measure $\Ibar$ within 10\% accuracy, but this amazing measurement will be difficult to accomplish in the \emph{near} future~\cite{kramer-wex}. This is because the effect of the moment of inertia (or equivalently, the spin-orbit coupling) in the motion of the binary is of $\mathcal{O}(10^{-5})$ relative to the leading-order contribution. This means that one needs to measure at least 3 post-Keplerian parameters to this accuracy, in order to determine the two masses and the moment of inertia. Of course, such a measurement is a big challenge, although not out of the question, as binary pulsar observations improve within the next ten years. 

\begin{figure}[htb]
\begin{center}
\begin{tabular}{l}
\includegraphics[width=8.0cm,clip=true]{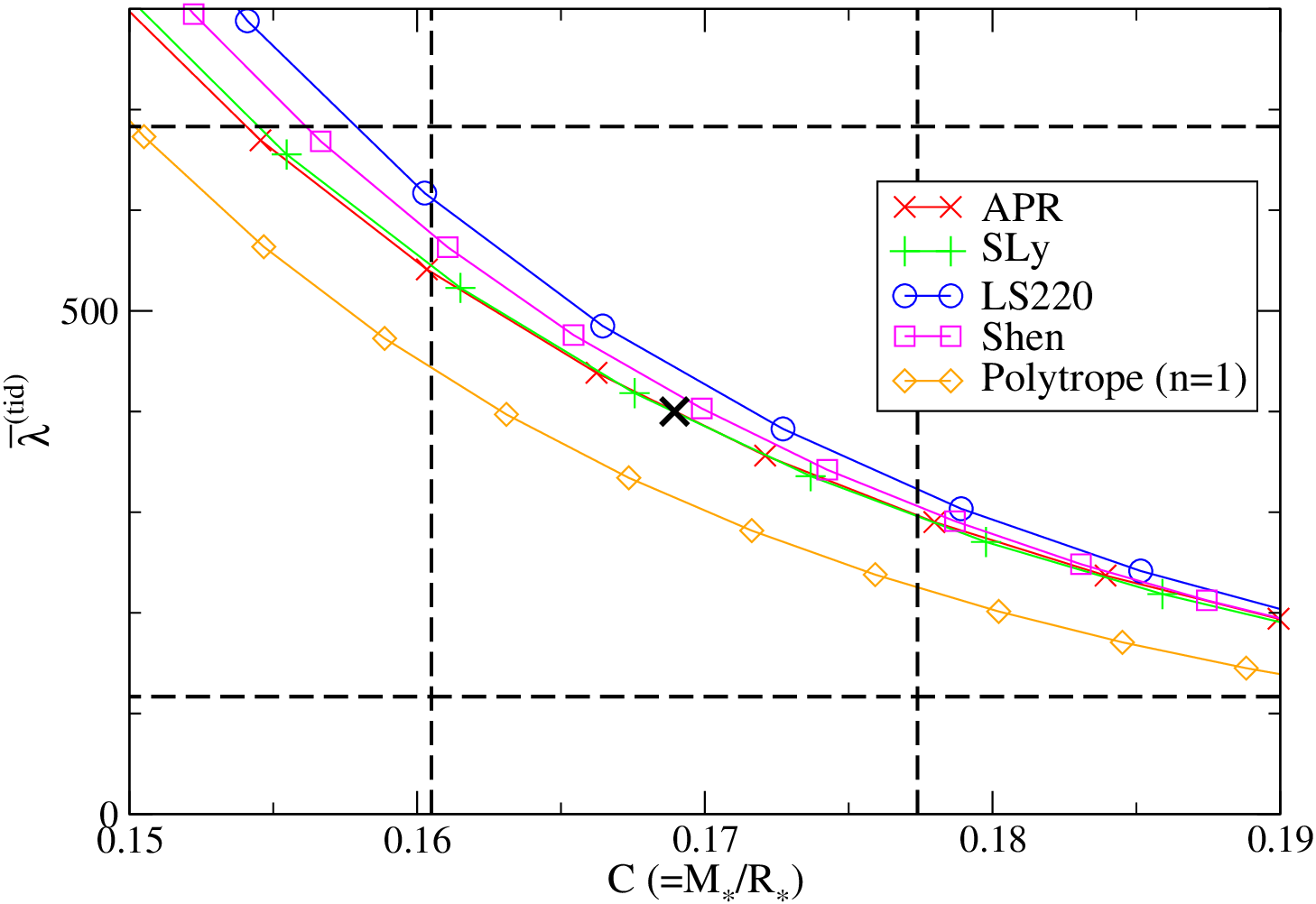}  
\end{tabular}
\caption{\label{fig:Love-C} (Color Online)
Love-C relations with realistic EoSs and the $n=1$ polytrope (solid lines), a fiducial measurement of $(\bar{\lambda}^\mrm{(tid)},C)$ (black cross) and projected measurement uncertainties (dashed black lines). We assume $\Delta C = 0.05$ for a NS with mass 1.4$M_\odot$, and a measurement of $\bar{\lambda}^\mrm{(tid)}$ with a roughly $70\%$ error. Observe that the Love-C relation loses some of the universality shown in the I-Love-Q relations. However, the error introduced due to EoS dependence is much smaller than the measurement error in the compactness or the tidal Love number.}
\end{center}
\end{figure}

Alternatively, one could use the NS compactness $C$ instead of $\Ibar$ to perform model-independent GR tests. Currently, $C$ has been measured to $\mathcal{O}(10)\%$ accuracy with low-mass X-ray binary observations~\cite{steiner-lattimer-brown,ozel-baym-guver,ozel-review,guver}. Figure~\ref{fig:Love-C} shows the $\lambdabartid$--$C$ relation for realistic EoSs and for the $n=1$ polytrope. We also show in this figure a fiducial measurement of $(C,\lambdabartid)$ (big black cross), as well as projected measurement accuracies (dashed lines). For the latter, we assume $\Delta C =0.05$ from electromagnetic observations of a NS with $M_\NS=1.4 M_\odot$ and a roughly $70\%$ measurement of $\lambdabartid$ from GW detection (equal-mass, non-spinning NS/NS binary with the individual NS mass of $M_\NS = 1.4M_\odot$ at $D_L=100$Mpc) with second-generation, ground-based detectors\footnote{The measurement error of $\Delta \lambdabartid \approx 0.7$ is slightly better than that shown in Fig.~\ref{fig:QLoveerror}. This is because we assumed parameter set A [Eq.~\eqref{parameter-A}] instead of B [Eq~\eqref{parameter-B}].}. Although the dependence on the EoS is relatively large compared to the universality of the I-Love-Q relations, the measurement errors are {\emph{larger}} than the uncertainties due to the EoS. This shows that one might be able to use the Love-C relation to perform model-independent tests of gravity.

\subsubsection{Example: Dynamical CS Gravity}

We now apply the results obtained above to see how testing a specific theory of gravity would go about. As an example, we choose dynamical CS gravity~\cite{jackiw,CSreview}, which is well-motivated from the Standard Model, superstring theory~\cite{polchinski1,polchinski2}, loop quantum gravity~\cite{alexandergates,taveras,calcagni} and inflation~\cite{weinberg-CS}. Dynamical CS gravity is a parity-violating, quadratic-curvature theory, where the Einstein-Hilbert action is modified through the Pontryagin density (the contraction of the Riemann tensor and its dual), coupled to a dynamical scalar field. This theory has a characteristic length $\xi^{1/4}$, which has been constrained by Solar System experiments, using Gravity Probe B~\cite{GPB} and LAGEOS~\cite{LAGEOS}, to $\xi^{1/4} < \mathcal{O}(10^8)$km~\cite{alihaimoud-chen}. Dynamical CS gravity should be treated as an {\emph{effective}} theory, and thus, one should work to leading-order in a small coupling expansion, ie.~to leading order in the dimensionless coupling constant $\zeta \equiv \xi M_\NS^2/\RNS^6$~\cite{kent-CSNS}.

NSs in dynamical CS gravity have been studied before. Reference~\cite{kent-CSNS} found that it would be difficult to meaningfully constrain this theory with binary pulsar observations in the standard fashion. This is because the largest CS correction appears in the rate of change of periastron advance at 1PN order, and thus it is suppressed by the ratio of the binary's mass to its separation (for J0737-3039, this is of ${\cal{O}}(10^{-6})$). The CS correction to the NS moment of inertia
was calculated in~\cite{yunespsaltis,alihaimoud-chen}, while the CS correction to the NS quadrupole moment was obtained in~\cite{kent-CSNS}. In the small coupling approximation, the CS corrections to $\bar{I}$ and $\Qbar$ scale linearly with $\zeta$. The leading-order CS correction to tidal effects enters through the gravitomagnetic tidal tensor (because of the parity)~\cite{quadratic}, and hence $\bar{\lambda}^\tid$ is not affected at leading-order.

Figure~\ref{fig:CS} shows the I-Love relation in dynamical CS gravity with a fixed value of the coupling constant $\xi = 1.85 \times 10^4 M_\NS^4$. Observe that the dependence on the EoS is stronger than that of the GR I-Love relation. We believe that this is because a compact object in dynamical CS gravity depends on the scalar-dipole charge, which encodes information on the internal structure of the body~\cite{kent-CSNS}. Given this reasoning, we expect that the I-Love-Q relations should be more sensitive to the NSs' internal structure in dynamical CS gravity than in GR. The bottom panel of Fig.~\ref{fig:CS} shows that there is indeed a loss of universality, but the latter is still preserved to a few \% level. 

With this in hand, let us estimate the projected bound that one could place on dynamical CS gravity using the I-Love relation. With $\xi=1.85 \times 10^4 M_\NS^4$, the I-Love curve in dynamical CS gravity barely crosses the error box. Since the larger $\xi$, the higher the CS curves, such an I-Love observation would automatically constrain $\xi < 1.85 \times 10^{4} M_{\NS}^{4}$, which corresponds to $\zeta = 0.0977$. Converting back to dimensional quantities, such a test would impose the constraint 
\be
\xi^{1/4} < \mathcal{O}(50) \mrm{km}\,.
\label{bound}
\ee
NS observations would then allow us to probe the theory within NS length scales, like the NS radius. Notice that the above bound is stronger than Solar System~\cite{alihaimoud-chen} and table-top~\cite{kent-CSBH} ones by more than six orders of magnitude. Notice, however, that this bound is slightly weaker than the proposed projected bound with GW observations of BH/BH binaries~\cite{kent-CSGW}. This is because the ``radius'' of a BH is smaller than that of a typical NS, and thus, with the former, we can probe shorter length scales. Notice also that the bound given above is dominated by the measurement error on the NS moment of inertia. This is reasonable because the tidal Love number is unaffected in dynamical CS gravity to the order of approximation considered here.

The measurement accuracy shown in Fig.~\ref{fig:CS} is obtained by assuming that the Shen EoS is the correct one, which gives the weakest bound on the theory among the realistic EoSs considered in this paper. This is because, with the NS masses fixed to $M_\NS=1.3382M_\odot$, the Shen EoS gives the largest $\bar{\lambda}^\tid$ (see Table~\ref{table:parameters}). Figure~\ref{fig:CS} shows that the deviation away from GR becomes larger for smaller $\bar{\lambda}^\tid$. This is because the compactness becomes larger for smaller $\bar{\lambda}^\tid$ (or smaller $\bar{I}$ and $\Qbar$) which allows us to probe stronger gravity. These studies suggest that the I-Love-Q relations can be very powerful in testing GR in the strong-field regime. 

\section{Future Directions}
\label{sec:conclusions}

We have derived relations between the moment of inertia, the quadrupole moment and the Love numbers, I-Love-Q relation, that are essentially independent of the EoS for uniformly and slowly-rotating NSs. These relations open the door to exciting applications in astrophysics, GW theory and fundamental physics. We have here carried out a preliminary study of a few applications, but our paper enables a lot more work. One example is a more detailed study of the measurement accuracy of binary parameters given a GW observation. We here carried out a Fisher analysis, but this is known to be inaccurate for signals with low SNR~\cite{vallisneri-fisher}, as initially expected with second-generation, ground-based detectors. Such an analysis could be improved on through a Bayesian study~\cite{vallisneri-fisher,cornish-PPE}. Another example is to repeat the analysis that uses the I-Love-Q relations to test GR to include systems with different mass ratios. This extension is particularly important, given that millisecond binary pulsars will probably not have exactly the same mass as the NSs observed through GWs emitted in the late inspiral. Our results suggest that the conclusions regarding tests of GR should be robust even when the masses differ by $10\%$, but a more detailed analysis would be desirable.

The analysis presented here has a few caveats that should be re-iterated here for completeness, although we have discussed this to some extent already in the Introduction. The framework in which the I-Love-Q relations have been found to be essentially EoS independent is one that employs a uniform rotation, slow-rotation and small tidal deformation approximation. Newly-born NSs are expected to be differentially rotating at very short periods, where the slow-rotation approximation would not be appropriate. Moreover, such NSs are expected to be much hotter than those in millisecond binary pulsars and those that emit GWs in the band of ground-based detectors. Temperature could introduce further deviations in the universality relations described here. 

Of these limitations, the slow-rotation approximation is perhaps the most severe, although such an approximation is reasonable for NSs in millisecond pulsars with periods comparable or larger than $1$ ms. We expect the EoS universality found here to persist even when including fast rotation, except that now there will be different universal relations for stars with different spin periods. The variation, however, should not exceed $10\%$~\cite{berti-stergioulas,berti-white,benhar,pappas-apostolatos}. A possible extension of this work would be to refine the universal relations to allow for rapidly spinning NSs. This could be achieved by numerically solving for rapidly rotating NSs and then extracting its multipole moments, as recently discussed in~\cite{berti-stergioulas,berti-white,benhar,pappas-apostolatos}. Let us reiterate, however, that for almost all NS that have been astrophysically observed, the spin period is sufficiently long that the slow-rotation expansion is an excellent approximation. From an academic point of view, however, it is also worth studying how differential rotation~\cite{chirenti} would change the I-Love-Q relations and their universality.

Another possible extension of our work would be to consider more generic NSs with anisotropic pressure~\cite{doneva} and large internal magnetic fields. Recent work has suggested that in fact the NS interior might be super-conducting and super-fluid (see e.g. Refs.~\cite{link1,link2} and references therein). The inclusion of these effects will certainly affect the EoS, but it is not clear that this will modify the I-Love-Q relations presented here. This would be particularly so if the I-Love-Q relations are truly only sensitive to the EoS far from the core, where super-fluidity and super-conductivity play a small role. 

Although we have investigated how the I-Love-Q relations change in dynamical CS gravity, it would be worthwhile to study such relations in other modified theories, such as scalar-tensor ones~\cite{fujii} and Einstein-\AE ther theory~\cite{AE-review}. Given any modified theory, one could investigate how the I-Love-Q relations change, whether universality still holds, and how strong one can constrain other theories with future observations. 

An interesting avenue to pursue would be to study whether universal relations exist between higher-order, multipole moments of the exterior gravitational field of isolated NSs. For BHs, the no-hair theorems guarantee that BH multipole moments can be written entirely in terms of the BH mass and spin angular momentum (assuming the charge is zero), leading to a unique relation that, of course, is independent of internal structure (BHs lack any). For NSs, such a relation does not exist, since the no-hair theorem does not apply. In this paper, however, we have found an interesting relation between the quadrupole and the dipole moment of the exterior gravitational field of an isolated NS that seems almost independent of the NS's internal structure. One might then naturally wonder whether similar relations hold for higher-order multipole moments, which may lead to a \emph{NS no-hair conjecture}, ie.~that NS multipole moments can be effectively expressed only in terms of the NS mass $M_\NS$, the NS angular velocity $\Omega_\NS$ and the NS moment of inertia $I$.

Finally, one could also investigate whether there are other universal relations between other NS quantities. Recent work has shown that there is indeed a relation between the f- and w-modes of NS oscillations~\cite{andersson-kokkotas,tsui-leung,lau}. One cannot help from asking whether these relations may also be related to the moment of inertia, quadrupole moment or Love number, thus yielding an I-Love-Q-f-w set of universal relations\footnote{A universal I-f relation has been shown to hold in~\cite{lau}.}. If so, one could also investigate whether these new quantities provide further insight into the fundamental reason for the existence of these universal relations. 

\acknowledgments
We would like to thank Eric Poisson, Luc Blanchet, Takahiro Tanaka, Hiroyuki Nakano and Kenta Hotokezaka for useful discussions in Japan. We would also like to thank Emanuele Berti, Vitor Cardoso, Neil Cornish, Michael Kramer, Lee Lindblom, Feryal $\ddot{\mrm{O}}$zel, Paolo Pani, Eric Poisson, Scott Ransom, Luciano Rezzolla and Masaru Shibata for additional comments on a shorter version of the manuscript and Tanja Hinderer for reading the current manuscript carefully and giving us valuable comments. NY acknowledges support from NSF grant PHY-1114374, as well as support provided by the National Aeronautics and Space Administration from grant NNX11AI49G, under sub-award 00001944. The authors thank the Yukawa Institute for Theoretical Physics at Kyoto University, where this work was initiated during the Long-term Workshop YITP-T-12-03  on ``Gravity and Cosmology 2012''. Some calculations used the computer algebra-systems MAPLE, in combination with the GRTENSORII package~\cite{grtensor}. Other calculations were carried out with the XTENSOR package for MATHEMATICA~\cite{2008CoPhC.179..586M,2009GReGr..41.2415B}.

\bibliography{master}
\end{document}